\documentclass[twoside,twocolumn,9pt]{article}
\usepackage{extsizes}
\usepackage[super,sort&compress,comma]{natbib} 
\usepackage[version=3]{mhchem}
\usepackage[left=1.5cm, right=1.5cm, top=1.785cm, bottom=2.0cm]{geometry}
\usepackage{balance}
\usepackage{mathptmx}
\usepackage{sectsty}
\usepackage{graphicx} 
\usepackage{lastpage}
\usepackage[format=plain,justification=justified,singlelinecheck=false,font={stretch=1.125,small,sf},labelfont=bf,labelsep=space]{caption}
\usepackage{float}
\usepackage{fancyhdr}
\usepackage{fnpos}
\usepackage[english]{babel}
\addto{\captionsenglish}{%

}
\usepackage{array}
\usepackage{droidsans}
\usepackage{charter}
\usepackage[T1]{fontenc}
\usepackage[usenames,dvipsnames]{xcolor}
\usepackage{setspace}
\usepackage[compact]{titlesec}
\usepackage{hyperref}
\usepackage{subfig}

\usepackage{epstopdf}

\definecolor{cream}{RGB}{222,217,201}

\usepackage{xcolor}
\usepackage[separate-uncertainty=true]{siunitx}
\DeclareSIPrefix\micro{\text{\textmu}}{-2}
\usepackage{textcomp}
\usepackage[separate-uncertainty=true]{siunitx}
\DeclareSIUnit{\framepersecond}{fps}
\usepackage{amsmath}
\usepackage{amssymb}  

\begin{document}

\pagestyle{fancy}
\thispagestyle{plain}
\fancypagestyle{plain}{
\renewcommand{\headrulewidth}{0pt}
}

\makeFNbottom
\makeatletter
\renewcommand\LARGE{\@setfontsize\LARGE{15pt}{17}}
\renewcommand\Large{\@setfontsize\Large{12pt}{14}}
\renewcommand\large{\@setfontsize\large{10pt}{12}}
\renewcommand\footnotesize{\@setfontsize\footnotesize{7pt}{10}}
\makeatother

\renewcommand{\thefootnote}{\fnsymbol{footnote}}
\renewcommand\footnoterule{\vspace*{1pt}%
\color{cream}\hrule width 3.5in height 0.4pt \color{black}\vspace*{5pt}} 
\setcounter{secnumdepth}{5}

\makeatletter 
\renewcommand\@biblabel[1]{#1}            
\renewcommand\@makefntext[1]%
{\noindent\makebox[0pt][r]{\@thefnmark\,}#1}
\makeatother 
\renewcommand{\figurename}{\small{Fig.}~}
\sectionfont{\sffamily\Large}
\subsectionfont{\normalsize}
\subsubsectionfont{\bf}
\setstretch{1.125} 
\setlength{\skip\footins}{0.8cm}
\setlength{\footnotesep}{0.25cm}
\setlength{\jot}{10pt}
\titlespacing*{\section}{0pt}{4pt}{4pt}
\titlespacing*{\subsection}{0pt}{15pt}{1pt}

\fancyfoot{}
\fancyfoot[LO,RE]{\vspace{-7.1pt}\includegraphics[height=9pt]{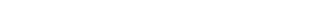}}
\fancyfoot[CO]{\vspace{-7.1pt}\hspace{13.2cm}\includegraphics{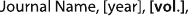}}
\fancyfoot[CE]{\vspace{-7.2pt}\hspace{-14.2cm}\includegraphics{head_foot/RF}}
\fancyfoot[RO]{\footnotesize{\sffamily{1--\pageref{LastPage} ~\textbar  \hspace{2pt}\thepage}}}
\fancyfoot[LE]{\footnotesize{\sffamily{\thepage~\textbar\hspace{3.45cm} 1--\pageref{LastPage}}}}
\fancyhead{}
\renewcommand{\headrulewidth}{0pt} 
\renewcommand{\footrulewidth}{0pt}
\setlength{\arrayrulewidth}{1pt}
\setlength{\columnsep}{6.5mm}
\setlength\bibsep{1pt}

\makeatletter 
\newlength{\figrulesep} 
\setlength{\figrulesep}{0.5\textfloatsep} 

\newcommand{\topfigrule}{\vspace*{-1pt}%
\noindent{\color{cream}\rule[-\figrulesep]{\columnwidth}{1.5pt}} }

\newcommand{\botfigrule}{\vspace*{-2pt}%
\noindent{\color{cream}\rule[\figrulesep]{\columnwidth}{1.5pt}} }

\newcommand{\dblfigrule}{\vspace*{-1pt}%
\noindent{\color{cream}\rule[-\figrulesep]{\textwidth}{1.5pt}} }

\makeatother

\twocolumn[
  \begin{@twocolumnfalse}
{\includegraphics[height=30pt]{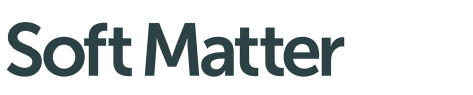}\hfill\raisebox{0pt}[0pt][0pt]{\includegraphics[height=55pt]{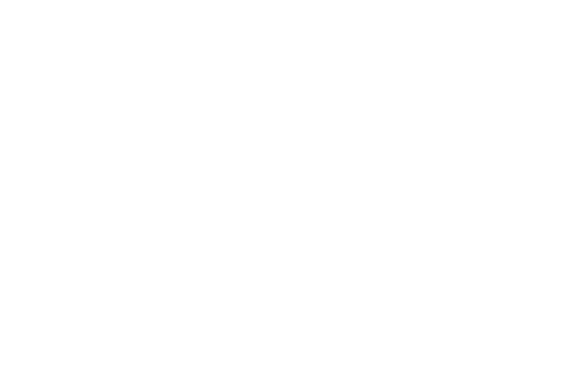}}\\[1ex]
\includegraphics[width=18.5cm]{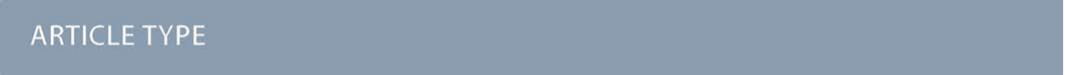}}\par
\vspace{1em}
\sffamily
\begin{tabular}{m{4.5cm} p{13.5cm} }

\includegraphics{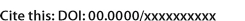} & \noindent\LARGE{\textbf{Dynamic wetting by concentrated granular suspensions$^\dag$}} \\
\vspace{0.3cm} & \vspace{0.3cm} \\

 & \noindent\large{Reza Azizmalayeri,\textit{$^{a}$} Peyman Rostami,\textit{$^{a}$} Thomas Witzmann,\textit{$^{a}$} Christopher~O.~Klein,\textit{$^{b}$} and Günter~K.~Auernhammer$^{\ast}$\textit{$^{a}$}} \\

\includegraphics{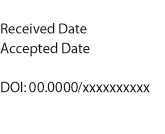} & 
\noindent\normalsize{Many functional materials, such as paints and inks used in applications like coating and 3D printing, are concentrated granular suspensions. 
In such systems, the contact line dynamics and the internal structure of the suspension interact through shear rate dependent viscosity and microstructural rearrangements. 
The local shear rate increases sharply near moving contact lines, leading to the non-Newtonian rheology of dense suspensions in this region. 
While hydrodynamic solutions can describe dilute suspensions, their applicability near advancing contact lines in dense suspensions remains unclear. 
This study quantifies the deviation from the Newtonian solution by systematically varying interparticle interactions through the choice of dispersion medium. 
We use silica particles suspended in two refractive index-matched fluids: (i) aqueous 2,$2^{\prime}$-thiodiethanol (weak interactions) and (ii) aqueous sodium thiocyanate solution (strong interactions). 
These systems exhibit substantially different rheological responses, shear-thickening and yield-stress behaviour, respectively. 
Using astigmatism particle tracking velocimetry (APTV), we resolve the three-dimensional trajectories of tracer particles within a drop driven over a substrate, in an arrangement enabling tracking the internal flows over a long travel distance of the drop. 
We observe distinct flow behaviours depending on the particle interactions and the resulting suspension rheology. 
The more the particle interactions play a role, i.e., the more pronounced the non-Newtonian effects are, the stronger the measured flow profiles differ from the Newtonian solution to the hydrodynamic equations. 
In case of the shear-thickening suspension, a notable deviation from Newtonian behaviour is observed. 
Conversely, the yield-stress suspension exhibits plug flow over the substrate, with Newtonian-like behaviour restricted to the yielded region near the substrate.
}

\end{tabular}

 \end{@twocolumnfalse} \vspace{0.6cm}

  ]

\renewcommand*\rmdefault{bch}\normalfont\upshape
\rmfamily
\section*{}
\vspace{-1cm}


\footnotetext{\textit{$^{a}$~Leibniz Institute of Polymer Research Dresden (IPF), Institute of Physical Chemistry and Polymer Physics, 01069 Dresden, Germany}}
\footnotetext{\textit{$^{b}$~Karlsruhe Institute of Technology, Institute of Chemical Technology and Polymer Chemistry (ITCP), 76131 Karlsruhe, Germany}}
\footnotetext{${\ast}$~Corresponding author:~auernhammer@ipfdd.de}

\footnotetext{\dag~Electronic Supplementary Information (ESI) available. See DOI: 10.1039/cXsm00000x/}



\section{Introduction}\label{sec:Introduction}

Understanding the dynamic wetting process of complex liquids requires a solid understanding of their rheology, surface tension, and dynamic wetting properties of substrates~\citep{lu2016critical, sudheer2023intertwined}. 
'Dynamic wetting' refers to the physical process in which the three-phase contact line (where the solid, liquid, and gas phases meet) moves across a substrate due to spontaneous or forced mechanisms. 
This motion causes the advancing/receding contact angles to deviate from their static values, resulting in advancing/receding dynamic contact angles ($\theta_{adv/rec}$)~\citep{snoeijer2013moving, roche2024complexity, drelichContactAnglesHistory2020}. 
Granular suspensions are one class of complex fluids of an especially high practical relevance. 
In this context, the term 'granular' pertains to particles that are significantly smaller than the characteristic fluid volume, yet sufficiently large, such that the effects of Brownian motion can be disregarded. 
Consequently, the Peclet number becomes large, indicating that particle motion is primarily controlled by advection due to external forces rather than diffusion. 
Such suspensions are widely employed in both industrial and scientific settings, including slot-die coating~\citep{fuaad2021simulation}, 3D-printing~\citep{m2017linking}, cosmetics~\citep{timm2012investigation}, painting~\citep{karlsson2019characterization}, dip-coating~\citep{jeong2022dip}, and drop impact on solid surfaces~\citep{jorgensen2020deformation}. 
For instance, in paint industry, pigments are added to the paint. 
The rheology must be optimized to prevent pigment sedimentation and enable easy re-dispersion through stirring. 
Additionally, it must display lower viscosity under shear and higher viscosity at rest, i.e., shear-thinning behaviour, facilitating smooth application and adherence to vertical surfaces without excessive dripping.

Hydrodynamic models describe fluid flow near moving contact lines within their range of validity. 
When extrapolated beyond this, solutions to these models are physically unrealistic due to the contradiction between contact line motion and the boundary condition, characterized by a singularity in the velocity gradient and/or pressure~\citep{huh1971hydrodynamic, moffatt1964viscous, Huh_Mason_1977}. 
To address this issue, a small slip length was introduced at the substrate, allowing finite velocity of the liquid on the solid substrate and resolving the velocity gradient singularity by permitting slip~\citep{dussan1979spreading}. 
However, the divergence in the pressure persists~\citep{Huh_Mason_1977}. 
A detailed mathematical analysis of this subject can be found in~\citep{fricke2020boundary}. 
The velocity gradient diverges proportionally to the inverse distance to the contact line~\citep{snoeijer2013moving}. 
Consequently, a high shear rate zone is created. 
Fluids having non-Newtonian properties, such as granular suspensions or polymer solutions, can exhibit even more complex behaviour near the contact line, affecting the distribution of viscous dissipation~\citep{li2023kinetic, grewal2015role, rostami2024spreading}.

\begin{figure}[t]
  \centerline{\includegraphics[width=0.45\textwidth]{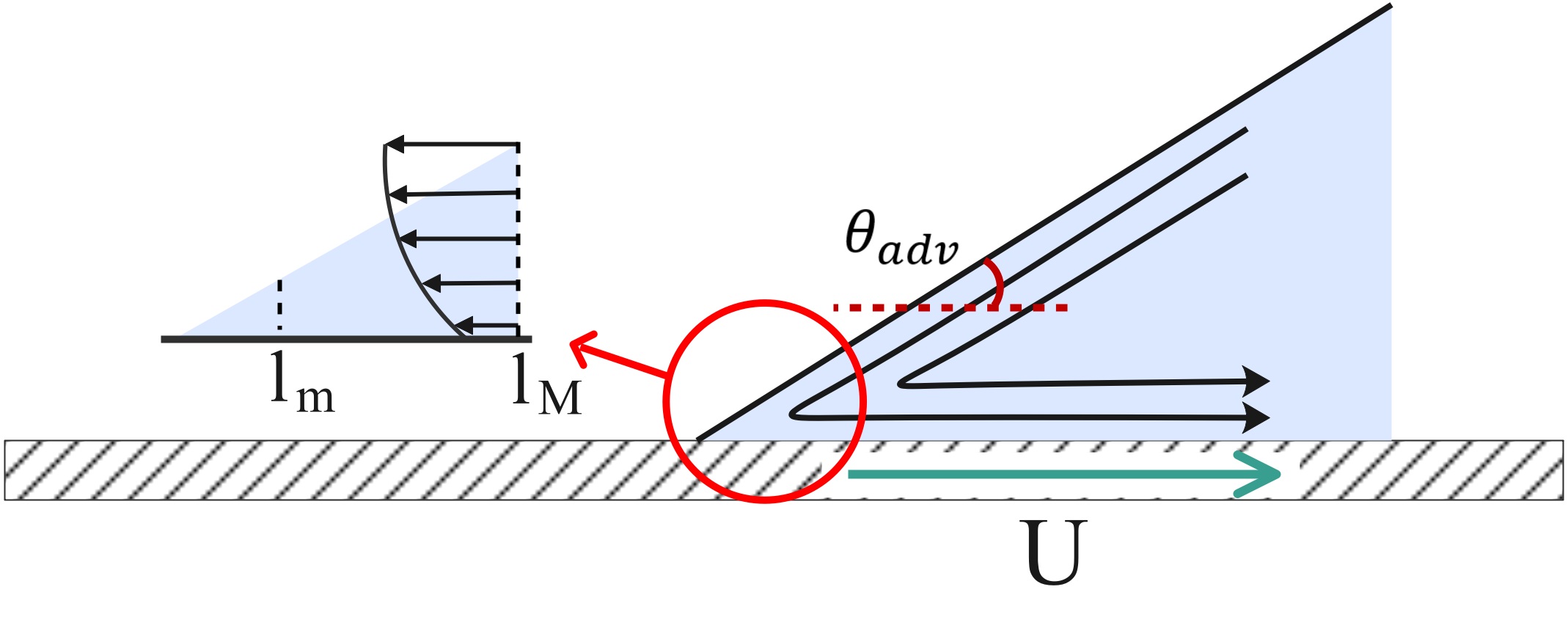}}
  \caption{Schematic representation of a droplet's advancing contact line and the lubrication flow near it.
The Stokes equation is solved within the lubrication approximation, following the individual works of Cox~\citep{cox1986dynamics} and Voinov~\citep{voinov1976hydrodynamics}.
  \label{fig:Schematic Cox-Voinov} }
\end{figure}

The classical hydrodynamic description of moving contact line dynamics involves the relationship between the dynamic contact angle and the velocity of the moving contact line. 
The Cox-Voinov relation (equation~\ref{my_first_eqn}) is generally used to predict the relationship between the dynamic advancing or receding contact angle $\theta_{adv/rec}$ and capillary number $Ca$ near the contact line~\citep{cox1986dynamics, voinov1976hydrodynamics, hocking1983spreading}. 
\begin{equation}\label{my_first_eqn}
\theta_{adv/rec}^3=\theta_{\;0,\;adv/rec}^3 \pm \,9 \mathbin {Ca} \log \left(\frac{\alpha \ell_M }{\ell_m}\right)
\end{equation}
The capillary number $Ca$ is defined as the ratio of viscous and capillary forces $Ca = \eta U / \gamma$, where $\eta$, $\gamma$, and $U$ are viscosity, surface tension, and contact line velocity, respectively. 
$\theta_{adv/rec}$ is the dynamic advancing or receding contact angle, $\theta_{\;0,\;adv/rec}$ corresponds to the static advancing or receding contact angle~\citep{butt2022contact}. 
Additionally, $\ell_M$ is established as the macroscopic length, encompassing the capillary length or the dimensions of a spreading droplet, whereas $\ell_m$ pertains to a microscopic length. 
The numerical constant $\alpha$ assumes a non-universal nature and depends on the intricacies within the microscopic and macroscopic boundary conditions. 
However, most substrates are not ideally smooth. 
The hydrodynamic model also summarizes microscopic effects, such as pinning on small length scales, using the logarithm of the microscopic term $\left(\alpha \ell_M / \ell_m\right)$ in equation~\ref{my_first_eqn}~\citep{henrich2016influence, fricke2020boundary}. 
The additional length scale arising from suspended particles lies between the microscopic $\ell_m$ and macroscopic $\ell_M$ length scales. 
A schematic representation of the Cox-Voinov solution near the advancing contact line is provided in Fig.~\ref{fig:Schematic Cox-Voinov}.

Beyond the continuum-level hydrodynamic description, the behaviour of dense suspensions is strongly influenced by the microscopic contact forces between individual particles. 
Granular particles experience various types of friction during suspension deformation, similar to their millimeter-scale counterparts~\citep{zhao2018rolling, santos2020granular}. 
These include sliding, rolling, and twisting motions, each associated with distinct frictional behaviours that have been documented at the single-particle level~\citep{Tomas:2004aa, Wenzl:2013aa}. 
Surface roughness contributes to resistance upon contact. 
Additional resistance can result from deformation of particle surfaces during interaction~\citep{fuchs2014rolling}. 
In addition to interparticle friction, normal forces arise through the mechanisms described by DLVO theory, equation~\ref{my_second_eqn}: 
Van der Waals (VdW) forces and electrostatic forces.
\begin{equation}\label{my_second_eqn}
V_\text{Total}=V_{VdW}+V_\text{Elect}
\end{equation}
The magnitude of VdW interactions is strongly influenced by the properties of the dispersing medium, which can either diminish or augment their effects. 
The Van der Waals attraction between two identical spheres of radius $R$, separated by a distance $h$, is proportional to the Hamaker constant ($A_H$). 
$A_H$ can be approximated by considering the differences in dielectric constants ($\varepsilon$) and refractive indices ($n$) between the particles and the dispersing medium, ignoring their frequency dependence ~\citep{butt2018surface, RANDALL20012733}. 
When the refractive indices of the particles and the surrounding medium are identical, $A_H$ becomes very small and so does the VdW interaction~\citep{yadav2007effective, wozniak2009highly, wozniak2011rheology, hopkins2014dielectric}.

Rheological measurements provide critical insight into how suspensions respond to applied shear forces. 
At low particle concentrations, suspensions exhibit Newtonian-like behaviour, with viscosity increasing as a function of volume fraction ($\phi$), consistent with the classical expressions of Einstein~\citep{einsteinNeueBestimmungMolekuldimensionen1906, einsteinBerichtigungMeinerArbeit1911} and Batchelor \& Green~\citep{batchelorDeterminationBulkStress1972, batchelorHydrodynamicInteractionTwo1972}. 
However, as the particle concentration approaches the dense regime, hydrodynamic interactions and interparticle contact forces become increasingly important~\citep{setoNormalStressDifferences2018}. 
These interactions give rise to pronounced non-Newtonian behaviours. 
Comprehensive reviews~\citep{guazzelli2018rheology, morris2020toward} outline the current understanding of these complex flow behaviours. 

Plotting the steady‐shear viscosity versus particle volume fraction reveals a characteristic divergence: as $\phi$ approaches the critical volume fraction $\phi_c$, the viscosity diverges as $\eta \sim (1 - \phi/\phi_c)^{-\beta}$. 
Numerous phenomenological expressions exist to describe this divergence. 
For example, the classical empirical model by Maron–Pierce assumes a fixed $\beta = 2$~\citep{maronApplicationReeeyringGeneralized1956}. 
Microscopically, this divergence corresponds to a transition from lubrication-mediated interactions to a frictional contact-dominated network that arrests particle motion~\citep{setoShearJammingFragility2019, Liu:1998aa}. 
The exact value of $\phi_{c}$ depends on particle shape, size distribution, and surface roughness~\citep{guazzelli2018rheology}. 
Near jamming, momentum transfer is dominated by particle contacts rather than by the viscous fluid, making the mechanics of these dense suspensions closely analogous to dry granular media~\citep{Liu:1998aa, trappe2001jamming}.

At a fixed particle concentration below $\phi_{c}$, as the shear rate increases, dense suspensions can undergo continuous or discontinuous shear-thickening, marked by a transition from fluid-like to solid-like behaviour. 
It typically begins with a first regime of Newtonian-like or slightly shear-thinning behaviour. 
This is followed by a second regime of shear-thickening, where viscosity rises sharply and stabilizes at a substantially elevated value~\citep{seto2013discontinuous, guyUnifiedDescriptionRheology2015, morrisShearThickeningConcentrated2020a, malbrancheScalingAnalysisShear2022}.

In oscillatory measurements, the storage modulus ($G^{\prime}$) represents the in-phase (elastic) component of the material response, corresponding to the energy reversibly stored during cyclic deformation. 
Conversely, the loss modulus ($G^{\prime \prime}$) corresponds to the out-of-phase (viscous) component, quantifying the energy dissipated as heat through internal friction as particles and fluid elements experience relative motion~\citep{mezger2020rheology}. 
In hard-sphere suspensions, $G^{\prime}$ arises from transient particle cages or force-bearing contacts. 
However, such elastic behaviour is significant only at high volume fractions and within small strain amplitudes that preserve the particle structure. 
At larger strains or lower concentrations, cage structures yield, and the rheology is dominated by viscous dissipation rather than elastic storage. 
Although yield-stress is commonly defined as the threshold below which a material does not flow, this transition is often complex and not captured by a single stress value~\citep{otsuki2018coupling, li2020numerical, ahujaTwoStepYielding2020}. 
The nature of these contacts, including their strength and geometry, can affect the transmission of forces between these force-carrying particles. 
The dynamic yield-stress ($\sigma_y^d$) refers to the constant steady-state stress observed at low shear rates; below this, some slow creeping motion might still be possible, but the corresponding viscosities can become extremely large.

In summary, considering suspensions as a combination of particles and carrier liquid rather than a single phase provides us with the complete picture to explain the non-Newtonian phenomena observed~\citep{morrisReviewMicrostructureConcentrated2009}. 
Other studies examining the correlation between rheology and suspension microstructure suggest that merely considering viscosity as a function of volume fraction is insufficient to fully understand suspension behaviour~\citep{edensShearStressDependence2021, singhStressactivatedConstraintsDense2022, 10.1063/5.0153614}. 
Particularly at higher particle volume fractions, it's crucial to account for shear-induced structures within the suspension.

Addition of particles to the droplet introduces the new length scale of the particle diameter into the dynamic wetting problem. 
The confined geometry due to the contact angle limits the accessibility of this region for the suspension particles~\citep{zhao2020spreading}. 
The visco-capillary region denotes the microscale zone adjacent to the advancing contact line, where viscous and capillary stresses govern the dynamics, and gravitational effects are negligible. 
Within a range of a couple of particle diameters, there is an interplay between the wetting dynamics and the non-Newtonian rheological behaviour of suspensions~\citep{pelosse2023probing}. 
Recent studies demonstrated that for particles entering the visco-capillary region at moderate concentrations (volume fraction $\phi=0.4$), the advancing contact line dynamics of the suspension is altered in response to variations in the viscosity~\citep{zhao2020spreading}. 
Still, the Cox-Voinov equation can be used to predict $\theta_{adv}^3$ in terms of capillary number by taking the effective viscosity of the suspension into account~\citep{pelosse2023probing}. 

Most studies of receding suspension contact lines focus on dip-coating, where a substrate is immersed in or withdrawn from a particle-laden pool. 
If the particles are smaller than the fluid film thickness, they become entrained, with entrainment governed by withdrawal velocity, particle concentration, and the receding contact angle. 
As velocity increases, the system transitions from no entrainment to a particle monolayer and eventually to thick films~\citep{jeong2022dip, gans2019dip, palma2019dip}. 
There is a stagnation point on the liquid-air interface near the receding contact line, separating shear flow into the film from recirculating flow into the drop bulk, thereby favoring particle accumulation at the meniscus~\citep{colosquiHydrodynamicallyDrivenColloidal2013}. 
When particles cluster together, their collective hydrodynamic drag and interfacial deformation enable them to overcome capillary barriers and penetrate into the film~\citep{colosquiHydrodynamicallyDrivenColloidal2013}.

Despite extensive research efforts, the complex wetting dynamics of dense granular suspensions remains poorly understood. 
The local rheology of these suspensions interacts with the shear stress distribution inside the drop, particularly within the high-shear region near the contact line. 
In this study, we investigate the dynamic wetting behaviour of two carefully designed suspension systems with contrasting rheological properties, achieved by varying the properties of the dispersion medium. 
One system exhibits viscous and shear-thickening behaviour, while the other displays yield-stress and shear-thinning characteristics. 
We systematically compare the resulting flow profiles in moving drops to classical hydrodynamic solutions. 
While the Cox–Voinov relation is a standard reference in the literature, our analysis of measured flow profiles highlights the deviations from such hydrodynamic models when applied to dense granular suspensions. 
We probe the coupling between contact line motion and rheology by tracking the three-dimensional trajectories of tracer particles near the advancing contact line. 
The trajectories are resolved using astigmatism particle tracking velocimetry (APTV). 
To support our interpretation of these flow fields, we perform a detailed rheological characterization of the samples.

\section{Materials, methods and sample characterization}\label{sec:Materials and methods}
\subsection{Sample preparation and characterization} \label{subsection: sample preparation and characterization}

\begin{figure}[t]
    \centering{\includegraphics[width=0.25\textwidth]{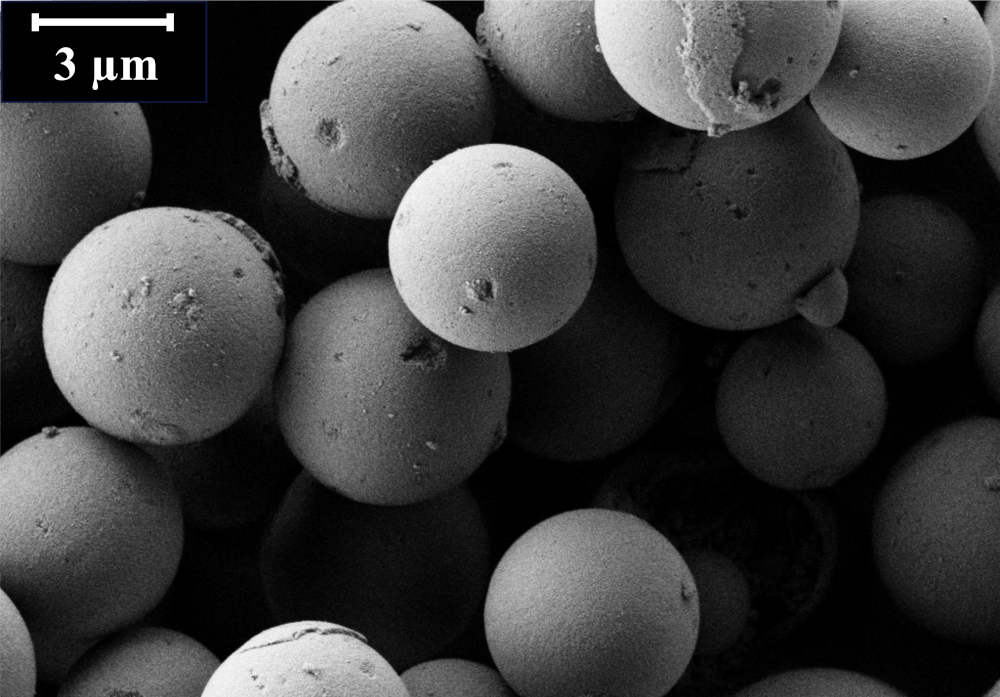}}
    \caption{Scanning electron microscopy image of spherical silica particles with a diameter of \SI{5}{\micro\metre}, used in the preparation of the suspensions.
    \label{fig:Particles}}
\end{figure}

%
For the APTV measurements, it is necessary to ensure that the refractive index of the dispersion medium is matched to that of the dispersed particles. 
This optical matching renders the silica particles invisible, allowing the analysis to focus on the fluorescently labelled tracer particles introduced into the system. 
The suspensions were prepared using \SI{5}{\micro\meter} diameter  silica particles (Orbit 100 Sil \SI{5}{\micro\metre}, MZ0930-BULK, MZ-Analysentechnik GmbH, Germany), Fig.~\ref{fig:Particles}. 
Monodisperse polystyrene \SI{4.47}{\micro\meter} diameter particles, red-fluorescent (PS-FluoRed): Ex/Em \SI{530}{\nano\meter}/\SI{607}{\nano\meter} (microParticles GmbH, Germany) have been used as tracer particles. 
The tracer particle concentration was around 0.0001 wt.\%, a negligible value at which any influence on the suspension’s hydrodynamics can be safely excluded. 
Two distinct methods were employed to achieve refractive index ($RI$) matching. 
Either a solution of 2,$2^{\prime}$-thiodiethanol, 99\% (TDE, Sigma-Aldrich) or a solution of sodium thiocyanate, 99\% (NaSCN, Sigma-Aldrich) is used. 
The specifications of the dispersion media are summarized in Table~\ref{tab:RI solutions}.

\begin{table*}[t]
\small
  \caption{Specifications of high refractive index solutions used as dispersion media}
  \label{tab:RI solutions}
  \begingroup
  \renewcommand{\arraystretch}{1.2}  
  \begin{tabular*}{\textwidth}{@{\extracolsep{\fill}}llllll}
    \noalign{\vspace{0.5em}}  
    \hline
    \noalign{\vspace{0.5em}}  
    Notation of the model fluid & Composition & $\rho\ (\si{\gram\per\cubic\centi\metre})$ & $\gamma\ (\si{\milli\newton\per\metre})$ & $\eta\ (\si{\milli\pascal\second})$ & RI\\
    \noalign{\vspace{0.5em}}  
    \hline
    \noalign{\vspace{0.5em}}  
    2,$2^{\prime}$-thiodiethanol (TDE) solution   & 69.75~wt.\% in water                & 1.14 & 56 & $8.3 \pm 0.1$ & 1.457 \\
    sodium thiocyanate (NaSCN) solution           & \SI{11.65}{\mol \per \liter} in water & 1.27 & 79 & $2.8 \pm 0.1$ & 1.457 \\
    \noalign{\vspace{0.5em}}  
    \hline
    \noalign{\vspace{0.5em}}  
  \end{tabular*}
  \endgroup
\end{table*}

\begin{figure*}[t]
    \centering
    \subfloat[\centering]
    {{\includegraphics[scale=0.40]{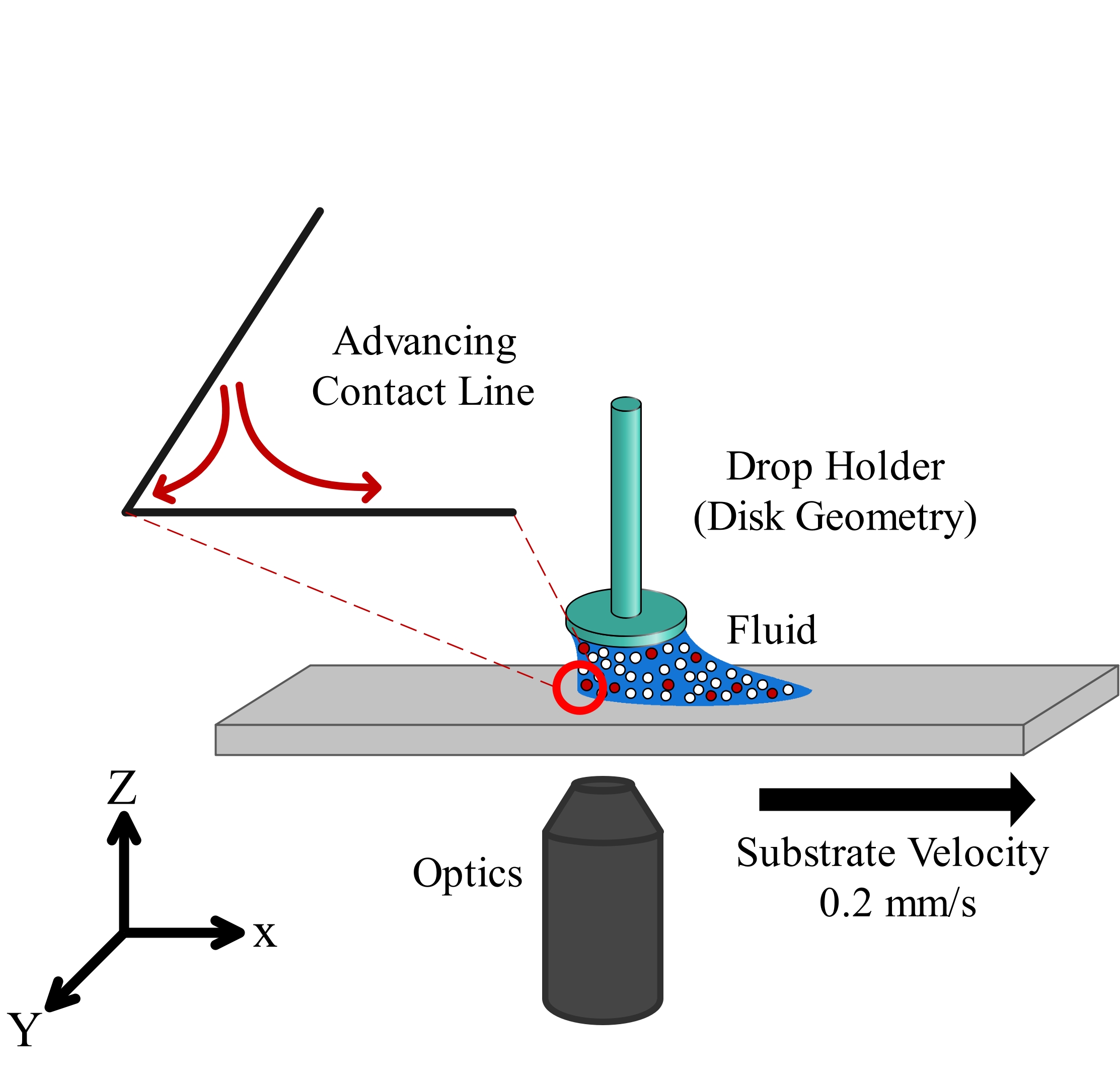}}}
    \qquad
    \subfloat[\centering]
    {{\includegraphics[scale=0.40]{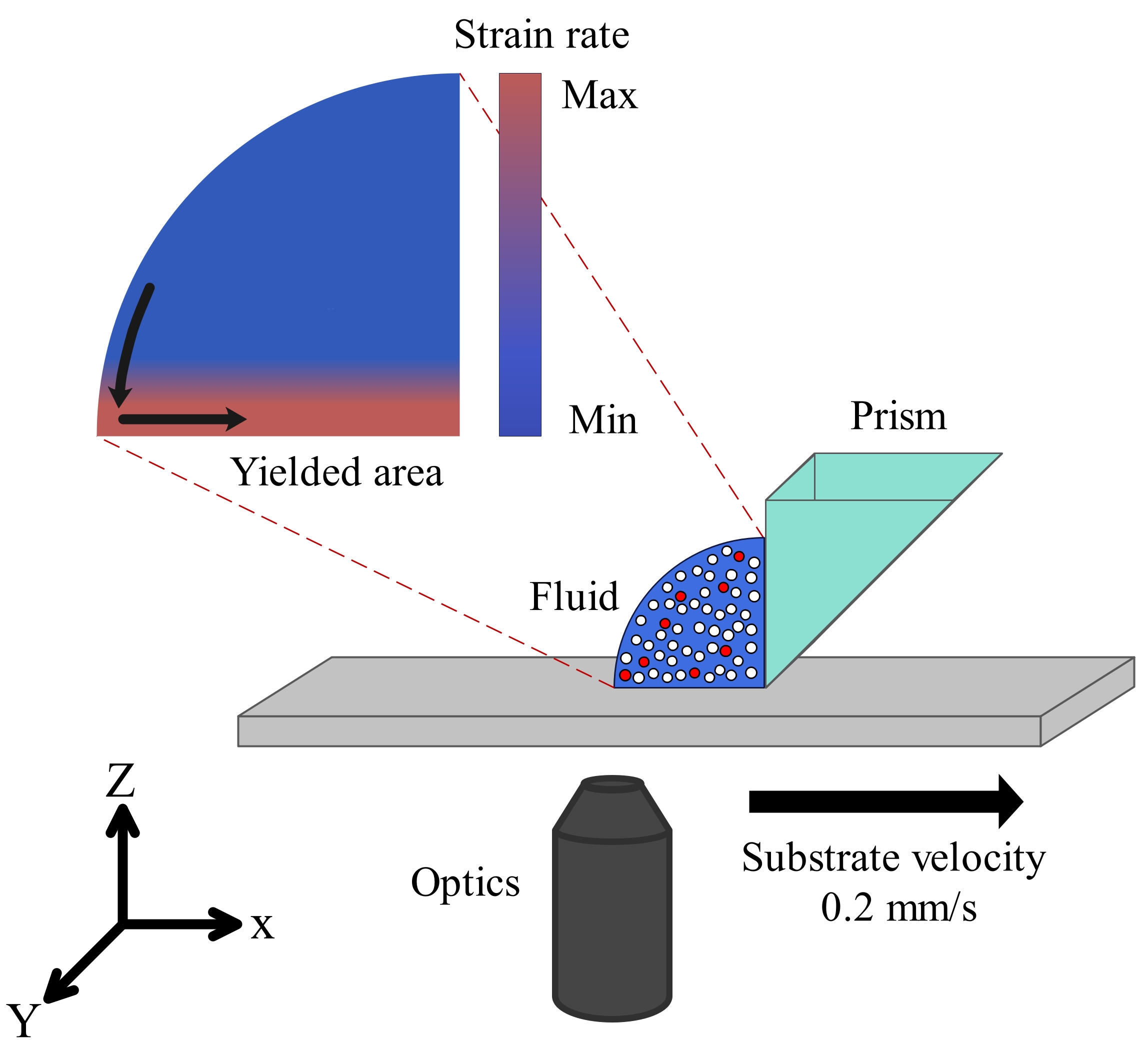}}}
    \caption{Schematic representation of the DropSlider configuration mounted on an inverted microscope for APTV measurements. 
    Two geometries are used: (\textit{a}) disk geometry for TDE-based suspensions, and (\textit{b}) prism geometry for NaSCN-based suspensions.}
    \label{fig:Geometries}
\end{figure*}

%
The density of each solution is determined using a \SI{25}{\milli \liter} Gay-Lussac pattern pycnometer made of borosilicate glass (BRAND\textregistered BLAUBRAND\textregistered). 
The surface tensions measured using OCA 35L (DataPhysics Instruments GmbH, Germany) and all the refractive index measurements were conducted at the D-Line wavelength of \SI{589.29}{\nano\meter} using a spectral-refractometer (SCHMIDT + HAENSCH GmbH, Germany). 
The comprehensive details of the refractive index matching approach can be found in other research papers~\citep{wiederseiner2011refractive, Wenzl:2013aa, wright2017review, auernhammer2020transparent}. 
Microscopy cover slides (thickness \SI{170}{\micro \metre}) were cleaned by sonicating in water, acetone, and ethanol, then left in a vacuum oven to dry for 12 hours. 
There was no additional coating applied to the substrates to keep the (dynamic) contact angles below \SI{90}{\degree}, the ideal range for APTV measurements near the contact line.

The suspensions were prepared by weighing the fluid into glass bottles, then adding the required particle mass to achieve the target weight fraction. 
Using hand mixing followed by gentle mixing on a rolling device ensures thorough mixing while minimizing air bubble entrapment. 
The porous silica particles used in this study exhibit a lower effective density ($\sim$\SI{1.9}{\gram\per\cubic\centi\meter}) than bulk fused silica (\SI{2.2}{\gram\per\cubic\centi\meter}) due to internal voids. 
Because of uncertainties associated with this estimation, particle concentrations are reported in terms of weight fraction rather than volume fraction (see Supplementary Information, section~\ref{Subsubsection silica density-Suppl}). 

\subsection{Measurement procedure and data analysis}
\subsubsection{DropSlider.}
The DropSlider setup~\citep{straub2021flow} ensures locally stable moving contact lines, enabling their visualization within the microscope's field of view, for an extended duration. 
The droplet was pinned while the substrate underwent controlled motion using a piezoelectric motor (LPS-45, Physik Instrumente, Germany). 
To fix the droplet, the geometry was selected based on the suspension rheology: a disk geometry for TDE-based and a prism geometry for NaSCN-based suspensions (Fig.~\ref{fig:Geometries}). 
The disk geometry has a diameter of \SI{5}{\milli \metre} and a gap of \SI{2}{\milli \metre} between the disk and the substrate. 
The prism geometry uses a right-angle prism with a face length and width of \SI{25}{\milli \metre}. 
For all experiments, a droplet volume of \SI{50}{\micro\litre} is used.

\begin{figure*}[t]
  \centerline{\includegraphics[width=0.75\textwidth]{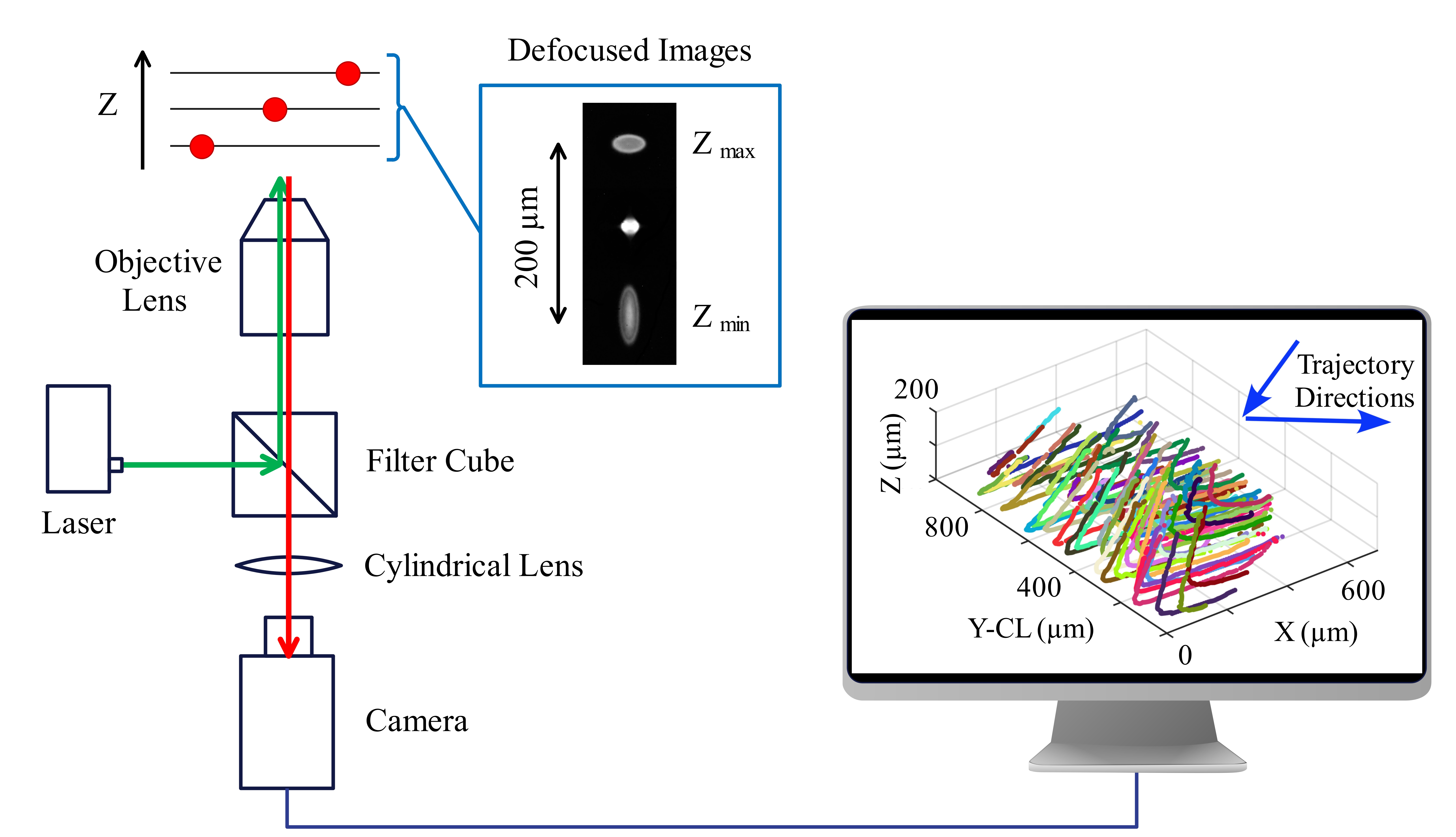}}
  \caption{Schematic representation of APTV setup utilized for three-dimensional tracking of fluorescently labelled tracer particles by integrating a cylindrical lens in front of the camera sensor. 
  The framed region highlights the deformation of particle shapes based on their relative distance to the focal planes. 
  \label{fig:APTV} }
\end{figure*}

\subsubsection{Astigmatism particle tracking velocimetry (APTV).}
\paragraph{Setup configuration.}
APTV stands out in scenarios with limited access to the sample, as it can measure the 3D flow field using a single objective and only 2D images. 
APTV operates with not just one, but two primary focal planes. 
For APTV measurements, the DropSlider configuration is positioned on the microscope table~\citep{straub2021flow}, above the objective lens of an inverted fluorescent microscope (Leica DMI6000 B). 
A combination of a 20$\times$ microscope objective and a \SI{150}{\milli\metre} cylindrical lens was employed for the measurements. 
The cylindrical lens was positioned between the microscope output and the camera sensor, close to the microscope output. 
The images were acquired using the FASTCAM Mini AX200 camera at a frame rate of \SI{50}{\hertz}, Fig.~\ref{fig:APTV}. 
The substrate travels at a constant velocity of \SI{200}{\micro\metre\per\second} covering a maximum range of \SI{22}{\milli\metre} in \SI{110}{seconds}.

\paragraph{Measurement procedure.}
A steady state and continuous flow was attained approximately \SI{20}{\second} after initiating the movement of the substrate. 
This steady state was maintained for at least \SI{60}{\second}. 
The initial 20 seconds period of transient dynamics of the droplet is excluded from the subsequent analysis. 
In our experiments, the droplet was allowed to rest for 5 minutes after deposition to minimize disturbances from particle sedimentation before starting the experiment. 
Consequently, there will be a denser layer of particles close to the substrate. 
In addition, each experiment was repeated at least three times to account for small, unavoidable variations introduced during droplet deposition. 
This ensured the reproducibility of the reported results.

\paragraph{Analysis procedure.}
The coordinate system is defined such that the $x$-axis aligns with the direction of droplet motion relative to the substrate, the $y$-axis lies parallel to the contact line, and the $z$-axis is normal to the $x$-$y$ plane, representing the out-of-plane direction. 
The deformation of particle images from circular to elliptical shapes is correlated with their axial ($z$) position relative to the focal planes. 
This relationship is quantified using the aspect ratio, based on calibration procedures described in~\citep{cierpka2010calibration, rossi2014optimization, straub2021flow}. 
A particle image shape can be reliably determined in the region between $z_{min}$ and $z_{max}$, shown in Fig.~\ref{fig:APTV}. 
This range is determined by the particle intensity being high enough for a reliable analysis. 
For all the data provided, the plane at $z=0$ corresponds to the substrate. 
The minor vertical displacement of the substrate was corrected by analyzing the trajectories of particles located on the substrate or confined at the contact line. 
To project data in a coordinate system that is co-moving with the contact line, the position of the contact line was determined in each image. 
This adjustment ensures that the resulting $x$ coordinate accurately represents the distance of the particles from the contact line, with $x=0$ being the contact line position. 
Further details of the experimental setup and analysis are published elsewhere~\citep{straub2021flow}. 
On average, out of 14360 particle velocities determined from tracking 58 particles, 0.6\% of these velocities exceeded the substrate velocity of \SI{200}{\micro \metre \per \second}. 
These are considered to be outliers due to the misdetection of the particle trajectories and are discarded from further analysis. 

\subsubsection{Side view imaging of droplet shape.}
Side-view images of the droplet were recorded using a FASTCAM Mini AX200 camera equipped with a Nativar 12X Zoom lens. 
A SCHOTT KL2500 LED light source with a diffuser was positioned behind the experimental setup to ensure uniform illumination. 
The droplet shapes were captured at a frame rate of \SI{50}{\hertz}. 
Droplet surface profiles were analyzed using edge-detection algorithms implemented in MATLAB.

\subsection{Particle-particle interactions}
Diverse methodologies are employed to elucidate particle-particle interactions and the resulting flow properties. 
AFM measurements enable direct quantification of nanoscale normal forces~\citep{Butt:2005aa, comtet2017pairwise}. 
Macroscopically, angle of repose measurements primarily reflect rolling friction~\citep{Al-Hashemil2018review}, while rheological measurements capture bulk flow behaviour. 
The following sections detail these measurements.

\subsubsection{Rheology measurements.} \label{subsubsection: Rheology measurements}
\paragraph{Protorheology, Dripping onto Substrate (DoS).} \label{Protorheology-Main}
Jamming in dense granular suspensions denotes the onset of a rigid, solid-like state, marked by a loss of flowability and a divergence in viscosity or the emergence of a finite yield stress. 
While this transition is typically associated with exceeding a critical particle volume fraction ($\phi_{c}$), it can also be dynamically induced at lower concentrations under sufficient shear stress~\citep{ morris2020toward, guazzelli2018rheology, morrisProgressChallengesSuspension2023}. 
To evaluate the proximity of each sample system to the jamming threshold, we employed dripping onto substrate (DoS) rheometry across a concentration series. 
DoS rheometry quantifies capillary driven pinch-off dynamics to extract rheological properties and is broadly applicable to Newtonian and complex fluids~\citep{dinicPinchoffDynamicsDrippingontosubstrate2017}.

The results of these measurements, shown in Fig.~\ref{fig:TDE_NaSCN_DOS}a and b of Supplementary Information, indicate that the maximum flowable concentration beyond which a transition to a jammed state occurs is system specific. 
For silica suspensions in TDE, the threshold is 35~wt.\%, while in NaSCN, the transition occurs at a lower concentration of 30~wt.\%. 
We attribute this shift to enhanced interparticle friction in NaSCN, as discussed in section~\ref{Subsubsec: Angle of repose}, which promotes the formation of stress bearing contact networks at lower solid loadings. 
These findings indicate that the concentration ranges used in subsequent rheometry and wetting experiments are close to the jamming threshold. 
Complementary rheological characterization using conventional rheometry was limited to 31~wt.\% in TDE and 28~wt.\% in NaSCN, the highest concentrations that could be reliably measured in the rheometer due to challenges at higher loadings. 
Our dynamic wetting experiments focus on suspensions containing 30 and 33~wt.\% silica in TDE and 30~wt.\% in NaSCN solutions. 
These concentrations represent the maximum amount of particles within the droplets that still allow their movement. 
An overview of all concentrations used throughout the study is provided in Table~\ref{tab:Concentrations} of the Supplementary Information.

\begin{figure}[t]
  \centerline{\includegraphics[scale=0.4]{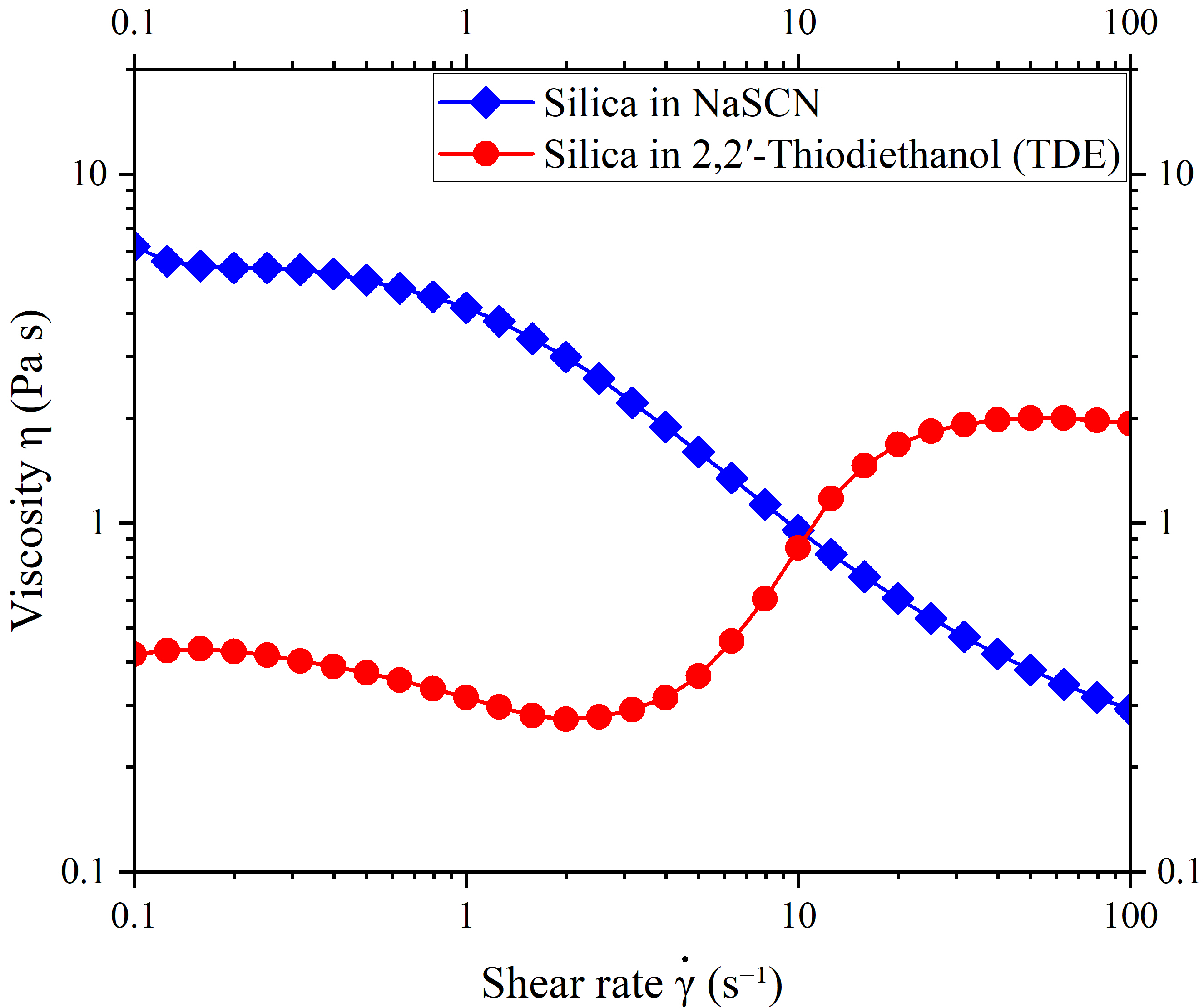}}
  \caption{Shear rate dependent viscosity of dense granular suspensions composed of \SI{5}{\micro\metre} silica particles dispersed in TDE and NaSCN solutions, measured at 31~wt.\% and 28~wt.\% respectively.}
\label{fig:rheology}
\end{figure}

\paragraph{Rheology measurements.}\label{Rheology measurements-Main}
Shear rate dependent viscosity and oscillatory strain sweep measurements were conducted to characterize the rheological response of the suspensions. 
The former is presented here, while the latter is detailed in Supplementary Information, section~\ref{Rheology measurements-Suppl}. 
The suspensions consisted of silica particles dispersed in either a TDE or NaSCN solution, with concentrations of 31~wt.\% and 28~wt.\%, respectively. 
During shear rate dependent viscosity experiments, the samples were subjected to continuous shear by gradually increasing the shear rate from \SI{0.1}{\per \second} to \SI{100}{\per \second}, while the corresponding stress response was monitored. 
The results of these measurements are presented in Fig.~\ref{fig:rheology}. 
Additional details regarding the measurement procedure and data reproducibility can be found in Fig.~\ref{SRS-Rheology} of Supplementary Information.

The dispersion of silica particles in TDE solution exhibits non-Newtonian behaviour, characterized by slight shear-thinning at low to intermediate shear rates followed by a pronounced shear-thickening at higher shear rates (Fig.~\ref{fig:rheology}). 
These rheological transitions arise from stress dependent interparticle interactions~\citep{morrisShearThickeningConcentrated2020a}. 
While two-body interactions are important in dilute suspensions~\citep{melrose2004continuous}, multi-body interactions dominate dense suspensions~\citep{morrisShearThickeningConcentrated2020a}. 
Prior to shear, particles are randomly distributed, and their inherent excluded volume interactions hinder flow~\citep{melrose2004continuous}. 
At low shear rates, weak hydrodynamic interactions dominate, allowing the particles to gradually rearrange and adopt more anisotropic configurations aligned with the flow direction. 
This restructuring reduces flow resistance and results in mild shear-thinning~\citep{wagner2009shear}. 
As the shear rate increases, particles are driven into closer proximity, narrowing the interstitial gaps~\citep{melrose2004continuous}. 
Once the applied stress surpasses a critical threshold, frictional contacts generate transient load-bearing networks that dominate the suspension’s rheology~\citep{seto2013discontinuous, gallier2014rheology, trulsson2017effect}. 
Rheological measurements were conducted at concentrations slightly below those used in wetting experiments. 
Nonetheless, for the TDE suspension, the concentration lies well within the regime where shear-thickening is clearly observed (Fig.~\ref{fig:rheology}). 
Notably, in non-Brownian suspensions, shear-thickening occurs at an approximately fixed critical stress across all investigated volume fractions, as established both experimentally and numerically~\citep{morrisShearThickeningConcentrated2020a, guyUnifiedDescriptionRheology2015, seto2013discontinuous, malbrancheScalingAnalysisShear2022}. 
Consequently, this trend is anticipated to hold even at higher concentrations.

On the other hand, the dispersion of silica particles in NaSCN salt solution exhibits a yield-stress fluid characteristic with a plateau at low frequencies, followed by considerable shear-thinning as it flows, Fig.~\ref{fig:rheology}. 
In granular suspensions, bulk stress arises from particle-level interactions, including adhesion and the frictional network of contacts between particles, which collectively govern the complex macroscopic behaviour~\citep{ancey1999theoretical}. 
For yield-stress granular suspensions where particle adhesion is weak (discussed further in AFM measurement section~\ref{AFM subsubsection}), yield-stress is mainly due to frictional and geometrical interactions between particles~\citep{fall2009yield, singh2023hidden}. 
Particles form a network based on their size, shape, and arrangement, with the friction creating resistance to deformation. 
Therefore, the yield stress emerges from the necessity to break and rearrange the interlocking particles. 
Shearing disrupts the network formed by interparticle interactions, allowing the suspension to flow, which results in shear-thinning.

\begin{figure*}[t]
  \centerline{\includegraphics[scale=0.35]{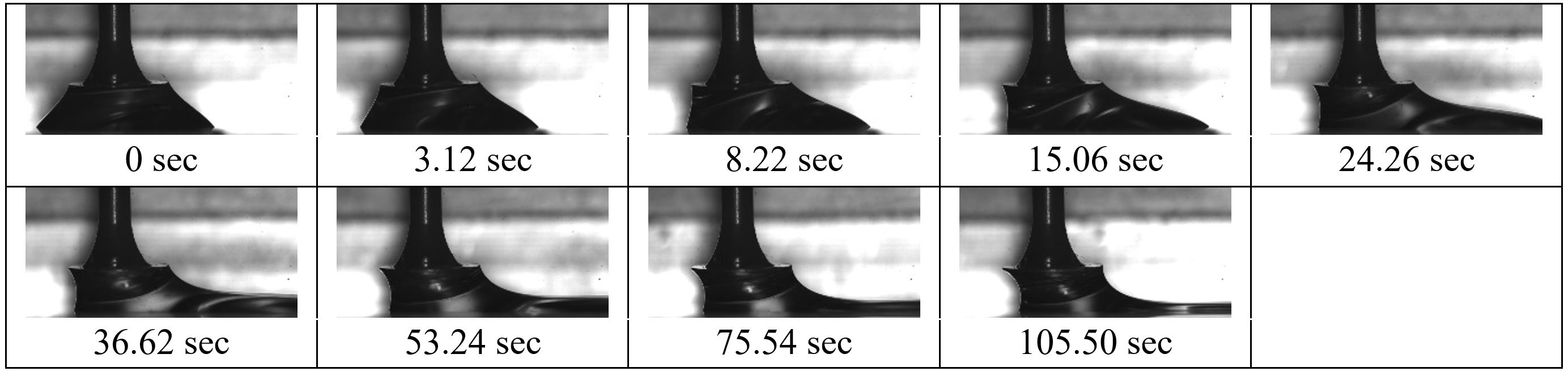}}
  \caption{
Side-view images of a suspension droplet containing 33~wt.\% silica particles with a diameter of \SI{5}{\micro\metre} dispersed in TDE solution, translating relative to the substrate at a constant velocity of \SI{200}{\micro\metre\per\second}.
  \label{fig:Side view}}
\end{figure*}

%
The observed differences in flow behaviour, characterized by viscous response with shear-thickening in the TDE-based suspension and yield-stress behaviour in the NaSCN-based system, are further substantiated by oscillatory measurements presented in Supplementary Information, section~\ref{Oscillatory strain sweep measurements-Suppl}. 
The TDE suspension consistently exhibits a dominant viscous response ($G'' > G'$) and a pronounced increase in both moduli beyond intermediate strain amplitudes, indicative of strain-hardening driven by the formation of stress-bearing microstructures. 
In contrast, the NaSCN suspension displays an early crossover between $G'$ and $G''$, followed by a monotonic decline in both, indicative of yielding and structural breakdown. 
These distinct nonlinear viscoelastic responses reflect a mechanistic difference in the underlying stress-bearing network that cannot be attributed solely to variations in effective packing fraction.

\subsubsection{Angle of repose measurements.} \label{Subsubsec: Angle of repose}
We examined how changing the medium between particles affects the friction coefficient by measuring the angle of repose, which indicates the rolling friction of the particles, with further details provided in Fig.~\ref{fig:Angle_of_Repose} of Supplementary Information~\citep{Al-Hashemil2018review, matsuo2014geometric, doan2023interactive}. 
In the TDE solution, the angle of repose was \SI{8 \pm 2}{\degree}, while in the NaSCN solution it was much higher at \SI{52 \pm 5}{\degree}. 
These results indicate that particle friction, particularly rolling friction, is significantly higher in the NaSCN solution. 
The electrostatic double layer functions as a lubricating interface, mitigating direct particle contact and thereby reducing friction~\citep{clavaud2017revealing}. 
In a high ionic strength solution such as the NaSCN solutions, the electrostatic double layer undergoes compression due to the ion screening~\citep{butt2018surface}, resulting in increased direct contacts and consequently higher friction. 
Refer to Supplementary Information, section~\ref{AFM-Suppl} for a more detailed discussion.

\subsubsection{Atomic force microscopy measurements.} \label{AFM subsubsection}
Particle-particle interactions in the normal direction were measured using Colloidal Probe Atomic Force Microscopy (CP-AFM), employing a silica colloidal probe from the same batch of silica particles used in the other experiments. 
Experimental details are given in the Supplementary Information, section~\ref{AFM-Suppl}. 
The two index-matched liquids, TDE solution and NaSCN solution, were used as dispersion media for the measurements. 
The results of adhesion force measurements are presented in Fig.~\ref{AFM-adhesion} and Fig.~\ref{fig:AFM} of the Supplementary Information. 
The adhesion forces measured by CP-AFM experiments are slightly higher in TDE solution (\SI{-5.1 \pm 4.8}{\nano\newton}) compared to NaSCN solution (\SI{-1.5 \pm 1.3}{\nano\newton}).

Modeling and simulation studies consistently demonstrate that contact forces, particularly the frictional forces arising from sliding and rolling, alongside normal forces, are pivotal in determining the suspension's response to applied shear stress~\citep{tomas2009energy, Seto:2012aa, Luding:2008ab}. 
The magnitude of the sliding friction is directly proportional to the normal force and is determined by the static and dynamic friction coefficients. 
Rolling resistance typically exerts a weaker influence compared to tangential sliding forces~\citep{Luding:2008ab}. 
However, the dominance of one type of force over another is contingent upon particle characteristics, including size, surface roughness, and electrostatic forces, which collectively determine the relative strength of these contact forces. 

Based on the characterization experiments, the rheological analysis reveals a significant difference in interparticle dynamics between the two granular suspension systems. 
Particles in the NaSCN solution exhibit notably enhanced interactions, including frictional contacts and adhesive forces. 
AFM measurements show no substantial difference in adhesive forces between the systems, with slightly higher adhesion in the TDE medium. 
Therefore, the increased resistance to particle motion in the NaSCN suspension is mainly due to higher sliding and rolling friction coefficients, as indicated by the angle of repose measurements. 
We categorize the suspension of silica particles in TDE as weakly interacting particles. 
Conversely, the suspension of silica particles in a NaSCN solution is described as strongly interacting particles.

\section{Results and discussion}
\subsection{Transient dynamics of receding contact line}\label{Transient receding-Main}

\begin{figure*}[t]
    \centering
    \subfloat[\centering]{\includegraphics[scale=0.34]{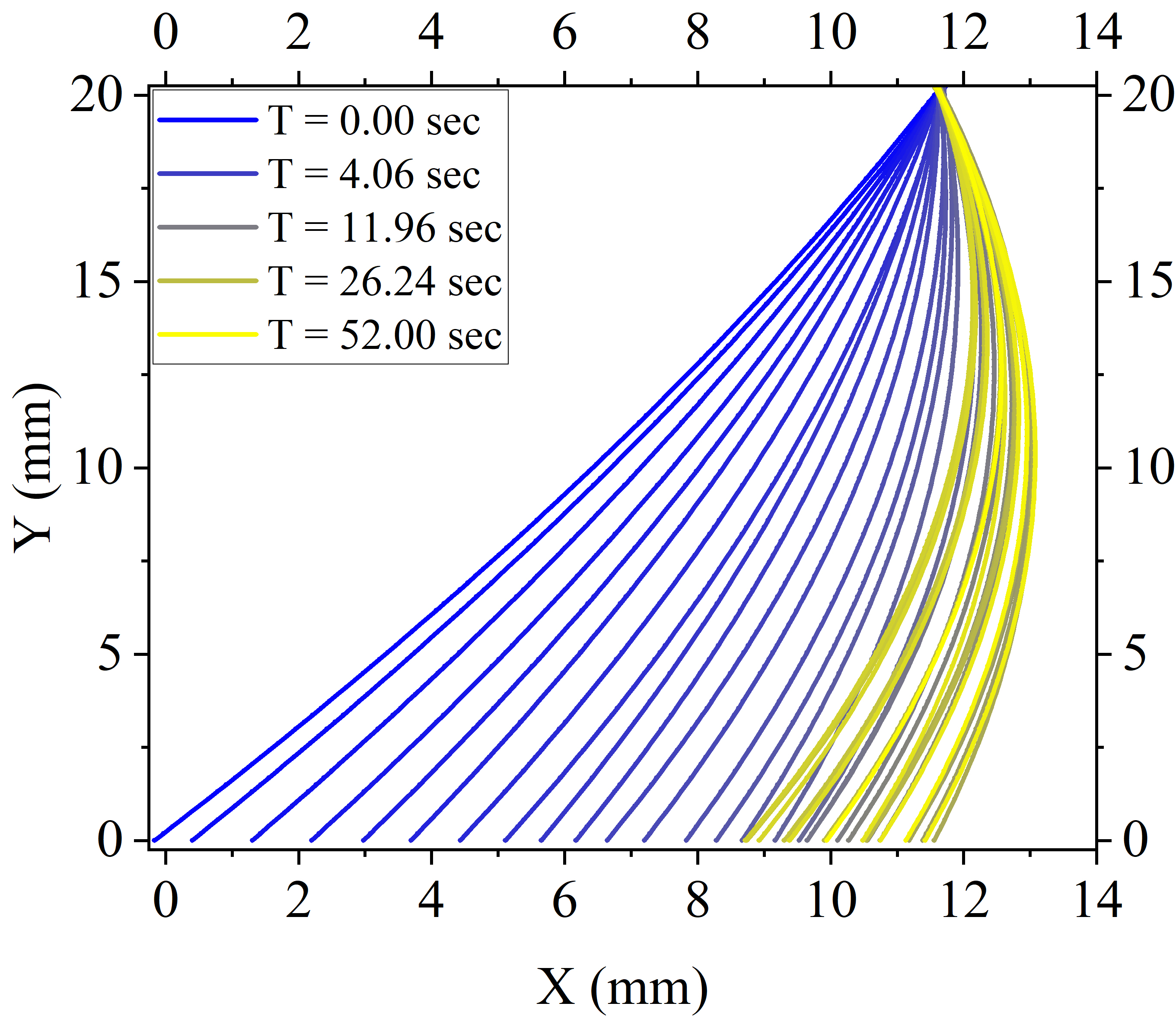}} \qquad
    \subfloat[\centering]{\includegraphics[scale=0.34]{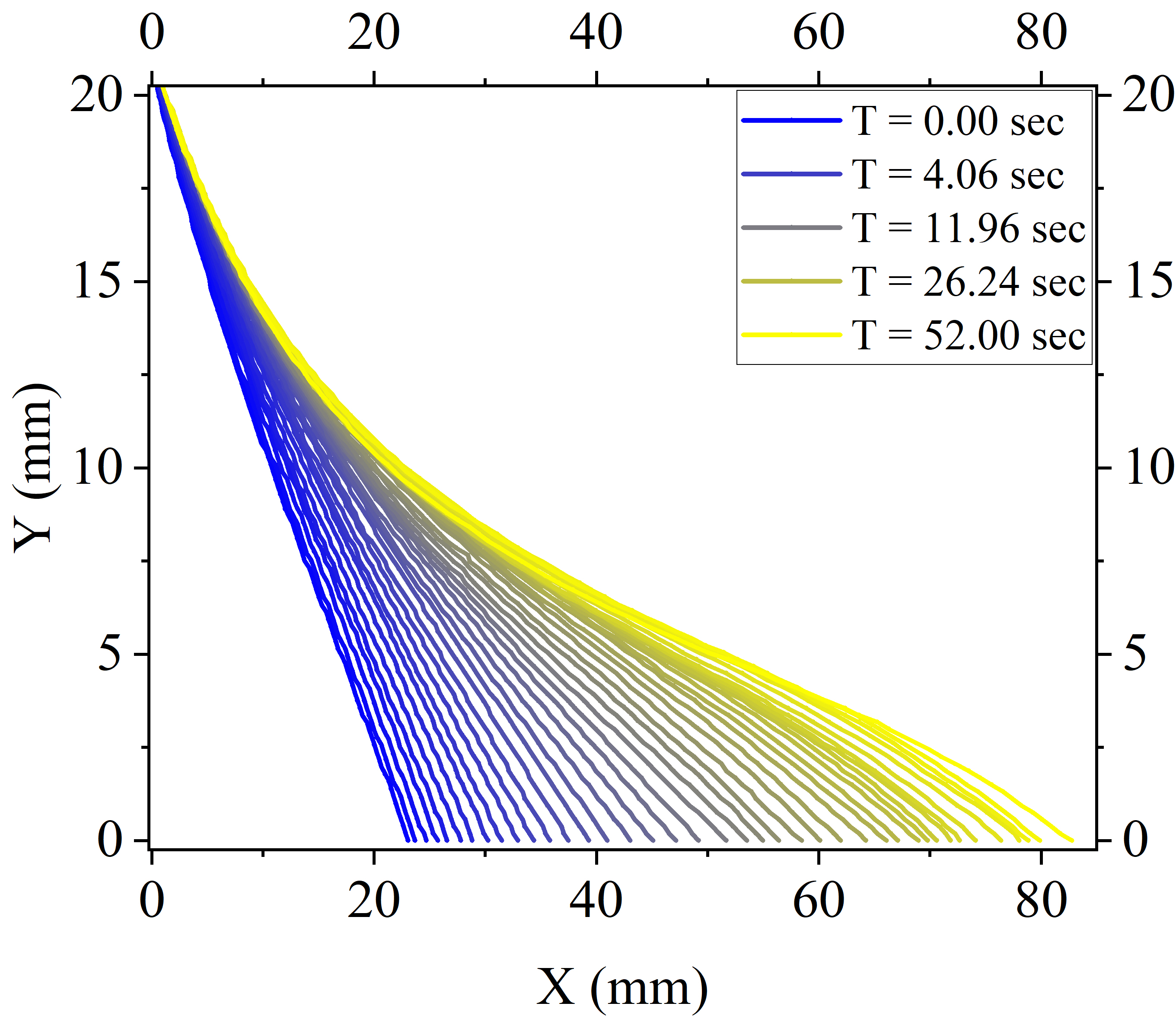}}\\[-1em]
    \subfloat[\centering]{\includegraphics[scale=0.34]{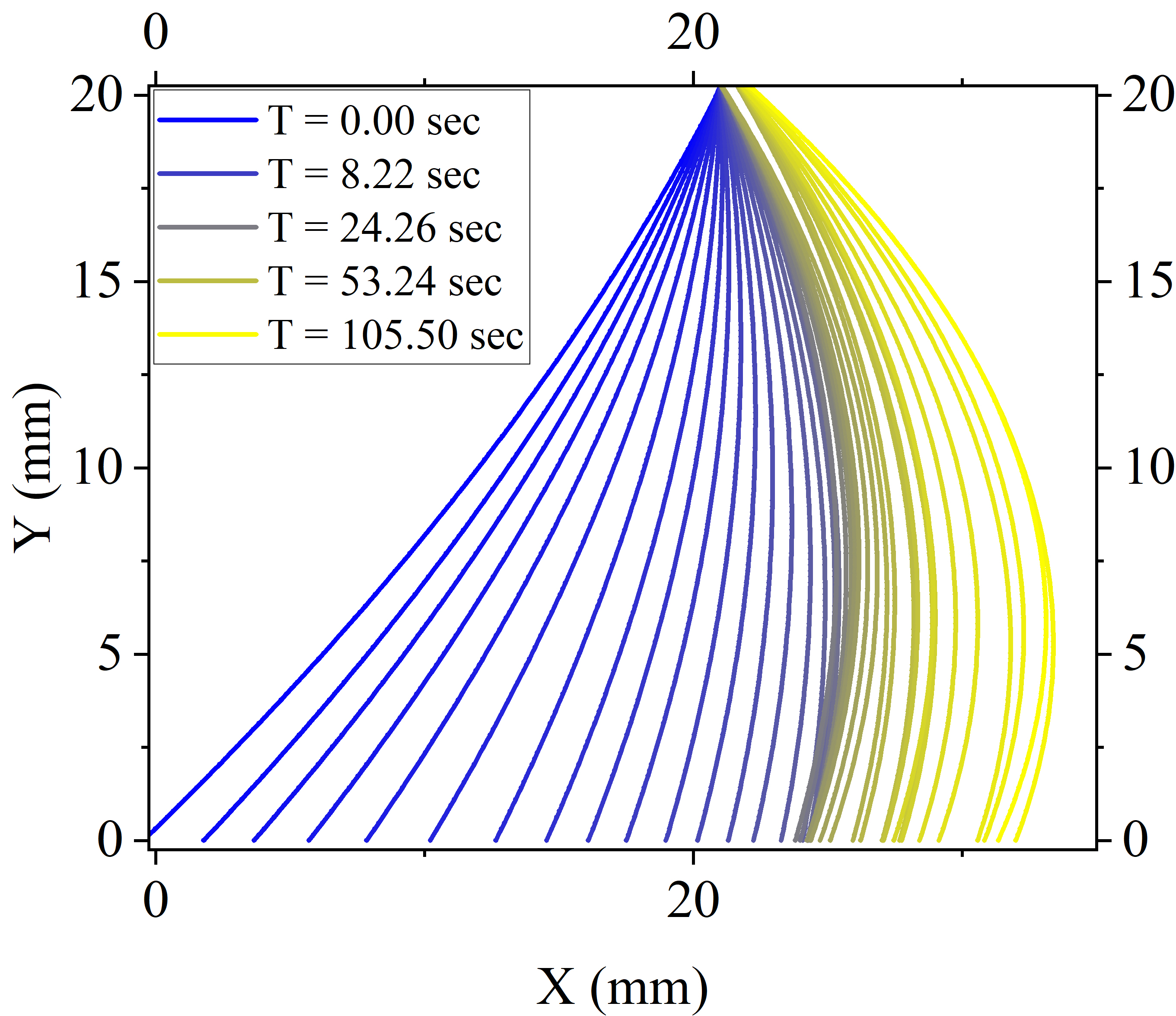}} \qquad
    \subfloat[\centering]{\includegraphics[scale=0.34]{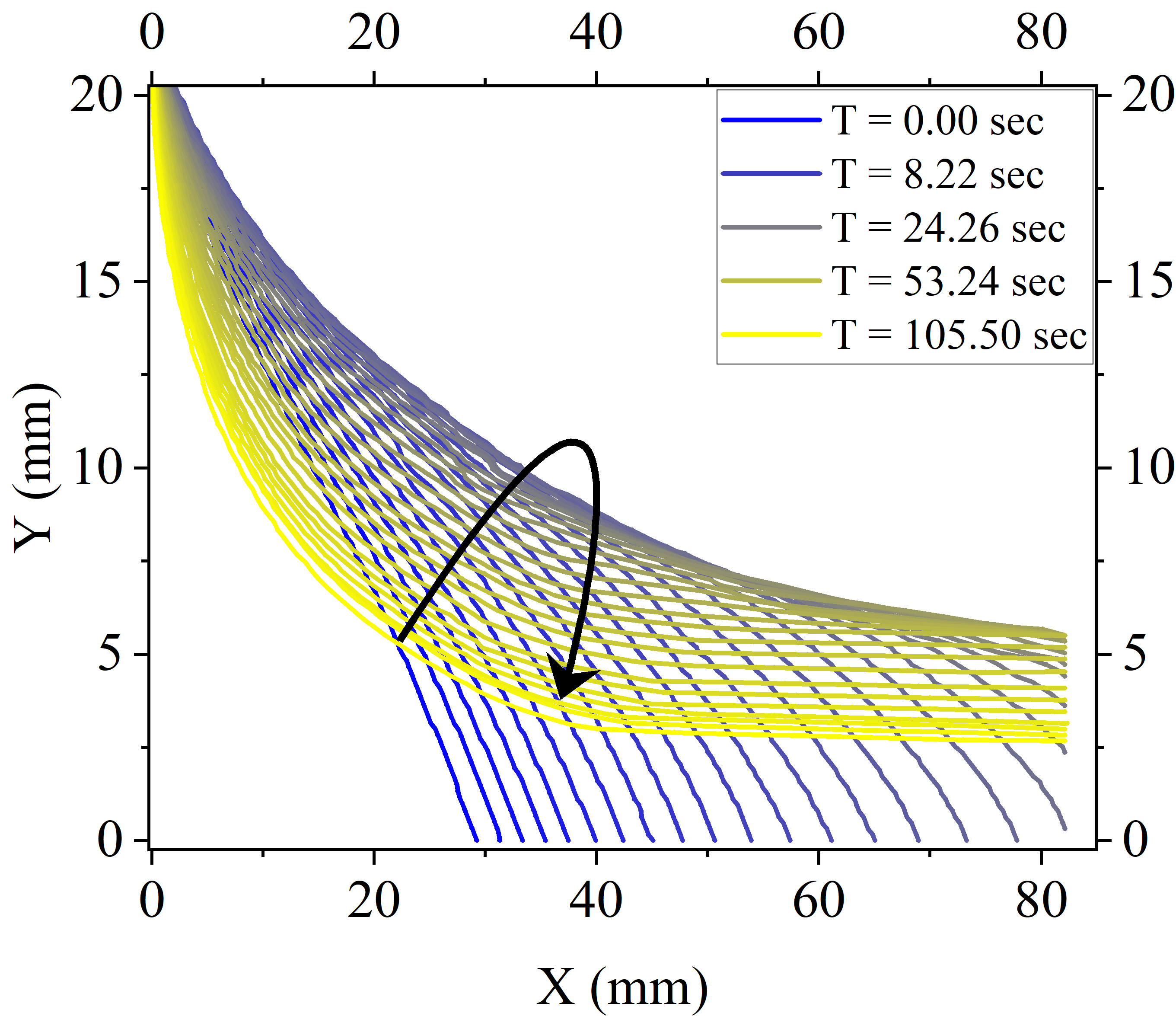}}\\[-1em]
    \caption{Time evolution of droplet surface profiles near the contact lines. 
    (\textit{a}, \textit{b}) Pure TDE droplet used as a Newtonian reference near the advancing and receding fronts. 
    (\textit{c}, \textit{d}) Suspension droplet containing 33~wt.\% silica particles dispersed in TDE. 
    The color gradient from blue to yellow encodes the temporal progression. 
    While the Newtonian case exhibits monotonic interface evolution, the suspension reveals a significant difference at the receding front, where a non-monotonic evolution of curvature emerges.}
    \label{fig:TransientDynamics}
\end{figure*}

%
To elucidate the interplay between the particulate phase and contact line dynamics, we analyze two representative cases using side-view imaging: (i) a reference case comprising the pure dispersion medium (TDE solution without particles), and (ii) a dense suspension containing 33~wt\% silica particles dispersed in the same medium (see Fig.~\ref{fig:Side view}). 
Receding and advancing surface profiles of the moving droplet are extracted from side view images and superimposed to visualize the temporal evolution of the droplet shape. 
These measurements distinguish the role of the suspending fluid from that of the particle-laden phase, particularly near the receding contact line, Fig.~\ref{fig:TransientDynamics}.

Side view experiments with the pure dispersion medium indicate clear dewetting close to the receding side. 
The contact line recedes without leaving a residual film on the substrate. 
The fluid alone does not form a film during recession. 
See also the receding contact angle measurements at very low velocities, provided in Fig.~\ref{fig:Receding_Contact_Angle_TDE_Solution} of the Supplementary Information.

In contrast, when employing a 33~wt.\% concentrated suspension in TDE solution, the dynamics changes fundamentally. 
At a substrate velocity of \SI{200}{\micro\metre\per\second}, we observe no dewetting. 
The receding contact angle diminishes steadily and eventually decays to near-zero. 
There is no classical signature of contact line recession. 
Instead, we observe a continuous particle-laden layer remaining on the substrate in the wake of the contact line. 
This layer is substantially thicker than the particle diameter (see Fig.~\ref{fig:Suspension_Layer_TDE_33} of Supplementary Information). 
As noted in the introduction, particle entrainment into thin films requires collective hydrodynamic assembly to overcome capillary resistance~\citep{colosquiHydrodynamicallyDrivenColloidal2013}. 
In our case, the suspension concentration lies well above the dilute limit. 
Moreover, interactions between particles and the substrate, as well as interparticle interactions at high concentration, likely contribute to the formation of a thick deposit. 
We did not further investigate the influence of individual mechanisms.

For the concentrated suspension, we observe a transient evolution of the droplet interface profile near the receding side Fig.~\ref{fig:TransientDynamics}d. 
The slope of the interface initially decreases, indicating a flattening or reduction in the steepness of the curve. 
This is followed by a subsequent increase in the slope, indicating a re-steepening or rise in the curvature. 
The droplet typically travels 1 to 1.5 times its initial footprint diameter before attaining a steady-state shape. 
The comparison of the advancing and receding surface profiles of the suspension in Fig.~\ref{fig:TransientDynamics}c and d shows that while the advancing side reaches a stationary state after a brief dynamic phase of approximately 10 seconds, the receding side requires over 30 seconds to stabilize. 
This difference arises from the droplet depositing a suspension layer as it recedes.

The deposition of a thin suspension layer as the droplet recedes is a well-documented phenomenon in similar systems~\citep{gans2019dip}. 
This behaviour is consistent with the Landau-Levich-Derjaguin relation, which predicts the thickness of the liquid film entrained during the withdrawal of a plate from a bath, accounting for the dependence of the suspension bulk viscosity on $\phi$~\citep{palma2019dip}. 
At the liquid–air interface near the receding contact line, the net interfacial stresses acting on fully wetted and neutrally buoyant particles are similar to the Laplace pressure in a pure liquid, thereby justifying the applicability of such classical capillary–hydrodynamic scaling to dense suspensions~\citep{chateau2018pinch}.

\captionsetup[subfloat]{justification=centering, position=bottom}

\begin{figure*}[t]
    \centering

    \subfloat[]{\includegraphics[scale=0.169]{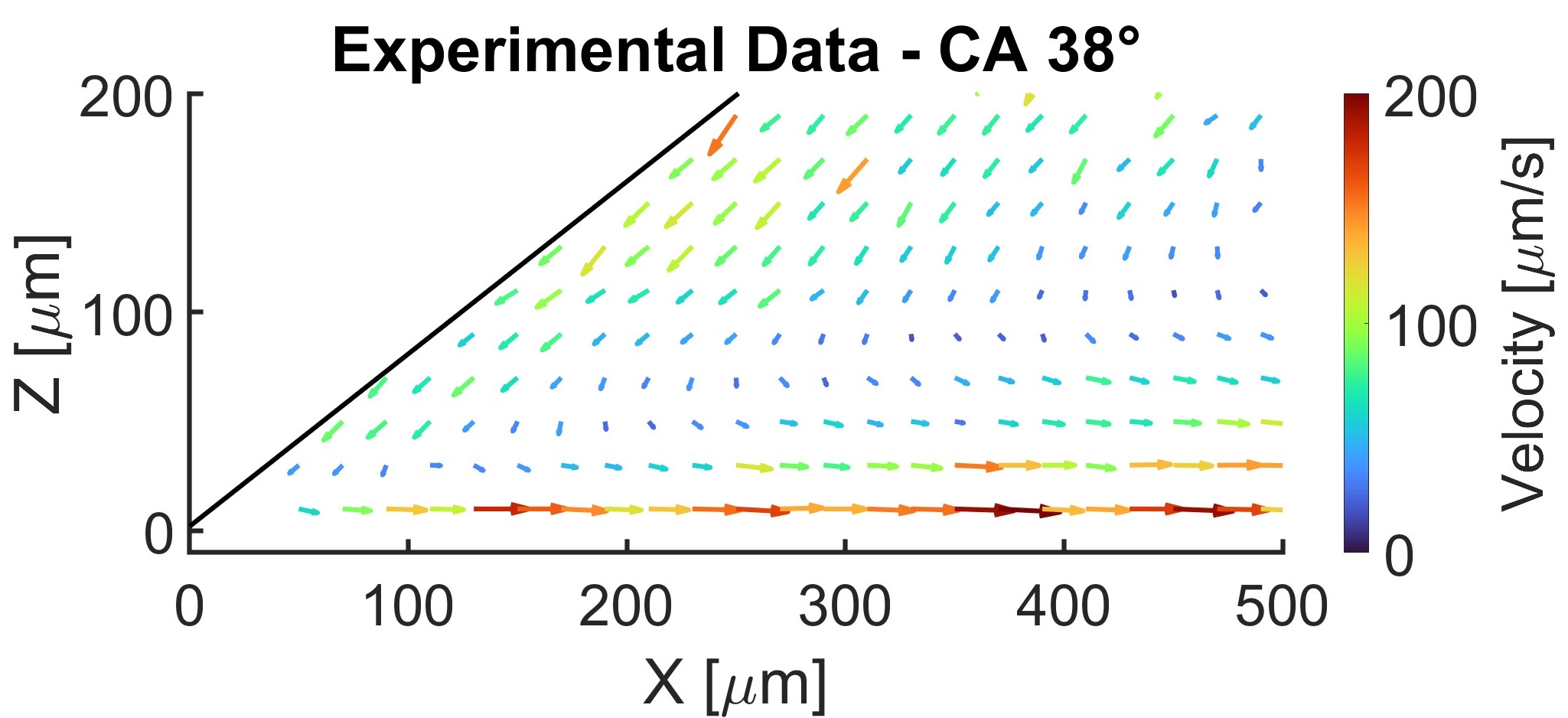}}
    \hspace{0pt}
    \subfloat[]{\includegraphics[scale=0.169]{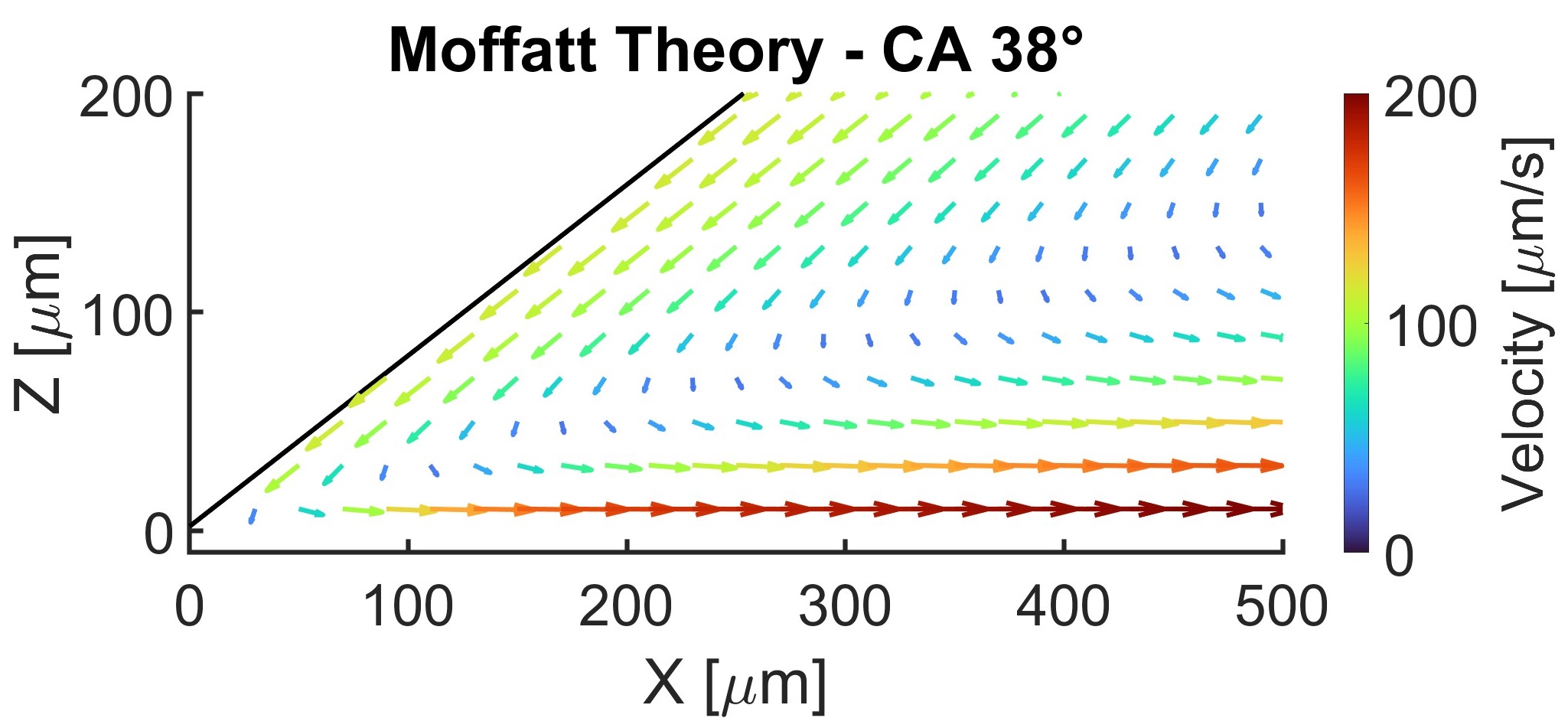}}
    \hspace{0pt}
    \subfloat[]{\includegraphics[scale=0.169]{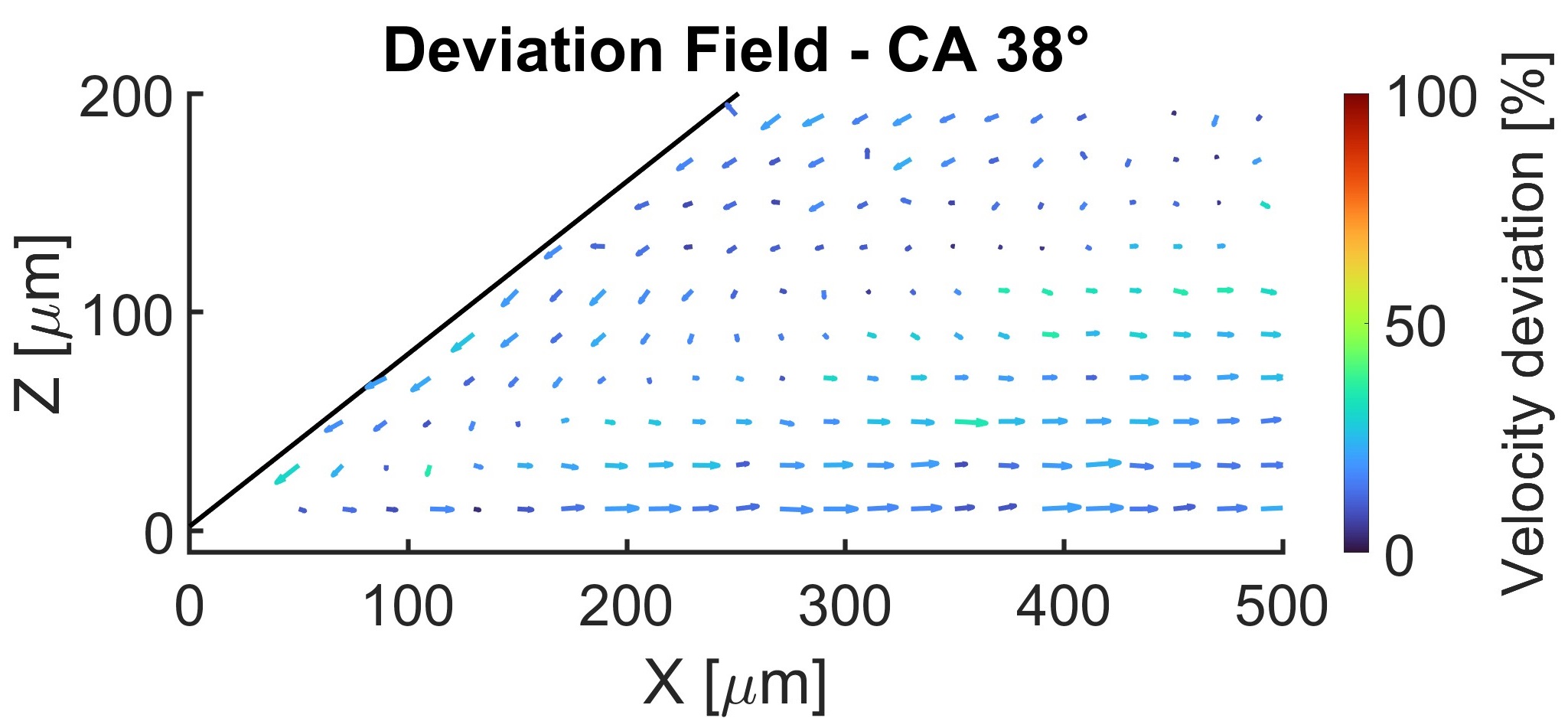}}
    \\[-0.5ex]

    \subfloat[]{\includegraphics[scale=0.169]{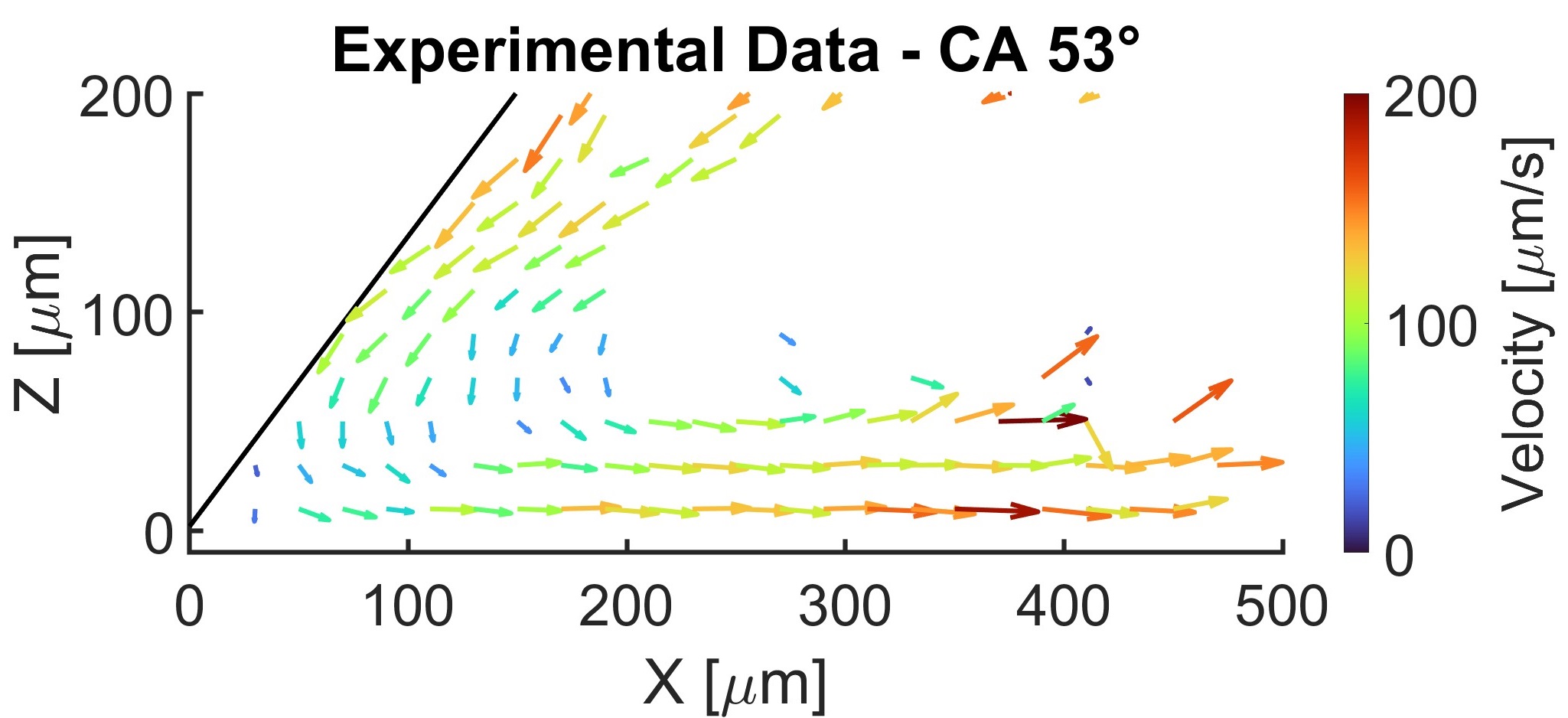}}
    \hspace{0pt}
    \subfloat[]{\includegraphics[scale=0.169]{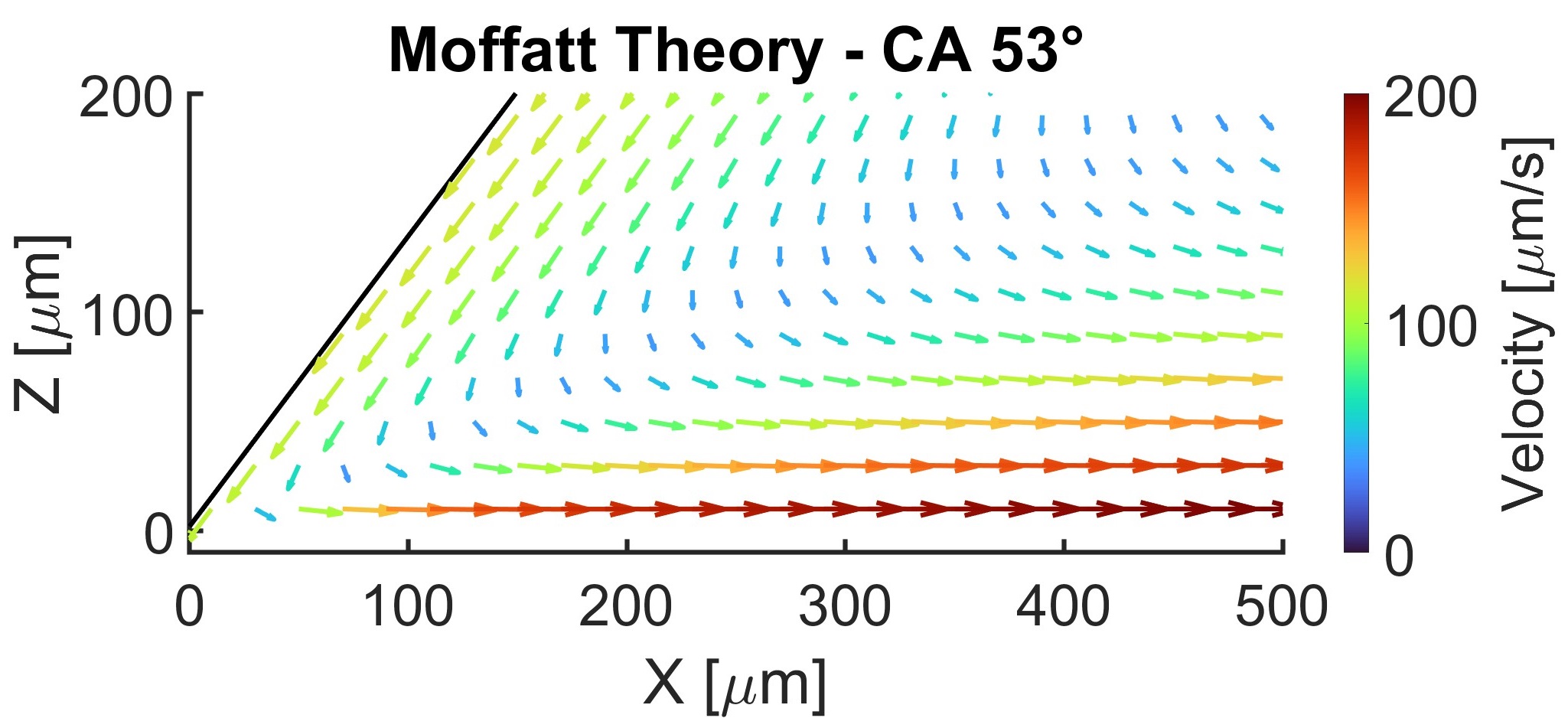}}
    \hspace{0pt}
    \subfloat[]{\includegraphics[scale=0.169]{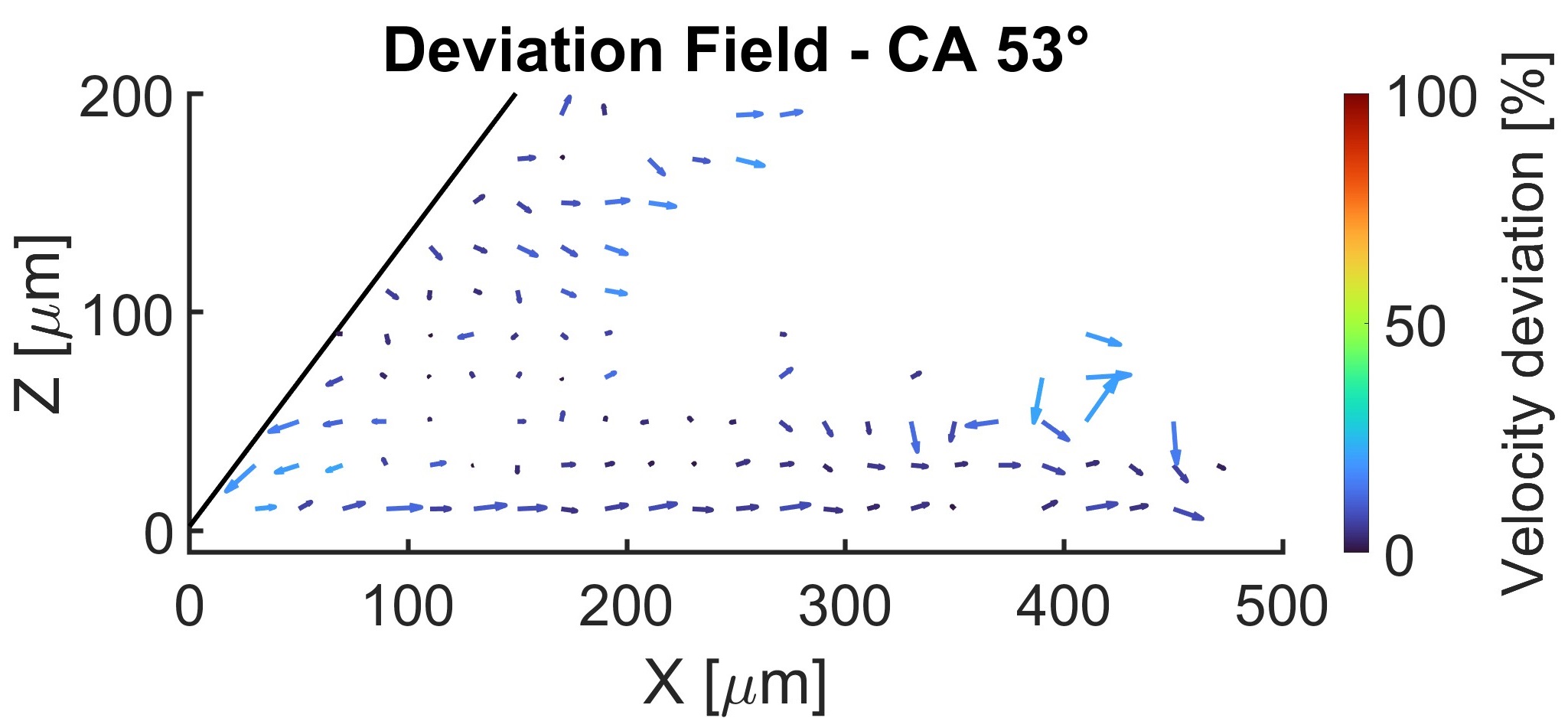}}
    \\[-0.5ex]

    \subfloat[]{\includegraphics[scale=0.169]{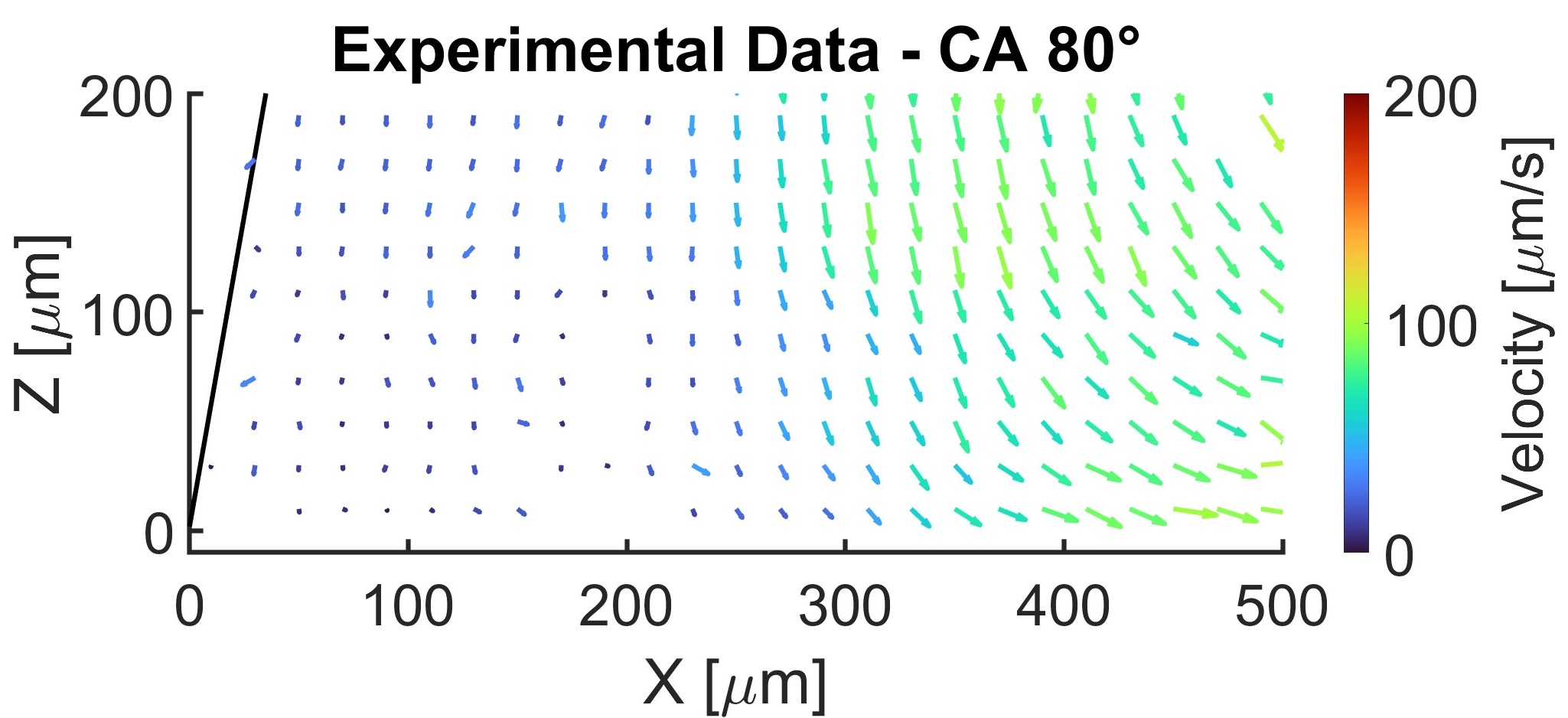}}
    \hspace{0pt}
    \subfloat[]{\includegraphics[scale=0.169]{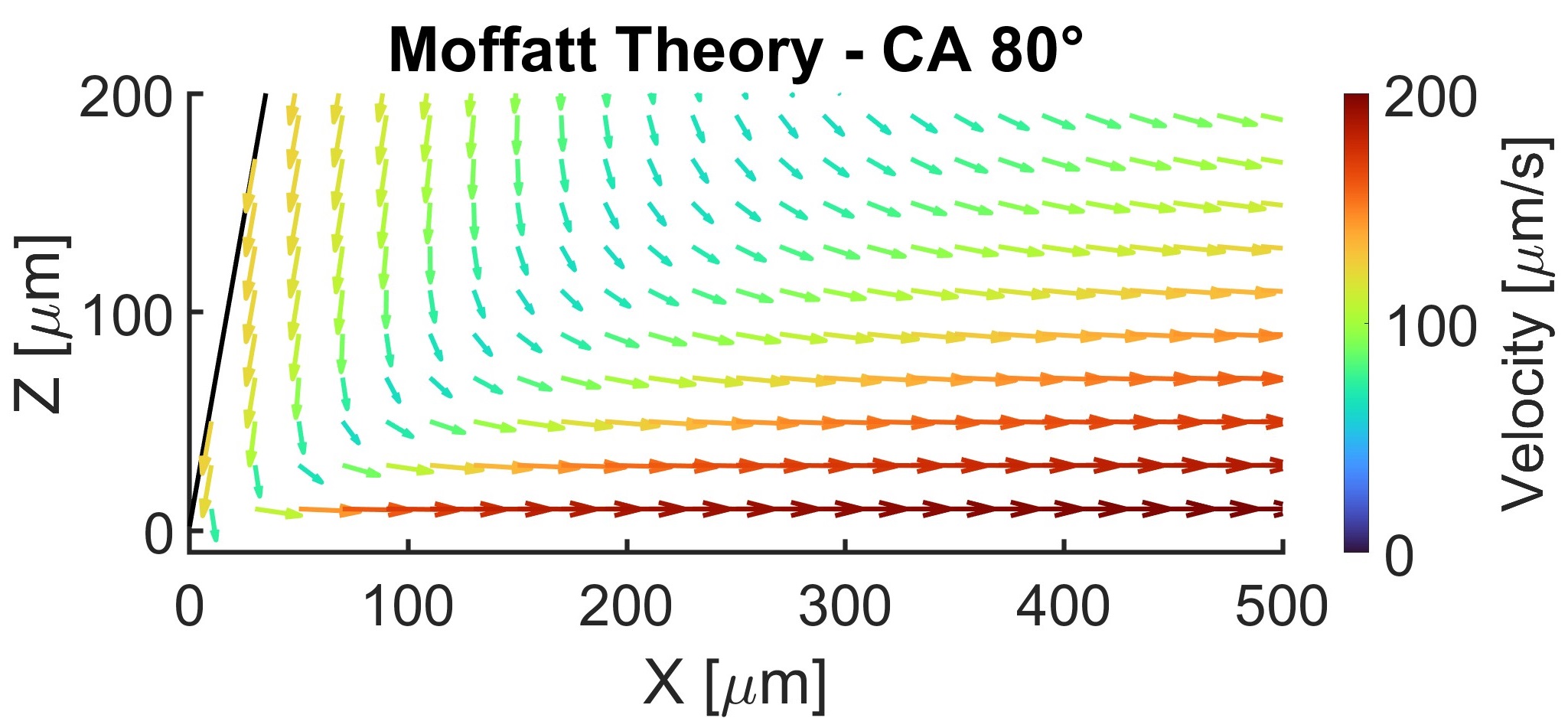}}
    \hspace{0pt}
    \subfloat[]{\includegraphics[scale=0.169]{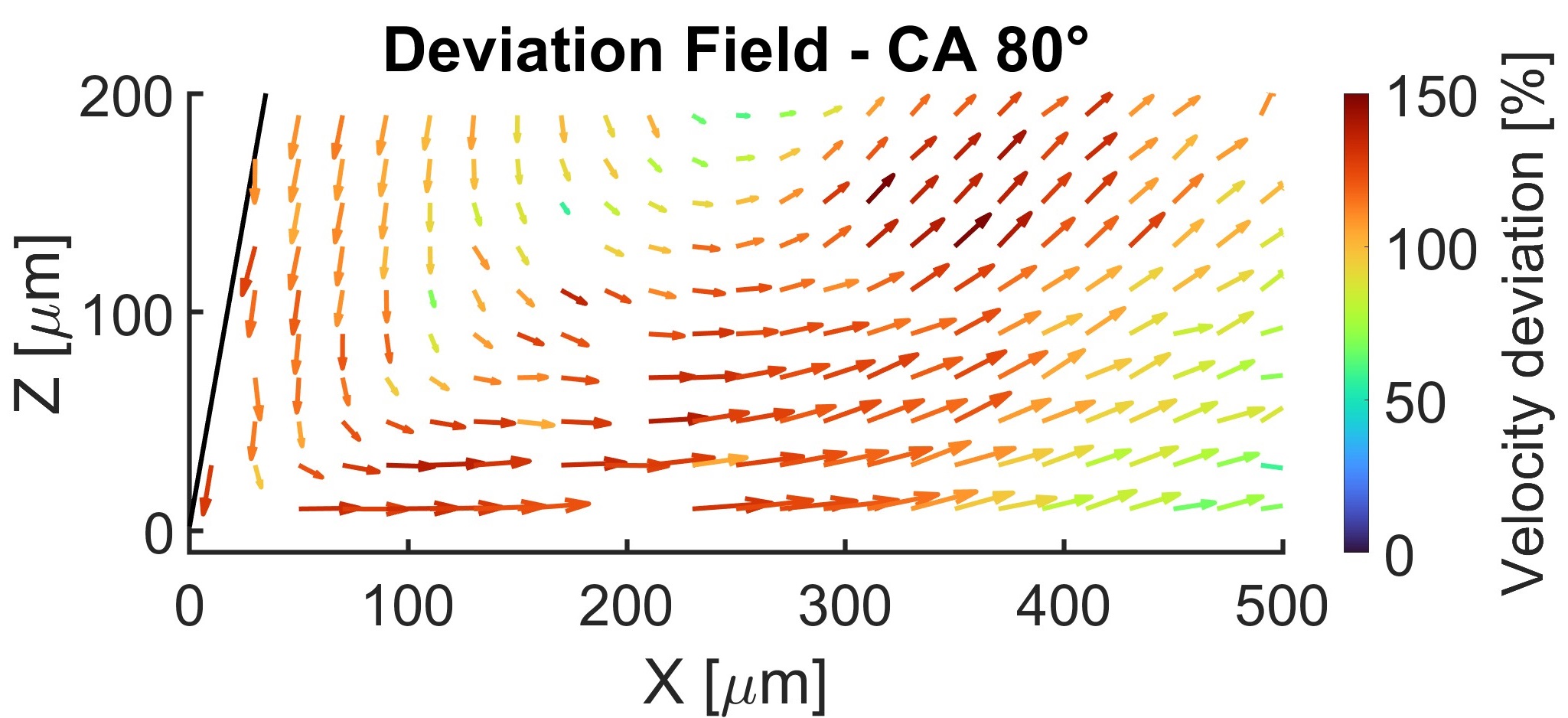}}

\caption{Comparison of experimental mean velocity fields obtained from APTV measurements (first column), flow fields predicted by the Moffatt solution (second column) and the deviation fields (third column), which quantify the discrepancy between experiment and theory at the same advancing contact angle. 
Shown are results for (a–c) the TDE dispersion medium, (d–f) the 30~wt\% moderately concentrated suspension in TDE, and (g–i) the 33~wt\% dense suspension in TDE, with corresponding contact angles of \SI{38}{\degree}, \SI{53}{\degree}, and \SI{80}{\degree}. 
For the experimental flow fields, the averaging is performed in the $x$–$z$ plane. 
Empty grid points indicate regions with insufficient tracer statistics for reliable averaging. 
For the theoretical flow fields, velocity vectors are plotted at the center of each \SI{20}{\micro\meter} $\times$ \SI{20}{\micro\meter} cell and are color-coded by their magnitude, calculated as $|\vec{v}_{xz}| = \sqrt{v_{x}^2 + v_{z}^2}$. 
The solid black line marks the advancing contact angle and defines the upper boundary of the wedge-shaped flow domain.}
    \label{fig:Average_Flowfield_Deviation}
\end{figure*}

\subsection{Advancing contact line of weakly interacting particles} \label{Advancing weakly interacting-Main}
In what follows, we examine the wetting dynamics of suspensions in TDE solution. 
We first describe the measured tracer trajectories and the resulting average flow fields, and then compare them with theory. 
Our analysis focuses on how the local flow structure near the advancing contact line interact with the non-Newtonian rheology of granular suspensions, rather than on testing the contact angle-capillary number relation.

We compare the average flow fields obtained from experiments across three representative cases: dispersion medium (TDE solution), moderately concentrated suspension (30~wt.\% silica in TDE), and dense suspension (33~wt.\% silica in TDE) with analytical predictions for Newtonian fluids near the advancing contact line. 
We interpret our experimental data using the Moffatt analytical solution for flow near a sharp corner~\citep{moffatt1964viscous}. 
Moffatt determined the solution to the Stokes equation in a wedge geometry and associated boundary conditions~\citep{moffatt1964viscous}. 
Both the Moffatt and Cox–Voinov solutions describe flow near a moving contact line, differing mainly in their treatment of the innermost region very close to the line, which typically spans from nanometers to a few micrometers~\citep{snoeijer2013moving}. 
Nevertheless, both descriptions are consistent within the outer, Stokes-flow region. 
Given that our measurements are performed using \SI{5}{\micro\metre} tracer particles, geometric constraints near the contact line set a lower bound on the accessible region. 
As the presence of the tracer particles perturbs the local flow, reliable velocity measurements can only be obtained at distances $l \gg D$, where $D$ is the tracer particle diameter, practically at least \SI{20}{\micro\metre} from the contact line. 
Our measurements therefore lie within the outer region, where the Moffatt solution provides an appropriate framework for interpreting the flow fields. 
Physical processes occurring closer to the contact line, from nanometers to a few micrometers, such as Van der Waals forces, viscous bending of the interface, and particle-induced interfacial deformations are not resolved in our measurements and lie beyond the scope of this analysis. 
The analytical expression of the Moffatt solution, along with the corresponding theoretical flow field used for comparison, is provided in the Supplementary Information, section~\ref{Moffatt theory-Suppl}.

Since the analysis involves comparison with a two-dimensional theoretical model, all flow fields are evaluated within the $x$-$z$ plane. 
Details of the mean flow field reconstruction from APTV data are provided in the Supplementary Information, section~\ref{Flow field reconstruction from APTV data-Suppl}. 
The resulting two-dimensional velocity fields are shown in the first column of Fig.~\ref{fig:Average_Flowfield_Deviation}. 
Each vector's color corresponds to its absolute velocity value, computed as $|\vec{v}_{xz,\;\mathrm{exp}}| = \sqrt{v_{x,\;\mathrm{exp}}^2 + v_{z,\;\mathrm{exp}}^2}$, where $v_{x,\;\mathrm{exp}}$ and $v_{z,\;\mathrm{exp}}$ are the experimentally measured velocity components in the horizontal ($x$) and vertical ($z$) directions, and the third component parallel to the contact line $v_{y,\;\mathrm{exp}}$ excluded.

To quantify the difference between experimental measurements and theoretical predictions, we compute a normalized deviation field. 
For each grid cell, the local deviation is defined as follows:
\begin{equation}
    v_{x,\;\mathrm{dev}} = v_{x,\;\mathrm{exp}} - v_{x,\;\mathrm{theo}}, \quad
    v_{z,\;\mathrm{dev}} = v_{z,\;\mathrm{exp}} - v_{z,\mathrm{theo}}
    \label{eq:component_deviation}
\end{equation}
where $v_{x,\;\mathrm{theo}}$ and $v_{z,\;\mathrm{theo}}$ denote the theoretical velocity components in the horizontal ($x$) and vertical ($z$) directions predicted by the Moffatt solution. 
$v_{x,\;\mathrm{dev}}$ and $v_{z,\;\mathrm{dev}}$ represent the corresponding deviations between experiment and theory.

The magnitude of deviation is then expressed as a normalized scalar quantity $S_{\mathrm{dev}}$,
\begin{equation}
    S_{\mathrm{dev}} = \frac{\sqrt{v_{x,\;\mathrm{dev}}^2 + v_{z,\;\mathrm{dev}}^2}}{\sqrt{v_{x,\;\mathrm{theo}}^2 + v_{z,\;\mathrm{theo}}^2}}\times 100  \;\; [\%],
    \label{eq:normalized_deviation}
\end{equation}
This formulation captures both magnitude and directional discrepancies. 
The resulting spatial map of $v_{\mathrm{dev}}$ and $S_{\mathrm{dev}}$ is shown in the third column of Fig.~\ref{fig:Average_Flowfield_Deviation}.

\begin{figure}[t]
    \centering
    \subfloat [\centering]
    {{\includegraphics[width=0.49\textwidth]{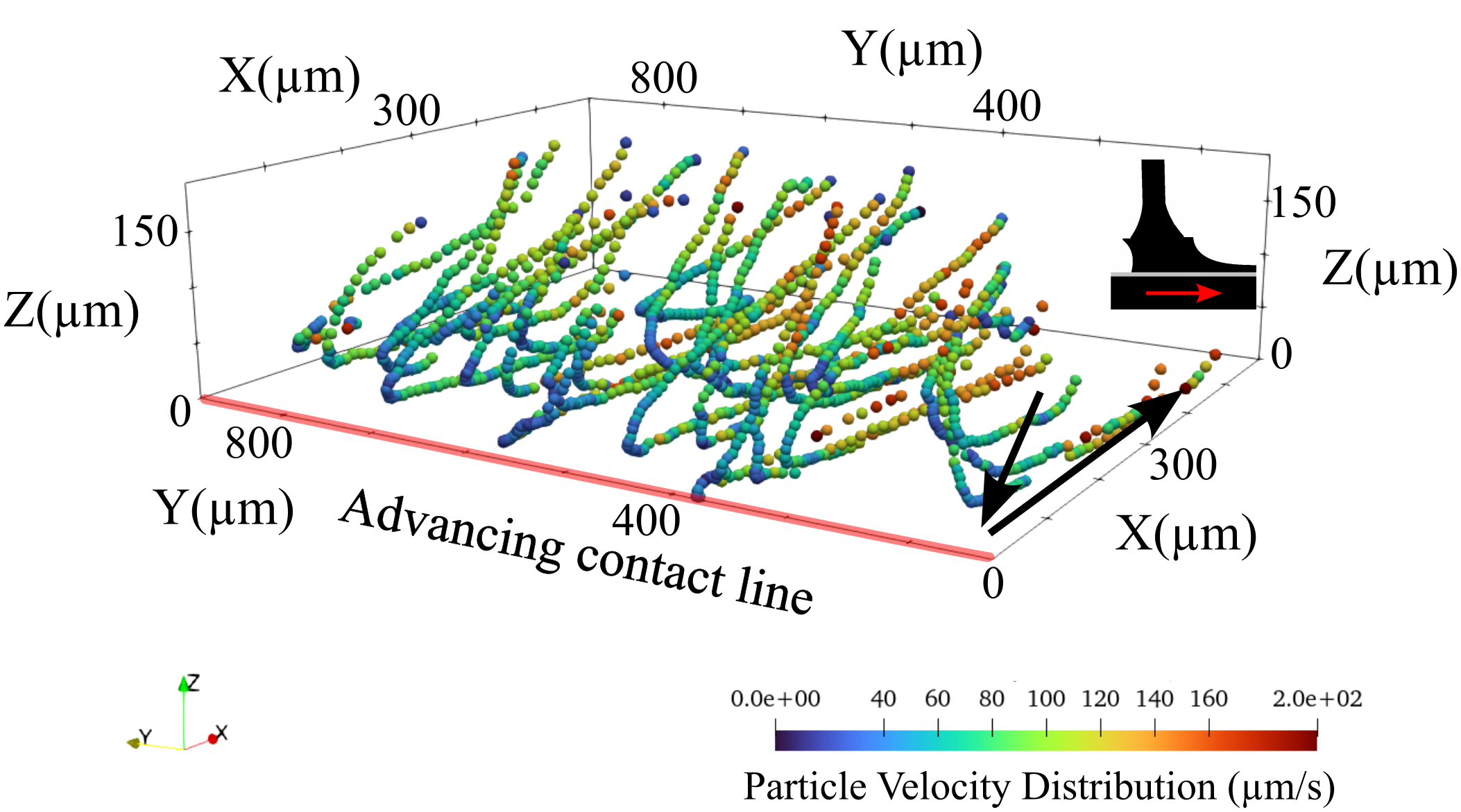}}}\\
    \subfloat [\centering]
    {{\includegraphics[width=0.49\textwidth]{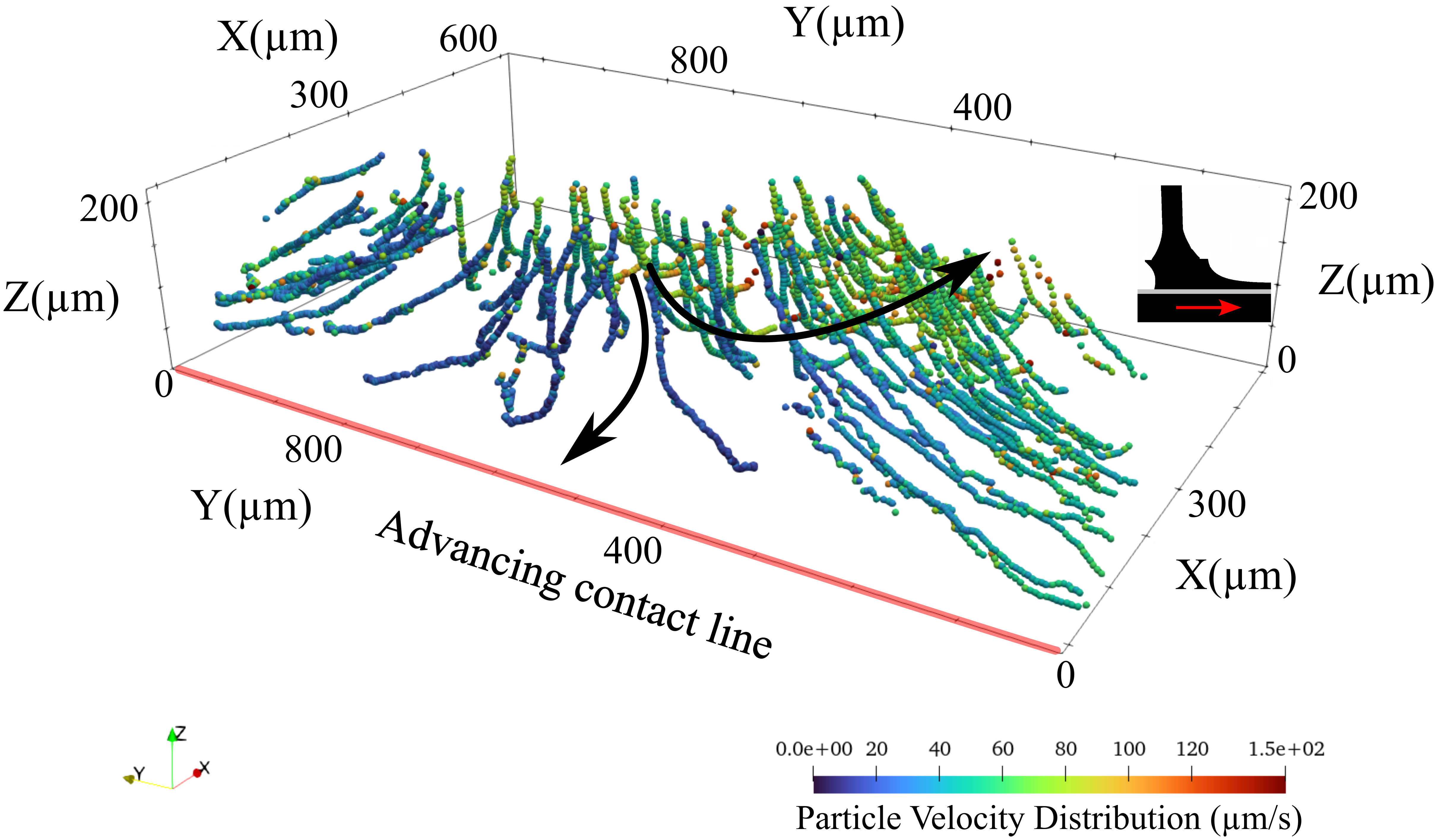}}}
\caption{Three-dimensional particle trajectories near the advancing contact line: (\textit{a}) 30~wt.\% silica particles in TDE solution, hydrodynamic solutions can predict the flow. 
(\textit{b}) 33~wt.\% silica particles in TDE solution, substantial deviation from hydrodynamic flow patterns. 
The data is shifted along the flow direction such that the $y$ axis coincides with the advancing contact line. 
The color bar indicates the particle velocity magnitude.
    \label{fig:flow field 30-33-TDE}}
\end{figure}

Fig.~\ref{fig:Average_Flowfield_Deviation}(a-f) illustrate the experimental flow fields, the theoretical predictions, and normalized deviation maps for the dispersion medium and the moderately concentrated suspension (30~wt.\%). 
In both cases, there is strong agreement between the measured velocity fields and the analytical Moffatt solution. 
The deviation magnitude remains generally below 25\%, confirming the applicability of classical hydrodynamic theory to these systems. 
Minor deviations arise from physical and experimental constraints. These include the finite spatial resolution imposed by the tracer particle size, the mismatch between the three-dimensional droplet curvature and the idealized two-dimensional wedge geometry of the Moffatt solution, and the local breakdown of the continuum assumption when approaching particle-scale dimensions (within approximately 50~\si{\micro\metre} of the contact line). 
Hence, three-dimensional trajectory visualizations are provided (see Fig.~\ref{fig:flow field 30-33-TDE}a) to capture the spatial complexity of granular suspension flows. 
Despite the increase in dynamic contact angle from 38$^\circ$ to 53$^\circ$ observed in the 30~wt.\% suspension, attributable to the enhanced effective viscosity at moderate particle loading, the measured flow field remains consistent with the hydrodynamic solutions.

As the particle concentration increases to 33~wt\%, entering the dense regime, we observe a clear alteration in the internal dynamics of the suspension near the contact line, a systematic departure from predictions of classical hydrodynamic models. 
Fig.~\ref{fig:Average_Flowfield_Deviation}(g–i) present the average velocity field and corresponding deviation map of the dense suspension. 
The advancing contact angle increases to 80$^\circ$. 
Two distinct types of tracer particle trajectories appear at approximately \SI{250}{\micro\metre} from the contact line: one moving with vanishing velocity toward the contact line, while the other moves toward the bulk. 
Within approximately \SI{250}{\micro\metre} to the contact line, tracer particles exhibit minimal displacement, with velocities remaining extremely low and approaching zero. 
This behaviour is indicative of a localized stagnation zone that persists throughout the duration of the experiment, which we interpret as the formation of a mechanically arrested structure. 
Over time, particles located within the stagnant region near the advancing contact line exhibit a gradual migration toward the edges of the droplet. 
Ultimately, the particles are either re-entrained into the flow or deposited onto the surface. 
This behaviour is clearly resolved in the three-dimensional trajectory data shown in Fig.~\ref{fig:flow field 30-33-TDE}b, where particles diverge away from the centerline, defined along the direction of relative motion. 
Additional evidence is provided by the top-view of the average flow fields presented in Fig.~\ref{Top view of the measurements-Suppl} of the Supplementary Information. 
Beyond this region, the flow resumes and particle trajectories show a circulation pattern. 
The particles initially move toward the advancing contact line. 
When they reach the proximity of the stagnation zone, they reorient parallel to the substrate and subsequently return to the bulk flow. 
The deviation field shown in Fig.~\ref{fig:Average_Flowfield_Deviation}(i) emphasizes the extent of mismatch between experimental data and continuum model predictions. 
3D trajectories for another run of the same experiment is provided in Fig.~\ref{Deviation-Hydrodynamics} of the Supplementary Information, reaffirming its reproducibility.

Another characteristic of Fig.~\ref{fig:flow field 30-33-TDE}b is the slower particle movement on the substrate in comparison to the substrate velocity.
Specifically, the particle's highest velocity within the TDE suspension droplet, measured close to the substrate, was \SI{150}{\micro\metre\per\second}. 
This is noticeably slower than the substrate's velocity, which was set to \SI{200}{\micro\metre\per\second}. 
This behaviour is consistent with particles rolling and sliding relative to one another and along the substrate, rather than moving affinely with the imposed substrate motion. 

Near the advancing contact line, the local shear rates are sufficiently high to trigger non-Newtonian effects. 
If we consider a distance $d$ in the range of \SI{5}{\micro\metre} (approximately one particle diameter) to \SI{1}{\milli\metre} from the contact line, and assume a characteristic contact line velocity $U = \SI{0.2}{\milli\metre\per\second}$, the local shear rate $\dot{\gamma}=U/d$ is estimated to be in the range $\sim$ \SIrange{0.2}{40}{\per\second}. 
The altered flow pattern reflects the strong coupling between local shear and the strong shear-thickening response observed in shear rate dependent viscosity measurements, Fig.~\ref{fig:rheology}. 
This interpretation is further supported by oscillatory strain sweep measurements, which reveal pronounced strain-induced stiffening in the TDE-based suspension, as evidenced by a sharp increase in both $G'$ and $G''$ beyond approximately 30\% strain amplitude (see Supplementary Information, Fig.~\ref{fig:Osillatory_Measurements}a). 
Furthermore, considering the viscosity-volume fraction relationship~\citep{maronApplicationReeeyringGeneralized1956}, the suspension concentration lies close to $\phi_c$, so that even modest external stresses can drive the system toward a shear-jammed, solid-like state~\citep{petersDirectObservationDynamic2016}.

When subjected to shear, granular suspensions respond differently depending on solid fraction. 
Dilute suspensions compact, whereas dense suspensions dilate. 
This dilation originates from the steric and mechanical constraints inherent to densely packed suspensions, where frictional contacts and geometric exclusion inhibit affine particle motion, thereby necessitating a local expansion of the network to accommodate shear-induced rearrangements~\citep{nessPhysicsDenseSuspensions2022, srivastavaViscometricFlowDense2021}. 
In our system with 33~wt.\% silica particles in TDE, the solid fraction is above the dilation threshold, placing the suspension in the dense regime (see Supplementary Information, section~\ref{Protorheology measurements-Suppl}). 
Within the proximity of the high shear zone near the advancing contact line, as the system attempts to dilate under shear, particles press against one another more strongly. 
This dilation is associated with an increase in particle pressure and the emergence of normal stress differences~\citep{deboeufParticlePressureSheared2009}. 
At the particle scale, shear is expected to develop an anisotropic contact and force network, and in the present dense regime this anisotropy is accompanied by shear-induced dilation~\citep{setoNormalStressDifferences2018}. 
Similarly, we expect such anisotropic stress states to influence local flow resistance and particle dynamics near the contact line, as observed in our APTV measurements.

Compared to Moffatt’s solution, the Cox–Voinov model is inherently more complex, as it employs matched asymptotic expansions to bridge the slip-dominated inner region near the contact line with the outer viscous flow. 
Previous studies have demonstrated that Cox–Voinov solution can predict the advancing contact line behaviour of granular suspensions for concentrations up to 75 \% of their critical jamming concentration $\phi_{c}$ when the suspension's effective viscosity is taken into account~\citep{Han2012Spreading, zhao2020spreading, pelosse2023probing}. 
In our experiments, however, we find that although the apparent contact angle increases with an increase in particle concentration, the average flow field near the advancing contact line of a dense suspension deviates significantly from the predictions of hydrodynamic models, for the outer region defined in Cox-Voinov framework.

The Cox–Voinov model was developed under the Stokes approximation, assuming small capillary number and viscous (Newtonian) flow where stress scales linearly with strain rate. 
Cox himself noted that non-Newtonian effects could also regularize the moving contact line singularity~\citep{cox1986dynamics}. 
While subsequent work has extended the framework to complex fluids by incorporating shear rate dependent viscosities, such as shear-thinning~\citep{bonn2009wetting}, no corresponding theory exists for concentrated granular suspensions. 
In such suspensions, momentum transfer is governed by frictional particle contacts rather than viscous stresses. 
These contact networks generate strong non-linearities that dominate the flow near the moving contact line, rendering the classical Cox–Voinov model inadequate for describing dense suspensions.

The Cox–Voinov solution assumes a quasi-steady state, with viscous dissipation as the sole energy dissipation mechanism. 
In dense granular suspensions, however, energy is dissipated not only via flow through the particle network but also at the interparticle contacts within the force-bearing structure~\citep{setoNormalStressDifferences2018}. 
This introduces localized flow variations and stress fluctuations, rendering the quasi-steady assumption inadequate for describing such systems.

Furthermore, the Cox–Voinov solution relies on the continuum hypothesis, which treats the liquid as a uniform medium across all scales. 
In our system with \SI{5}{\micro\metre} particles, the discrete nature of the suspended phase generates localized flow disturbances. 
At low to moderate particle concentrations, the Cox–Voinov description remains applicable, as the disturbances induced by individual particles are effectively averaged out. 
Our spatial resolution (\SI{20}{\micro\metre}), approximately four times larger than the particle size, effectively smooths over the particle scale heterogeneities. 
Thus, the suspension can still be considered as an effective continuum. 
As the suspension approaches the jamming threshold, particle motion becomes increasingly collective, and the associated microstructural length scales grow substantially. 
In this regime, the continuum approximation breaks down, and our flow measurements exhibit pronounced deviations from Cox–Voinov predictions near the advancing contact line.

\subsection{Advancing contact line of strongly interacting particles}\label{Advancing strongly-Main}

\begin{figure}[t]
  \centerline{\includegraphics[width=0.35\textwidth]{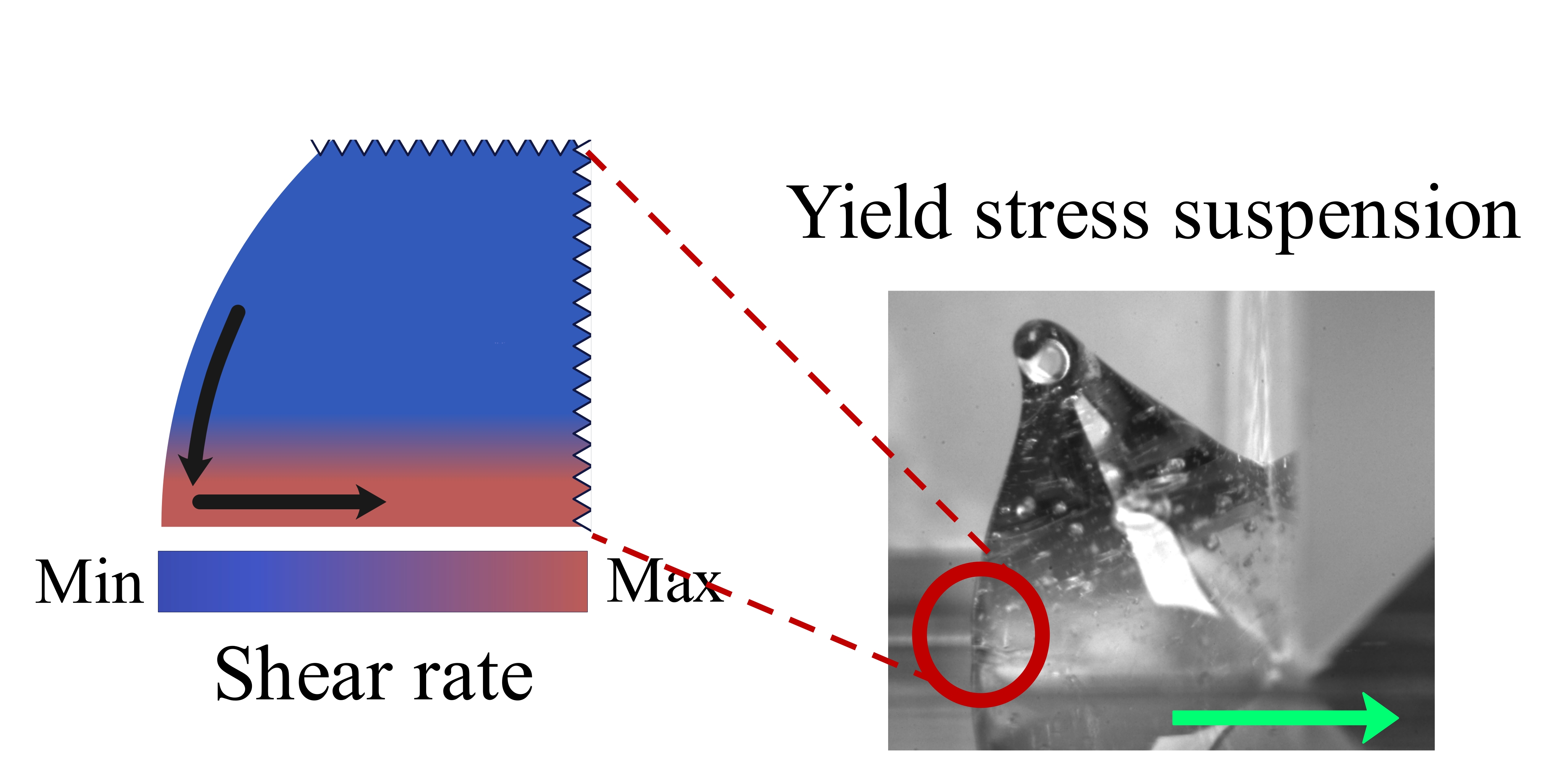}}
\caption{Side view of a droplet containing 30~wt.\% silica particles in NaSCN solution pushed over a substrate. 
The blunted tip of the droplet, along with the entrapped air bubbles, indicates a high yield stress. 
The red circle marks the distribution of shear rate within the droplet near the advancing contact line.
  \label{fig:Yielded region}}
\end{figure}

\begin{figure*}[t]
    \centering
    \subfloat[\centering]
    {{\includegraphics[scale=0.52]{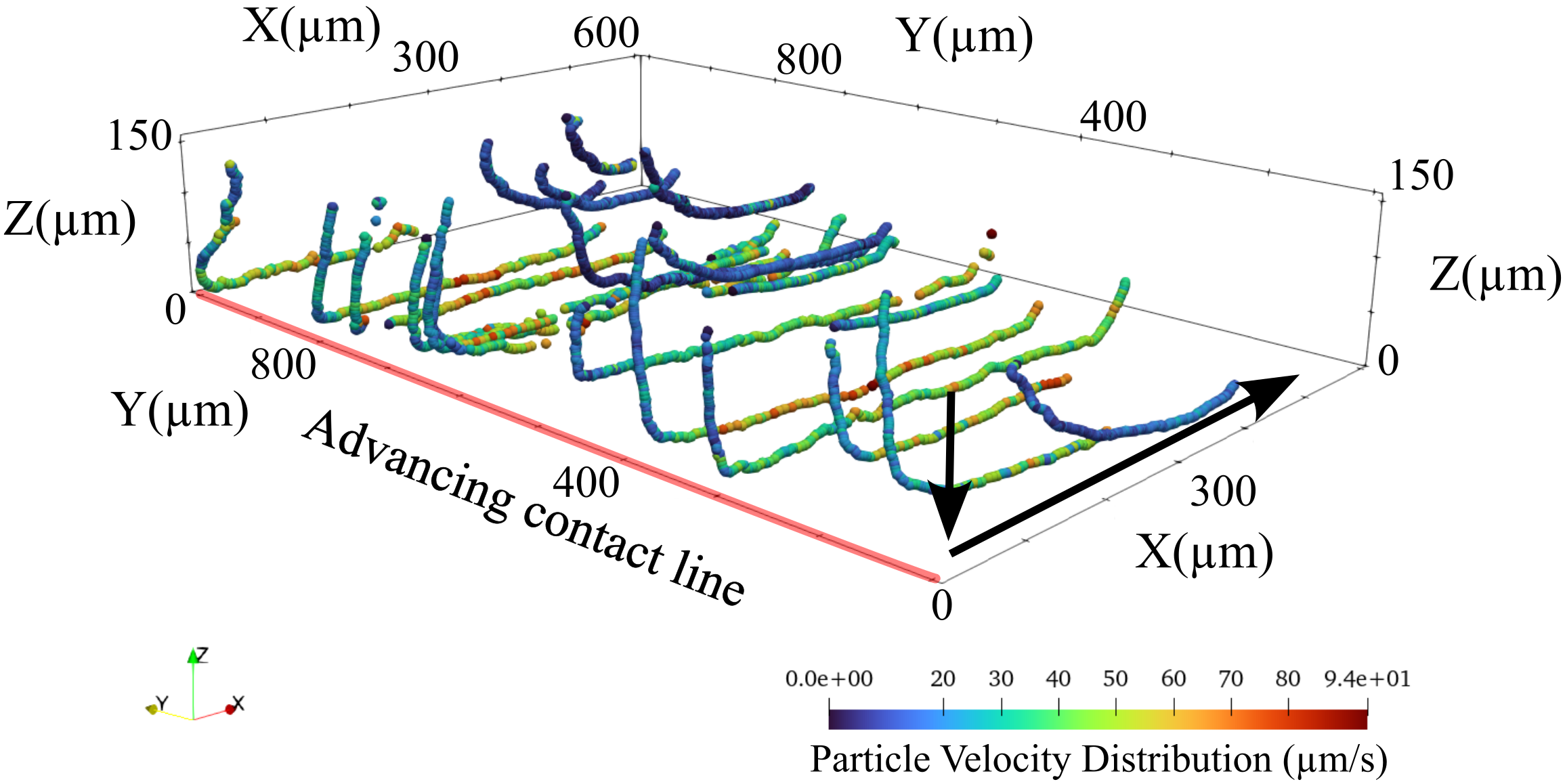} }}
    \qquad
    \subfloat[\centering]
    {{\includegraphics[scale=0.30]{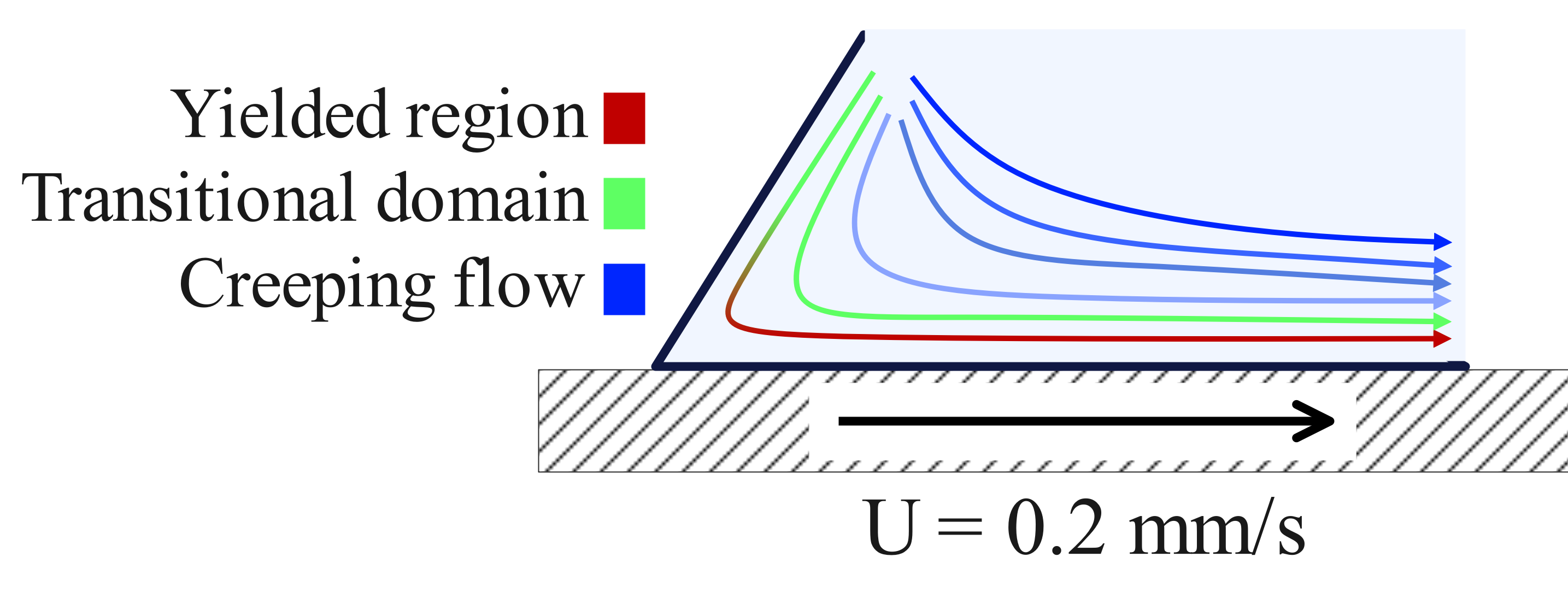}}}
\caption{Three-dimensional trajectories of tracer particles near the advancing contact line of a suspension containing 30~wt.\% silica particles in NaSCN solution. 
(\textit{a}) Experimental measurements, where color bar indicates particle velocity magnitude. 
A yield-stress suspension is formed; upon initiation of relative motion, the droplet bulk undergoes creep deformation. 
Particles migrate toward the contact line and, upon entering the high-shear region close to the substrate, they move parallel to it. 
(\textit{b}) Schematic illustration highlighting the distinct flow regimes: creeping flow (blue), transitional domain (green), and yielded region (red). The imposed substrate velocity is \SI{0.2}{\milli\metre\per\second}.
	\label{fig:Flow field-NaSCN}}
\end{figure*}


In what follows, we first present the droplet shapes and particle trajectories for the NaSCN-based suspension, and then discuss how these observations relate to its yield-stress behaviour. 
Unlike Newtonian fluids, yield-stress droplets spread spontaneously only when capillary stresses exceed the yield threshold~\citep{martouzetDynamicArrestSpreading2021}. 
Capillary stresses may induce limited flow near the substrate, but the upper region remains solid-like, and spreading eventually halts at a finite contact angle~\citep{martouzetDynamicArrestSpreading2021}. 
In our experiments we consider forced wetting, where substrate motion imposes shear stresses exceeding the yield stress, thereby sustaining localized shear-driven flow near the advancing contact line.

For silica suspensions in NaSCN solution, the droplet retains a solid-like character in the absence of substrate motion. 
This behaviour is consistent with strong interparticle rolling and sliding friction that maintains a persistent contact network, effectively inhibiting flow. 
This is consistent with the high yield-stress observed in shear rate dependent viscosity measurements (Fig.~\ref{fig:rheology}) and further supported by oscillatory strain sweep measurements (Supplementary Information, Fig.~\ref{fig:Osillatory_Measurements}b). 
As a result, the droplet retains its deposited shape and resists deformation, Fig.~\ref{fig:Yielded region}. 
The circular disk cannot exert sufficient force to hold the droplet in place during substrate motion. 
To overcome this limitation, a prism geometry is used to hold the droplet while the substrate moves. 
A side-view image of the yield-stress droplet in motion is provided in Fig.~\ref{fig:Yielded region}. 
The shape of the droplet tip following deposition with a pipette reflects a balance between yield-stress and surface tension, where elevated yield-stress suppresses capillary-driven thinning~\citep{luu2009drop, aytounaDropFormationNonNewtonian2013}. 

Once the substrate starts moving, particle layers near the substrate experience a high shear stress. 
When the local stress exceeds the yield threshold, the suspension in proximity to the substrate undergoes yielding and initiates flow. 
This is consistent with the shear-thinning observed in our rheological measurements (Fig.~\ref{fig:rheology}). 
The resulting particle velocity distribution indicates the presence of three different regions, as shown in the schematic representation of particle trajectories in Fig.~\ref{fig:Flow field-NaSCN}b. 
The suspension near the substrate transitions into a liquid-like state, whereas the rest of the droplet experiences creep deformation, with a transitional region occurring between these two regions. 
Here, creep deformation refers to the very slow, gradual rearrangement of particles under stresses below the yield threshold~\citep{houssaisAthermalSedimentCreep2021} and should not be confused with creeping flow (Stokes flow), which describes inertialess, viscosity-dominated motion in Newtonian fluids at low Reynolds numbers. 
Creep deformation happens in regions where localized stresses are not high enough to cause a strong flow but slightly exceed the local frictional thresholds between particles. 
The particle structure rearranges itself, allowing gradual deformation through minute movements of particles relative to each other. 
Particles situated within the bulk of the droplet, where the prevailing shear stress has yet to exceed the yielding threshold, demonstrate a tendency to gradually migrate towards the transitional zone. 
Upon entry into this transitional domain, a change in their type of motion takes place, causing them to adopt a faster motion parallel to the substrate. 
The system reaches a dynamic equilibrium, where particle rearrangement balances the applied shear, resulting in steady flow near the high-shear region adjacent to the substrate. 
Consequently, when observed from the outside, the motion seems akin to a mobile plug traversing the surface. 
This coexistence of yielded and non-yielded regions within the droplet is consistent with earlier observations~\citep{martouzetDynamicArrestSpreading2021}, which also reported a non-flowing region far from the substrate.

The measured trajectories of tracer particles near the advancing contact line of the yield-stress suspension are shown in Fig.~\ref{fig:Flow field-NaSCN}a. 
Tracer particles move at velocities consistently below that of the substrate. 
The maximum tracer velocity measured is a factor of two smaller than the substrate velocity, suggesting persistent rolling and sliding along the substrate. 
Additionally, the average particle velocity in the NaSCN solution is notably lower than in the TDE solution. 
Specifically, the maximum particle velocity in the TDE suspension is \SI{150 \pm 10}{\micro\metre\per\second}, compared to \SI{94 \pm 10}{\micro\metre\per\second} in the NaSCN suspension. 
The reduced velocity in the NaSCN solution highlights the profound influence of solvent properties on the rolling and sliding of particles relative to each other and the substrate.

A quantitative estimate of the locally fluidized layer thickness is obtained from bulk rheology data and compared with the flow-field measurements (Supplementary Information, section~\ref{Section: Yielded-layer thickness in NaSCN-based wetting experiments}). 
Fitting the NaSCN-based suspension rheology to a Herschel–Bulkley constitutive relation yields a characteristic yield-onset shear rate $\dot{\gamma}_0 \approx \SI{1.9}{\per\second}$; together with the imposed substrate velocity $U = \SI{0.2}{\milli\metre\per\second}$, this implies a characteristic yielded-layer thickness $\delta \sim U/\dot{\gamma}_0 \approx \SI{100}{\micro\metre}$. 
Consistently, the height–velocity histogram of the tracer velocity magnitude (Supplementary Information, Fig.~\ref{fig:Vxy-density}) shows a narrow band of elevated velocities confined to the first $\approx \SI{50}{\micro\metre}$ above the substrate, beyond which the velocity rapidly decays into a low-value plateau.

The emergence of both partially and fully yielded regions within the droplet aligns with the two step yielding framework~\citep{ahujaTwoStepYielding2020}. 
In this framework, jammed suspensions yield sequentially through an initial rupture of weak interparticle bonds (such as adhesive contacts), followed by a second yielding event associated with larger scale structural rearrangements, such as cage escape or aggregate breakdown. 
In our system, both the droplet bulk and the region near the substrate undergo the first yielding stage. 
However, only the high-shear region adjacent to the substrate experiences the second yielding stage, wherein the constraints imposed by densely packed particle cages are overcome and the material transitions into a fully fluidized state.

\section{Conclusion}
This study elucidates the complex interplay between contact line dynamics and the rheology of dense granular suspensions. 
By employing APTV to resolve three-dimensional flow fields near the advancing contact line and complementary rheological measurements, we directly probed the shear rate dependent internal flow fields of suspensions comprising \SI{5}{\micro\metre} silica particles dispersed in two distinct refractive index-matched media. 
This allowed us to systematically explore how particle interactions, due to the different dispersion media, influence the wetting dynamics and the corresponding suspension flow behaviour. 
Near the advancing contact line, there is a localized high-shear zone where the non-Newtonian nature of these granular suspensions emerges. 
For instance, at extremely high concentrations of particles close to the jamming concentration, silica suspensions in TDE solutions exhibited a transition from hydrodynamic lubrication-dominated flow to frictional contact-dominated behaviour as particle concentration increased, with strong deviations from the flow pattern predicted by conventional solutions of hydrodynamic models near the contact line. 
In contrast, silica suspensions in NaSCN solution, characterized by yield-stress and shear-thinning behaviour, showed distinct flow states; the droplet showed a plug flow over the substrate with a Newtonian-like behaviour within the boundary layer adjacent to the substrate. 
These differences underscore the importance of particle-scale interactions, which were validated through complementary CP-AFM and angle of repose measurements. 
These results highlight the limitations of traditional hydrodynamic models in capturing the complex dynamics of dense suspensions. 
The assumptions underlying classical relations, such as the Cox-Voinov relation, become invalid under conditions of high particle concentration or significant interparticle friction. 
Instead, the discrete nature of particles, shear-induced structural transitions, and non-Newtonian rheological responses must be integrated into theoretical frameworks to accurately predict contact line behaviour. 
This understanding has far-reaching implications for industrial applications, where controlling wetting dynamics and suspension flow behaviour is essential for achieving uniform deposition, structural stability, and reliable performance in processes such as 3D printing, coatings, and cosmetics.


\section{Supplementary information (SI)}
\subsection{Protorheology measurements} \label{Protorheology measurements-Suppl}

\begin{figure*}
    \centering
    \subfloat [\centering]
    {{\includegraphics[scale=0.37]{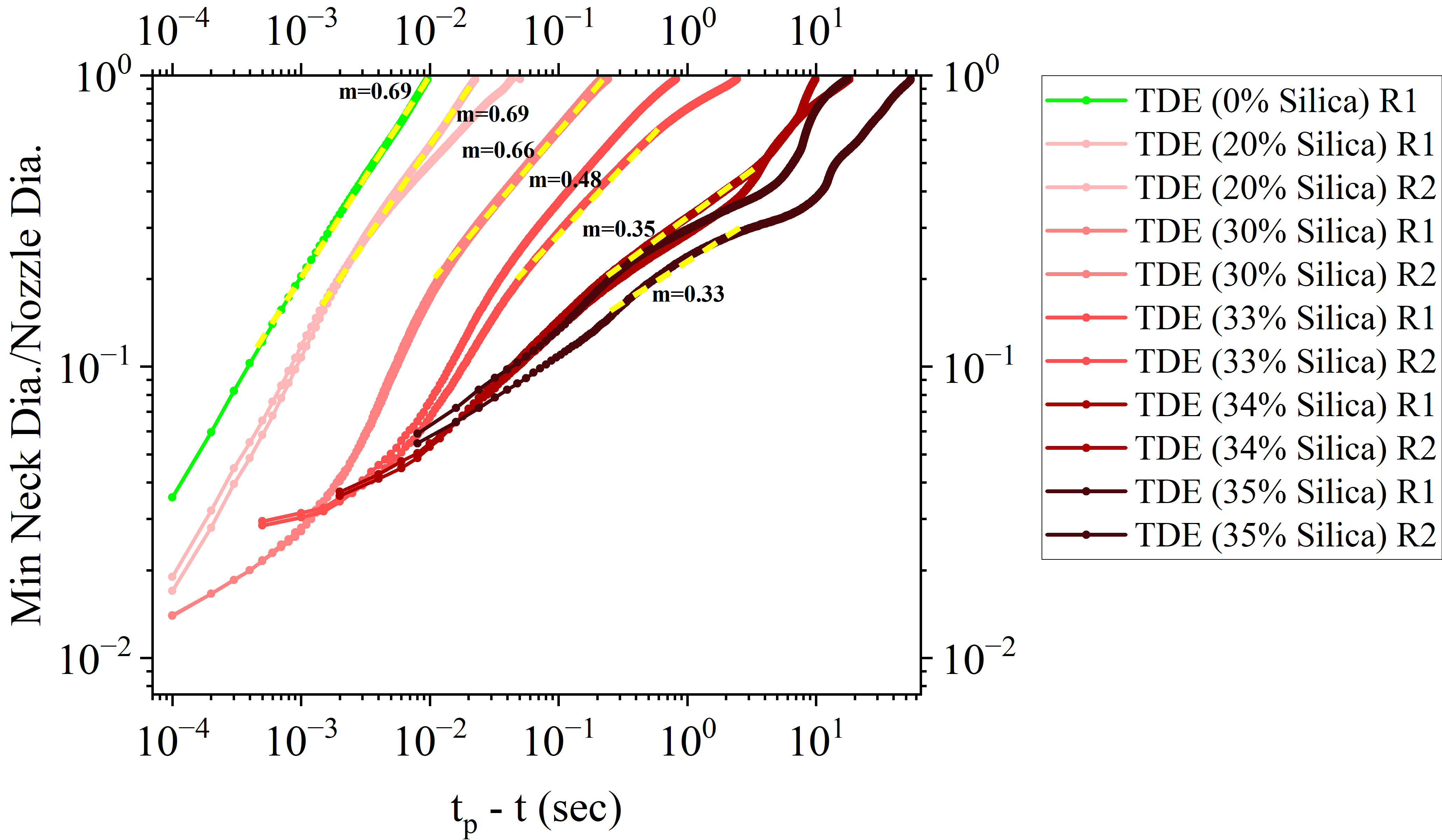}}}\\
    \hspace*{2mm} 
    \subfloat [\centering]
    {{\includegraphics[scale=0.37]{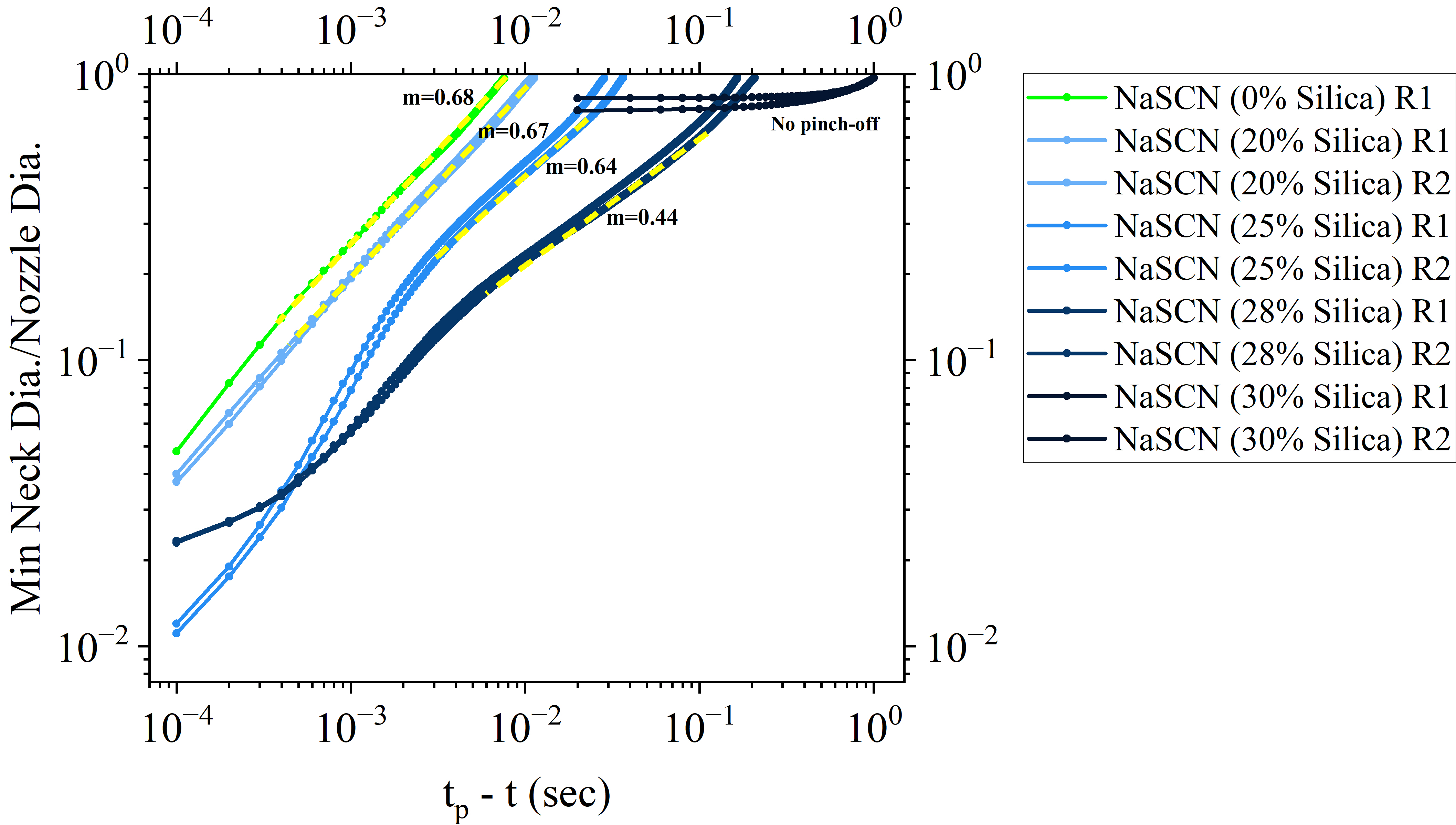}}}
\caption{Minimum neck diameter normalized by nozzle diameter plotted as a function of time to pinch-off $t_p - t$ on a log-log scale for dripping onto substrate (DoS) experiments. 
The green curves in both panels represent the pure dispersion media without any particles. 
(\textit{a}) TDE-based suspensions are shown in red and (\textit{b}) NaSCN-based suspensions in blue, with increasing color intensity indicating increasing silica concentration in each system. 
Data from two independent runs (R1, R2) are shown for each suspension. 
The annotated slopes represent the power-law exponents $m$ obtained from fits of the form $h(t) \sim (t_p - t)^m$, evaluated within the early stage of neck thinning after its dynamics become decoupled from the preceding wetting dynamics.}
    \label{fig:TDE_NaSCN_DOS}
\end{figure*}

\begin{figure*}[t]
    \centering
    \subfloat [\centering]
    {{\includegraphics[scale=0.33]{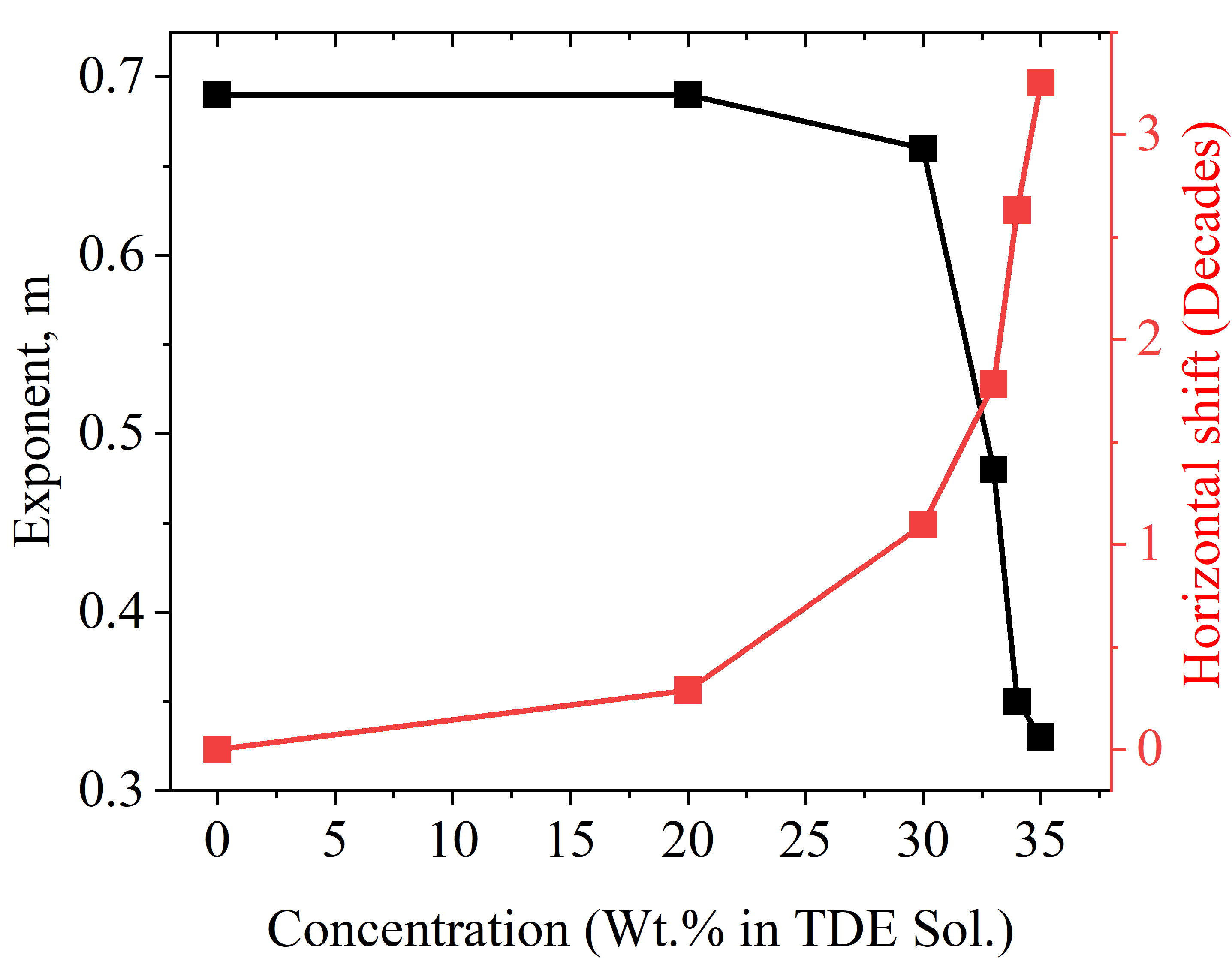}}}\:
    \subfloat [\centering]
    {{\includegraphics[scale=0.33]{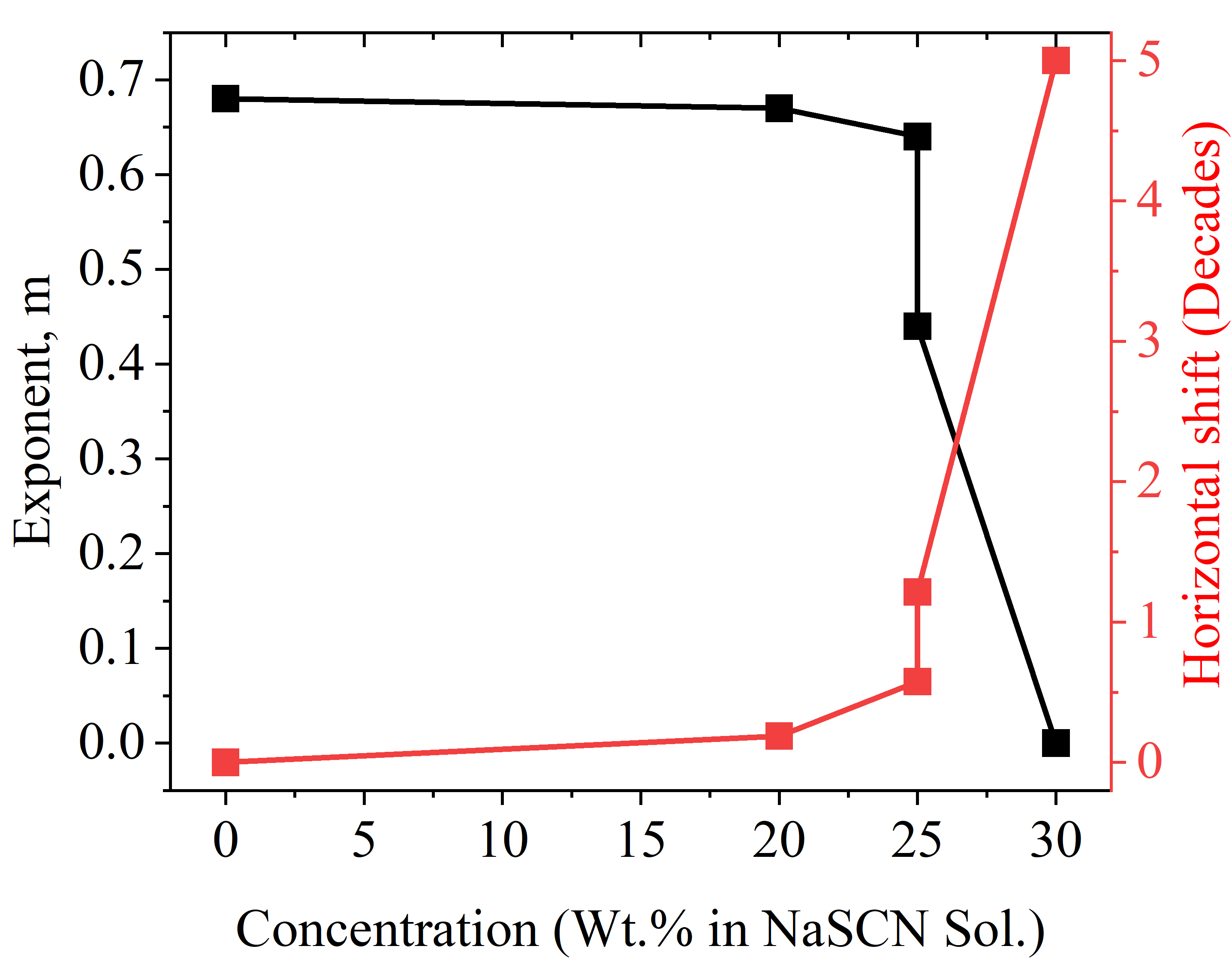}}}
\caption{Thinning exponent $m$ (black, left axis) and horizontal shift in time to pinch-off (red, right axis) as functions of particle concentration for (a) TDE-based suspensions and (b) NaSCN-based suspensions. 
The exponents were obtained from power-law fits to the log–log evolution of the normalized minimum neck diameter during droplet pinch-off. 
The horizontal shift quantifies the temporal displacement of the fitted power-law curves, calculated as the difference in $\log_{10}(t_p-t)$ relative to the solvent at the normalized neck diameter $=$ 0.3, expressed in decades. 
For NaSCN suspensions at 30~wt.\%, the thinning exponent approaches zero and the temporal shift diverges. 
For plotting purposes, a large finite value was assigned to the temporal shift at this concentration. 
}
    \label{fig:Exponent_Shift_TDE_NaSCN_DOS}
\end{figure*}

To determine the relative proximity to the critical concentration of each suspension system, we performed dripping onto substrate (DoS) experiments over a broad range of concentrations. 
The concentration series spanned \SIrange{0}{35}{}~wt.\% for TDE-based suspensions and \SIrange{0}{30}{}~wt.\% for NaSCN-based suspensions.
These experiments were motivated by the protorheological approach, wherein the pinch-off dynamics of a liquid bridge is used to infer rheological behaviour~\citep{dinicPinchoffDynamicsDrippingontosubstrate2017}. 
In our experimental configuration, a pendant drop of \SI{18}{\micro\liter} was slowly formed at the tip of a stainless steel 14G nozzle (outer diameter $=$ \SI{2.1}{\milli\meter}) at a controlled flow rate of \SI{18}{\micro\liter\per\second}. 
A cleaned and dried hydrophilic glass substrate was gently raised until it made contact with the droplet, initiating the spreading of the droplet on the glass substrate and formation of a capillary bridge between the nozzle and the substrate. 
Subsequent capillary driven thinning and pinch-off of the droplet neck were captured using a FASTCAM Mini AX200 high speed camera operating at frame rates ranging from \SI{500}{\framepersecond} to \SI{10000}{\framepersecond}, depending on the dynamics of the sample under investigation. 
The recorded image sequences were analyzed using custom MATLAB scripts to extract the minimum neck radius as a function of time. 
The resulting radius evolution data were plotted on a log–log scale and are presented in Fig.~\ref{fig:TDE_NaSCN_DOS}a and b for the TDE-based and NaSCN-based suspensions, respectively.  
In all cases, the time origin is defined as the instant of droplet pinch-off ($t_p$), corresponding to the final frame in which a continuous liquid bridge is observed.

%
We analyze the early stage thinning dynamics of capillary bridges in DoS experiments, specifically focusing on the regime where the thinning dynamics is decoupled from drop dynamics on the substrate and the suspension remains macroscopically homogeneous. 
The results, reveal systematic trends in the thinning dynamics as a function of concentration. 
For TDE-based suspensions Fig.~\ref{fig:TDE_NaSCN_DOS}a, the evolution of the minimum neck radius can be described by a power-law scaling $h(t) \sim (t_p - t)^m$, where $h(t)$ denotes the minimum neck radius,  $t_p - t$ is the time to the pinch-off moment, and $m$ is the scaling exponent.

At low to moderately concentrated suspensions (\SIrange{0}{30}{}~wt.\%), the observed exponents remain nearly constant across concentration: $m = 0.69$ for the pure TDE solution, $m = 0.69$ for the 20~wt.\%, and $m = 0.66$ for the 30~wt.\% suspension. 
These values closely follow the dynamics observed in the pure solvent, albeit with a discernible temporal shift in the evolution curves to longer timescales, consistent with enhanced bulk viscosity, Fig.~\ref{fig:Exponent_Shift_TDE_NaSCN_DOS}a. 
This consistency in the exponent suggests that the initial pinch-off dynamics are predominantly governed by an effective Newtonian response, wherein the increased viscosity arises from the presence of suspended particles, as reported in capillary bridge studies of non-Brownian suspensions~\citep{alexandrouBreakupCapillaryBridge2010, bonnoitAcceleratedDropDetachment2012}.

A markedly different regime emerges for more concentrated systems (\SIrange{33}{35}{}~wt.\%). 
Here, a pronounced reduction in the power-law exponent is observed, with $m = 0.48$, $0.35$, and $0.33$ for 33~wt.\%, 34~wt.\%, and 35~wt.\%, respectively. 
This transition to a slower thinning regime suggests the onset of nontrivial effects beyond a simple increase in effective viscosity. 
The progressive reduction in slope with increasing concentration, which reaches a minimum indicates a growing resistance to flow, likely arising from frictional contacts, and hindered particle mobility, and the emergence of local microstructural heterogeneities. 
This is in agreement with prior reports that identify such behaviour as indicative of a transition from a continuum like regime to a disordered, particle dominated regime where local fluctuations in volume fraction increasingly dictate the dynamics, approaching the jamming transition~\citep{thievenazOnsetHeterogeneityPinchoff2022, chateau2018pinch}.

The 35~wt.\% concentration, was the highest flowable limit within our experimental configuration. 
This is shown in Fig.~\ref{fig:Exponent_Shift_TDE_NaSCN_DOS}a, where the corresponding curve exhibits the lowest thinning exponent and the largest temporal shift. 
We interpret this threshold as an operational estimate of the critical concentration, $\phi_c$, for the TDE suspension. 
While the jamming transition itself is not directly measurable under these conditions, the pronounced suppression of thinning rates and the convergence of the curves onto a low-exponent thinning regime strongly indicate that the system approaches the jammed state.

\begin{table*}[t]
\small
  \caption{Silica particle concentrations used in different experiments}
  \label{tab:Concentrations}
  \begingroup
  \renewcommand{\arraystretch}{1.2}  
  \begin{tabular*}{\textwidth}{@{\extracolsep{\fill}}llll}
    \noalign{\vspace{0.5em}}  
    \hline
    \noalign{\vspace{0.5em}}  
    Experiment type & Dispersion medium & Silica concentration & Notes \\
    \noalign{\vspace{0.5em}}  
    \hline
    \noalign{\vspace{0.5em}}  
    Protorheology (DoS pinch-off)   & TDE Sol.  & 0--35~wt.\% & $\phi_c$ $\approx$ 35~wt.\% \\
                                    & NaSCN Sol. & 0--30~wt.\% & $\phi_c$ $\approx$ 30~wt.\% \\
    Rheology (shear \& oscillatory) & TDE Sol.   & 31~wt.\%   & -- \\
                                    & NaSCN Sol. & 28~wt.\%   & -- \\
    Flow field measurements         & TDE Sol.   & 30~wt.\%, 33~wt.\% & -- \\
                                    & NaSCN Sol. & 30~wt.\%   & -- \\
    \noalign{\vspace{0.5em}}  
    \hline
    \noalign{\vspace{0.5em}}  
  \end{tabular*}
  \endgroup
\end{table*}

\begin{figure*}[t]
    \centering
    \includegraphics[scale=0.4]{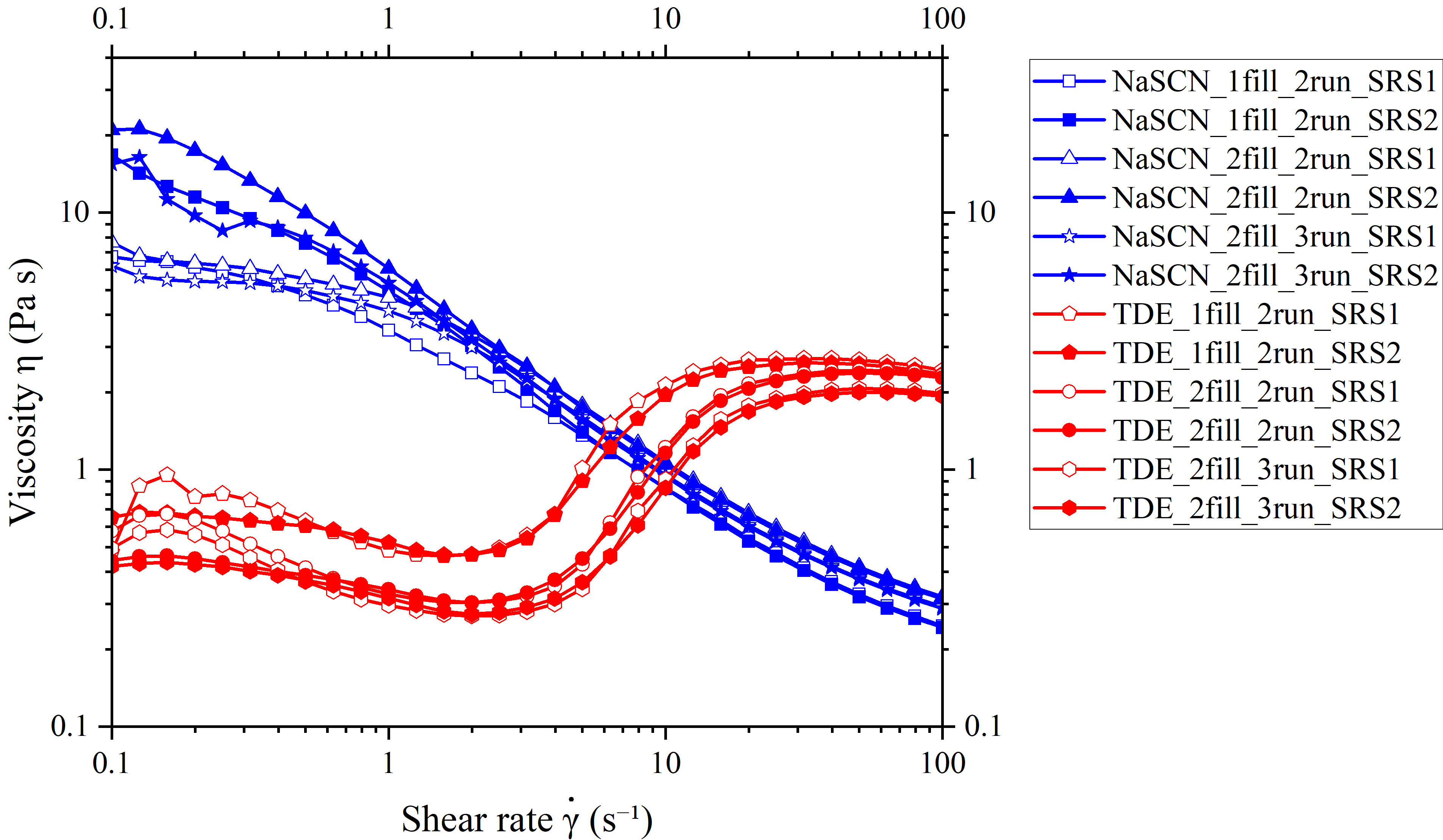}
    \caption{Shear rate dependent viscosity measurements for dense granular suspensions comprising \SI{5}{\micro\metre} silica particles dispersed in 31~wt.\% TDE and 28~wt.\% NaSCN solutions. 
    Measurements were conducted using a parallel cylinder (Couette) geometry with a fixed inner cylinder and rotating outer cylinder, across a shear rate range of \SI{0.1}{\per\second} to \SI{100}{\per\second}. 
    The viscosity is plotted as a function of shear rate. 
    Multiple curves per suspension correspond to repeated runs performed over multiple sample fillings to ensure data reproducibility.}
    \label{SRS-Rheology}
\end{figure*}

%
The interpretation of the DoS plot for NaSCN suspensions Fig.~\ref{fig:TDE_NaSCN_DOS}b is guided by insights from our rheometry measurements, which exhibit a yield-stress characteristic for the NaSCN suspension (see section~\ref{Rheology measurements-Suppl}), and angle of repose measurements, which indicate a higher friction between particles dispersed in NaSCN solution (see section~\ref{Angle of repose-Suppl}). 

The NaSCN-based suspensions, similar to the TDE-based suspensions, exhibit a strong dependence on particle concentration, yet with distinct rheological signatures from the TDE system. 
At lower concentrations (\SIrange{0}{25}{}~wt.\%), the evolution of the minimum neck radius during the early stage of pinch-off, when the dynamics are decoupled from substrate effects, follows a power-law scaling, with exponents $m = 0.68$, $0.67$, and $0.64$ for the 0, 20, and 25~wt.\% suspensions, respectively. 
The modest reduction in $m$ and the systematic shift of the thinning curves to longer time scales are indicative of a gradual increase in effective viscosity with increasing particle content, Fig.~\ref{fig:Exponent_Shift_TDE_NaSCN_DOS}b. 
This behaviour aligns with previous studies on dilute to semi-dilute suspensions, where the fluid remains below the yield-stress threshold and microstructural effects are subdominant~\citep{louvetNonuniversalityPinchOffYield2014}.

A notable deviation emerges at 28~wt.\%, where the scaling exponent sharply drops to $m = 0.44$. 
This marked reduction in thinning rate suggests the onset of yield-stress like behaviour, wherein the stress required to drive the pinch-off becomes comparable to the capillary stresses. 
In contrast to the TDE system, where jamming manifests through a gradual reduction in the thinning exponent, the NaSCN suspensions exhibit a more abrupt transition, Fig.~\ref{fig:Exponent_Shift_TDE_NaSCN_DOS}. 
At a high concentration of 30~wt.\%, a cylindrical filament forms at the nozzle tip in place of a Laplacian droplet, indicating yield-stress behaviour. 
No thinning or pinch-off is observed. 
Upon contacting the substrate, the droplet spreads slightly and arrests, with no further evolution. 
This lack of thinning implies the suspension no longer yields under capillary stress, consistent with our rheological measurements (see section~\ref{Rheology measurements-Suppl}). 
Since no pinch-off occurs, the zero time reference cannot be defined, and the time axis for this curve is set arbitrarily as the start of image acquisition. 
This arrested state marks the flowability limit under deformation and provides a practical estimate of approaching the critical concentration, $\phi_c$.

Together, the absence of thinning at 30~wt.\% and the sudden reduction in thinning exponent at 28~wt.\% highlight the emergence of a yield-stress in NaSCN suspensions. 
Notably, the maximum flowable concentration in this system is reached at a lower particle loading compared to the TDE suspensions. 
This is consistent with the higher rolling and sliding friction between the particles denoted by angle of repose measurements (see section~\ref{Angle of repose-Suppl}). 

An overview of the silica particle concentrations employed across the different experiments is provided in Table~\ref{tab:Concentrations}.

\begin{figure*}[t]
    \centering
    
    \subfloat[\centering]{
        \includegraphics[scale=0.32]{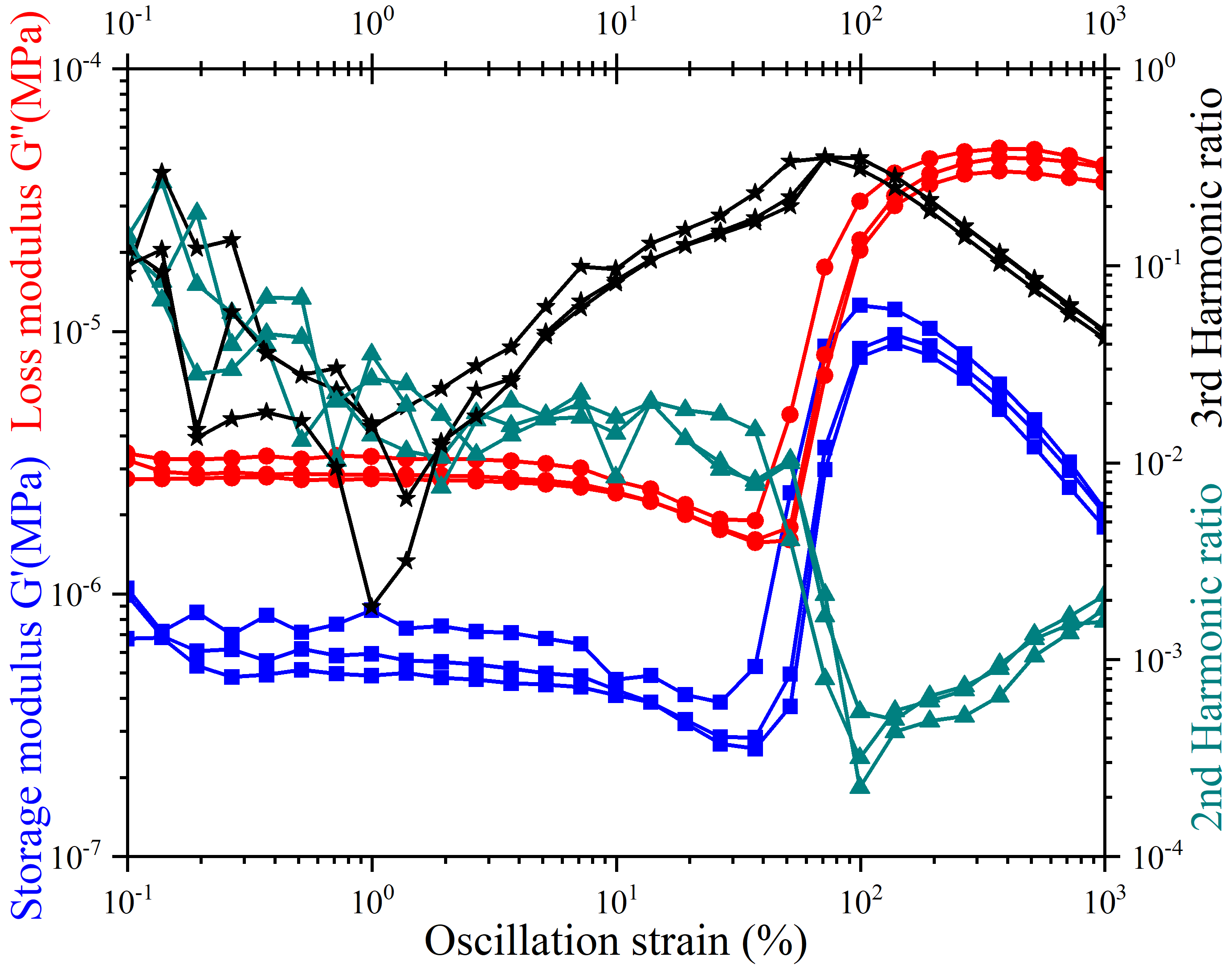}
    }\quad
    \subfloat[\centering]{
        \includegraphics[scale=0.32]{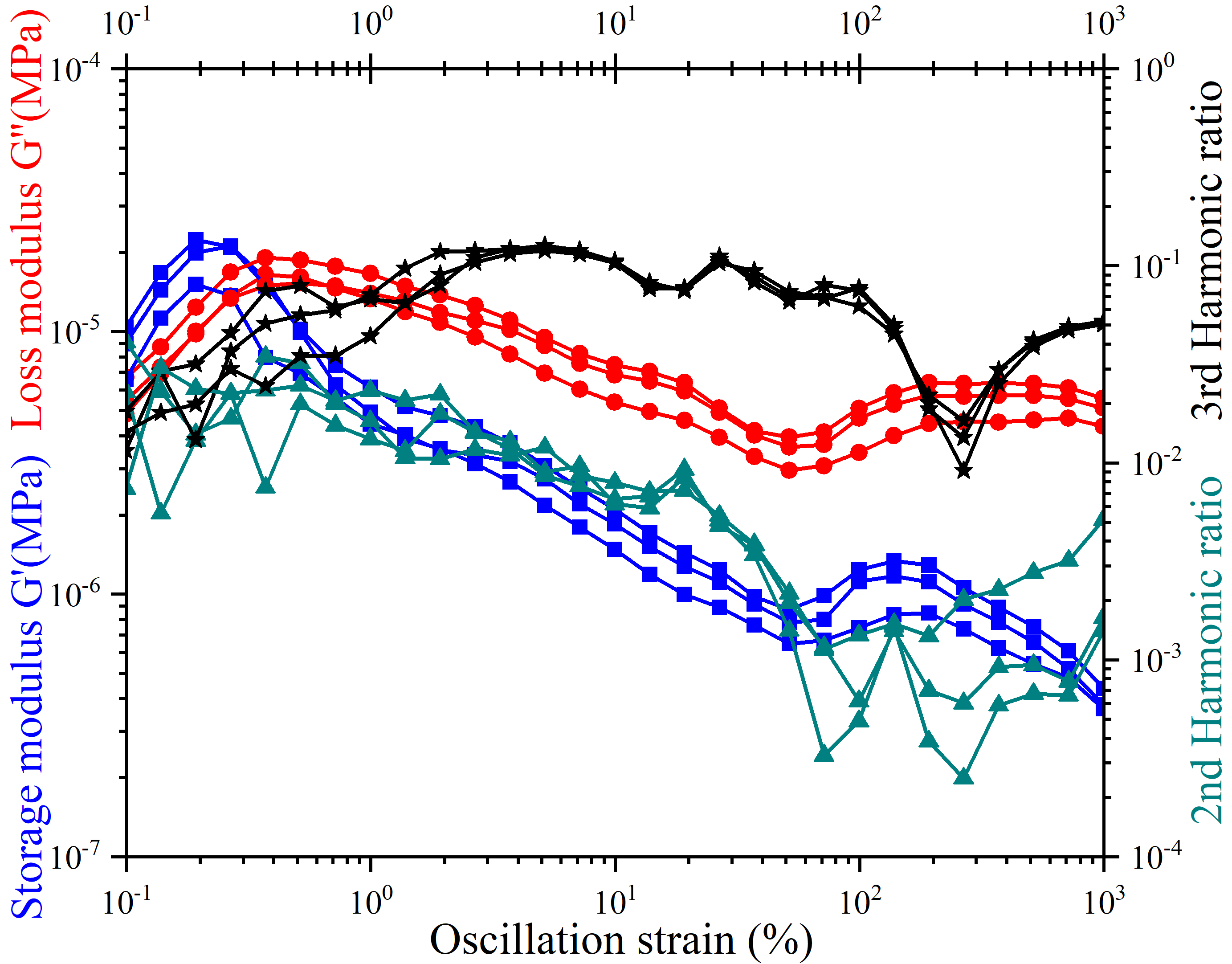}
    }
    \caption{Oscillatory strain sweep measurements for granular suspensions of \SI{5}{\micro\metre} silica particles dispersed in (a) 31~wt.\% TDE and (b) 28~wt.\% NaSCN, conducted at a fixed oscillation frequency of \SI{3}{\hertz} over a strain amplitude range of \SI{0.1}{\percent} to \SI{1000}{\percent}. 
    The storage modulus ($G'$, blue dash-square) and loss modulus ($G''$, red dash-circle) are plotted against oscillatory strain using the left vertical axis, capturing the elastic and viscous components of the mechanical response, respectively. 
    The second and third harmonic intensity ratios ($|I_2/I_1|$, green dash-triangle; and $|I_3/I_1|$, black dash-star) are plotted using the right vertical axis to characterize nonlinearity and slip behaviour. 
    Multiple curves of the same color represent measurements obtained from multiple runs over multiple sample fillings, ensuring data reproducibility.}
    \label{fig:Osillatory_Measurements}
\end{figure*}

\subsection{Rheology measurements} \label{Rheology measurements-Suppl} 
\subsubsection{Shear rate dependent viscosity measurements} \label{Shear rate dependent viscosity measurements-Suppl}

Granular suspensions are prepared by dispersing \SI{5}{\micro\metre}  silica particles in a TDE or NaSCN solution. 
These suspensions contain 31~wt.\% and 28~wt.\% silica particles, respectively. 
A series of rheological experiments was performed using a high-end ARES G2 rheometer from TA-Instruments, to assess the response of suspensions to the applied shear rate. 
A concentric cylinder geometry (Couette cell) was utilized. 
In a Couette geometry, the inner cylinder remains stationary, while the outer cylinder rotates. 
The dimensions of the geometry are as follows: an inner diameter of the cup of \SI{19.99}{\milli\metre}, an outer diameter of the bob of \SI{18.6}{\milli\metre}, and a height of the bob of \SI{27.93}{\milli\metre}. 
With an operating gap of \SI{0.7}{\milli\metre} this translates to a required sample volume of \SI{2.3}{\centi\metre\cubed}.

To initiate each experiment, the sample is filled into the cup of the Couette geometry. 
The bob was then carefully inserted into the filled cup, ensuring that no air bubbles were trapped and that the suspension occupied the entire annular gap. 
The measurements were started immediately to prevent sedimentation of the particles. 
The time required for filling the geometry, inserting the bob, and initiating the measurement procedure was sufficiently long to allow temperature equilibration. 
Due to the significantly greater mass of the cylinder, the sample has the same temperature as the cylinder. 

To ensure reproducibility, different runs were performed for each filling of the Couette cell. 
The first run was always discarded, and only the second and third runs were taken into account. 
Each measurement sequence consisted of five consecutive steps: 
(i) a first frequency sweep, (ii) an amplitude sweep, (iii) a second frequency sweep, (iv) a first shear rate sweep, and (v) a second shear rate sweep. 
The results of the two shear rate sweeps (steps iv and v) are shown in Fig.~\ref{SRS-Rheology}, and the amplitude sweep (step ii) is presented in Fig.~\ref{fig:Osillatory_Measurements}. 
The frequency sweeps (steps i and iii) were not considered in the subsequent analysis. 

Additional shear rate sweep data in Fig.~\ref{SRS-Rheology} confirm the high reproducibility across multiple runs and different fillings of the Couette cell.

\begin{figure*}[t]
    \centering
    \subfloat[\centering]
    {{\includegraphics[width=0.415\textwidth]{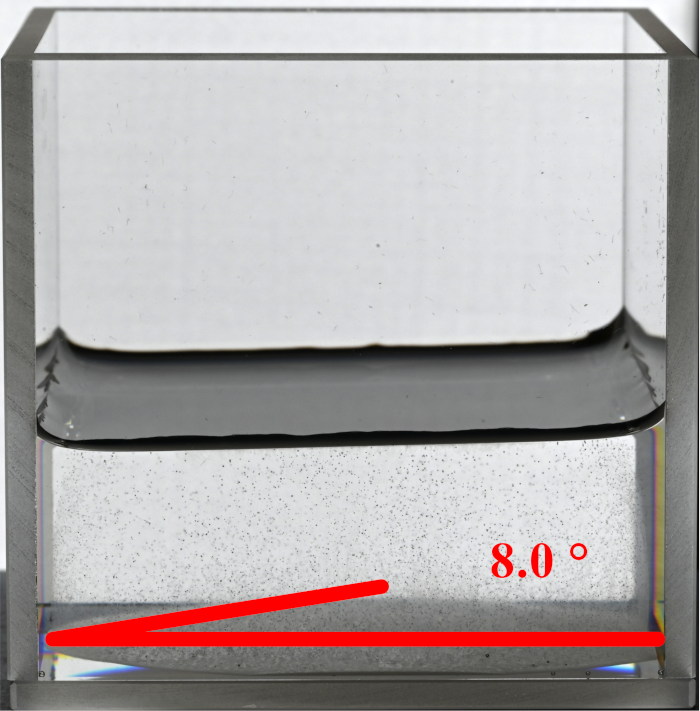} }}
    \qquad
    \subfloat[\centering]
    {{\includegraphics[width=0.44\textwidth]{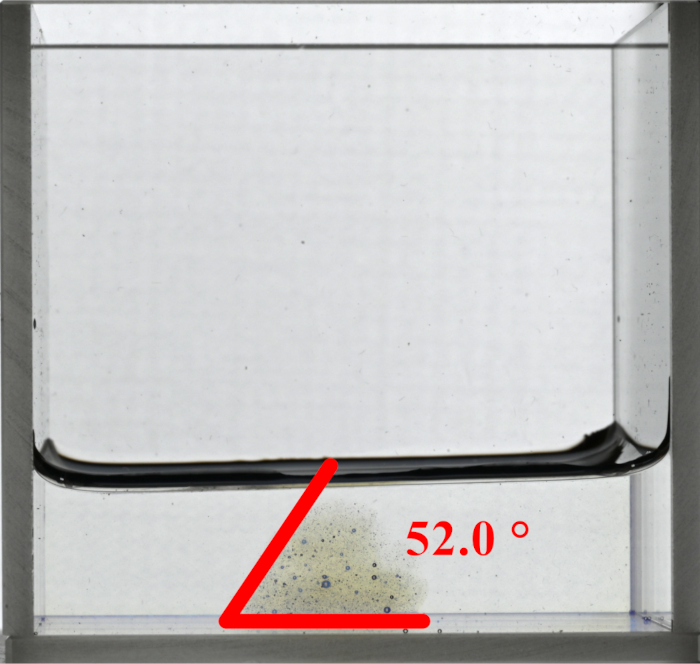}}}
    \caption{A comparison of the angle of repose for submerged silica particles in (\textit{a}) TDE solution and (\textit{b}) NaSCN solution, illustrating the greater friction between particles immersed in NaSCN solution.
\label{fig:Angle_of_Repose}}
\end{figure*}

\subsubsection{Oscillatory strain sweep measurements} \label{Oscillatory strain sweep measurements-Suppl}

Oscillatory strain sweep measurements, presented in Fig.~\ref{fig:Osillatory_Measurements}a and b, were performed at a fixed oscillation frequency of \SI{3}{\hertz}, with strain amplitudes ranging from 0.1\% to 1000\%. 
This broad range enables a detailed assessment of how the mechanical response of two granular suspensions evolves with increasing deformation. 
The viscoelastic behaviour is characterized by the storage modulus ($G'$), which quantifies the elastic response, and the loss modulus ($G''$), which quantifies viscous dissipation. 
Both moduli are reported in \si{\mega\pascal} as functions of oscillatory strain (\%), and each system was evaluated over multiple runs and independent cell fillings to assess reproducibility.

In the TDE-based suspension (Fig.~\ref{fig:Osillatory_Measurements}a), $G'$ and $G''$ scale proportionally across the entire strain range, so that $\tan\delta = G''/G'$ remains approximately constant. 
$G''$ consistently exceeds $G'$ by roughly half an order of magnitude. 
At small to intermediate strain amplitudes, both moduli remain relatively constant with a slight decreasing trend, suggesting that the underlying particle network undergoes minimal structural evolution under modest deformations. 
Beginning at strains around ($\sim$30-50\%), both $G'$ and $G''$ increase sharply, indicating strain induced stiffening and enhanced energy dissipation, likely arising from the formation of interparticle contacts and transient force chains under oscillatory shear. 
This behaviour aligns with the shear-thickening observed in the corresponding shear rate dependent viscosity measurement (see Fig.~\ref{SRS-Rheology}). 
At higher strains ($\sim$100-200\%), $G'$ reaches a pronounced maximum before declining, indicative of the onset of structural degradation or yielding. 
In contrast, $G''$ plateaus and remains elevated, suggesting that while the elastic framework weakens, energy dissipation remains substantial. 

The second and third harmonic responses of the TDE suspension under oscillatory strain are presented in Fig.~\ref{fig:Osillatory_Measurements}a. 
Typically, elevated second harmonic ratios, $|I_2/I_1|$, indicate asymmetric stress responses. 
Even harmonics are generally suppressed under no-slip conditions~\citep{saint-michelLocalOscillatoryRheology2016}. 
The third harmonic ratio, $|I_3/I_1|$, reflects bulk nonlinearity in the material ~\citep{atalikOccurrenceEvenHarmonics2004}. 
The interpretation of harmonic responses can be more nuanced in specific cases. 
At low strain amplitudes (approximately \SIrange{0.1}{2}{}~\%), the second harmonic response remains within the noise-dominated regime and is not interpreted. 
As the strain increases from approximately \SIrange{2}{30}{}~\%, the second harmonic plateaus, while the third harmonic ratio, $|I_3/I_1|$, rises steadily from $\sim$2\% to $\sim$80\% strain, marking the emergence of pronounced bulk nonlinearities. 
This increase in the third harmonic coincides with slight decreases in both the storage and loss moduli. 
Beyond $\sim$30\% strain, $|I_2/I_1|$ drops sharply, pointing to reduced slip contributions. 
Simultaneously, both $G'$ and $G''$ rise sharply, indicating the emergence of a mechanically reinforced structure. 
This transition highlights a shift from boundary-influenced behaviour at lower strains to a response governed by internal particle interactions and structural rearrangements at higher strains.

In the NaSCN-based suspension, Fig.~\ref{fig:Osillatory_Measurements}b, the mechanical response exhibits features characteristic of a yield-stress fluid. 
At low strain amplitudes, the storage modulus $G'$ exceeds the loss modulus $G''$, indicating a predominantly elastic behaviour and the presence of a microstructure capable of sustaining stress. 
As the strain amplitude increases from approximately \SIrange{0.1}{1}{}~\%, a crossover of $G'$ and $G''$ is observed. 
This crossover is indicative of yielding, marking the transition from an initially solid like structure to a flowable, viscous dominated regime. 
Beyond the crossover, $G''$ remains larger than $G'$, and both moduli exhibit a gradual decline with increasing strain amplitude. 
The monotonic decrease in $G'$ reflects the progressive breakdown of the stress bearing particle network, while the sustained magnitude of $G''$ suggests continued energy dissipation through viscous mechanisms. 
The broad, gradual nature of this transition, without evidence of strain induced stiffening, implies a yielding process. 

The second and third harmonic trends in the NaSCN suspension suggest the absence of wall slip. 
The second harmonic ratio, $|I_2/I_1|$, is slightly noisy yet reliable and decreases steadily with increasing strain, with variations remaining within the noise level across the entire strain amplitude range. 
In parallel, the third harmonic ratio, $|I_3/I_1|$, remains nearly constant, features indicative of symmetric, bulk dominated nonlinear behaviour.

\subsection{Angle of repose measurements} \label{Angle of repose-Suppl} 

When granular materials are poured through a funnel onto a flat substrate, they form a conical pile. 
The friction coefficient between particles is determined by the tangent of the inclination angle~\citep{andreotti2013granular}. 
Although the concept is traditionally applied to dry granular systems, it can be extended to fully submerged particles, provided that hydrodynamic forces do not dominate~\citep{clavaud2017revealing}. 
Angle of repose measurements primarily capture the rolling friction between particles.
In our study, the angle of repose serves as a proxy for assessing the microscopic friction coefficient of silica particles dispersed in different media.

In our experiments, a cylindrical tube containing a 30~wt.\% silica suspension was gently withdrawn from a \SI{5}{\centi\metre} $\times$ \SI{5}{\centi\metre} cuvette filled with the same fluid, resulting in a conical pile of particles within the fluid. 
Analysis of side-view images indicated that the angle of repose was \SI{8 \pm 2}{\degree} in the TDE solution and significantly higher \SI{52 \pm 5}{\degree}, in the NaSCN solution (Fig.~\ref{fig:Angle_of_Repose}).

This pronounced difference highlights the stronger frictional interactions between particles in the NaSCN medium. 
The electrostatic double layer normally acts as a lubricating interface, reducing direct particle contact and thereby lowering friction. 
In a high ionic strength solution such as NaSCN, the electrostatic double layer is compressed by ion screening~\citep{butt2018surface}, leading to increased direct contacts and thus higher friction.

\subsection{Atomic force microscopy measurements.} \label{AFM-Suppl}

\begin{figure*}[t]
\centering
\includegraphics[scale=0.6]{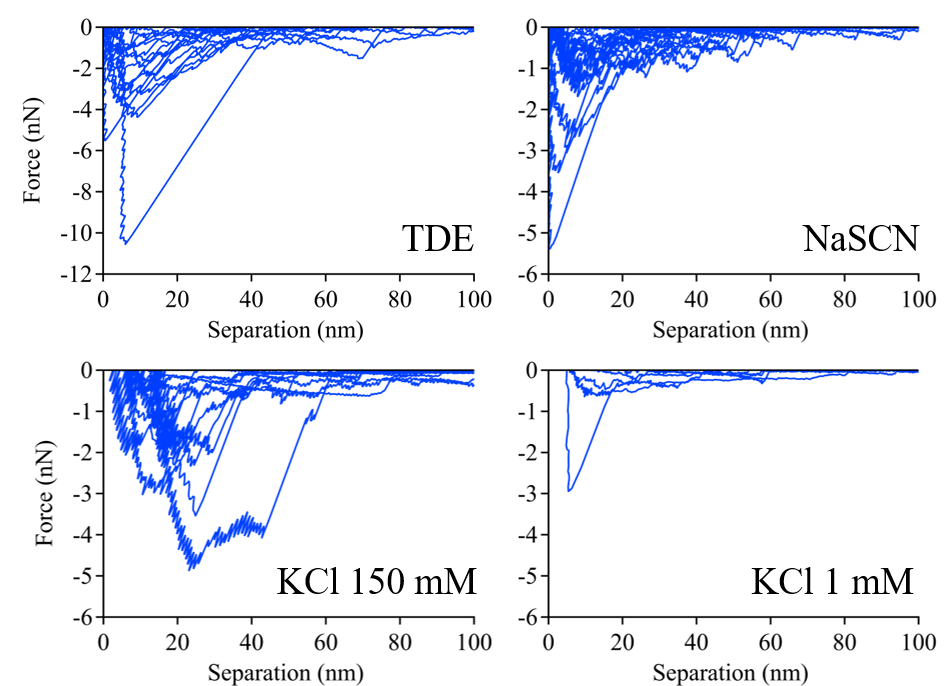}
\caption{Adhesion force curves from direct force measurements with CP-AFM. The experiments were conducted in different solutions from left to right, TDE 69.75~wt.\%, NaSCN \SI{11.65}{M}, KCl \SI{150}{\milli M} and \SI{1}{\milli M}.}
\label{AFM-adhesion}
\end{figure*}

\begin{figure*}[t]
  \centerline{\includegraphics[scale=0.35]{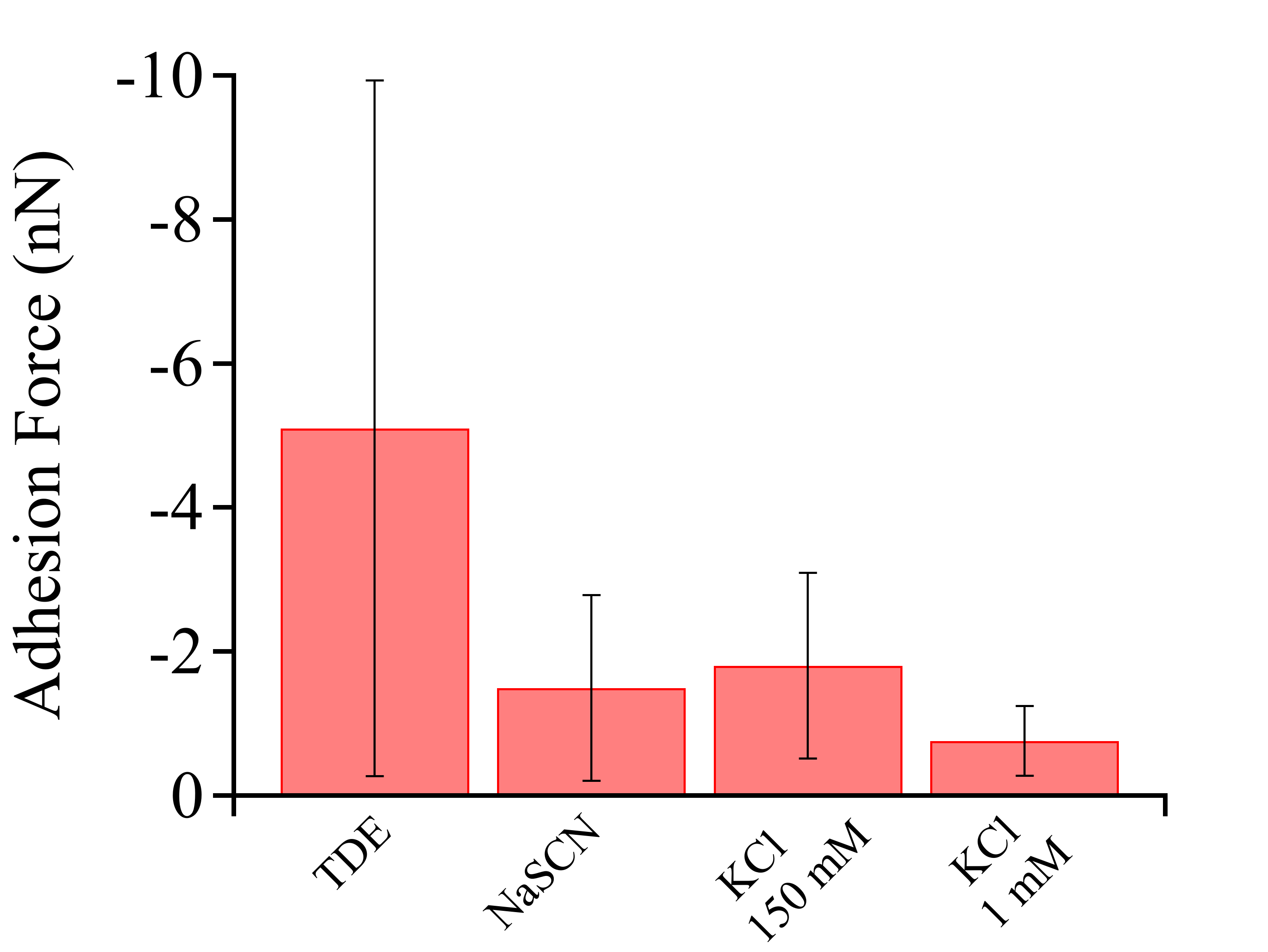}}
\caption{Adhesion forces of silica particles measured in four different media, shown from left to right: TDE (69.75~wt.\%), NaSCN (\SI{11.65}{M}), KCl (\SI{150}{\milli M}), and KCl (\SI{1}{\milli M}).
\label{fig:AFM}}
\end{figure*}

\begin{figure*}[t]
    \centering
    \subfloat[]{
        \raisebox{2.4ex}{\includegraphics[scale=0.18]{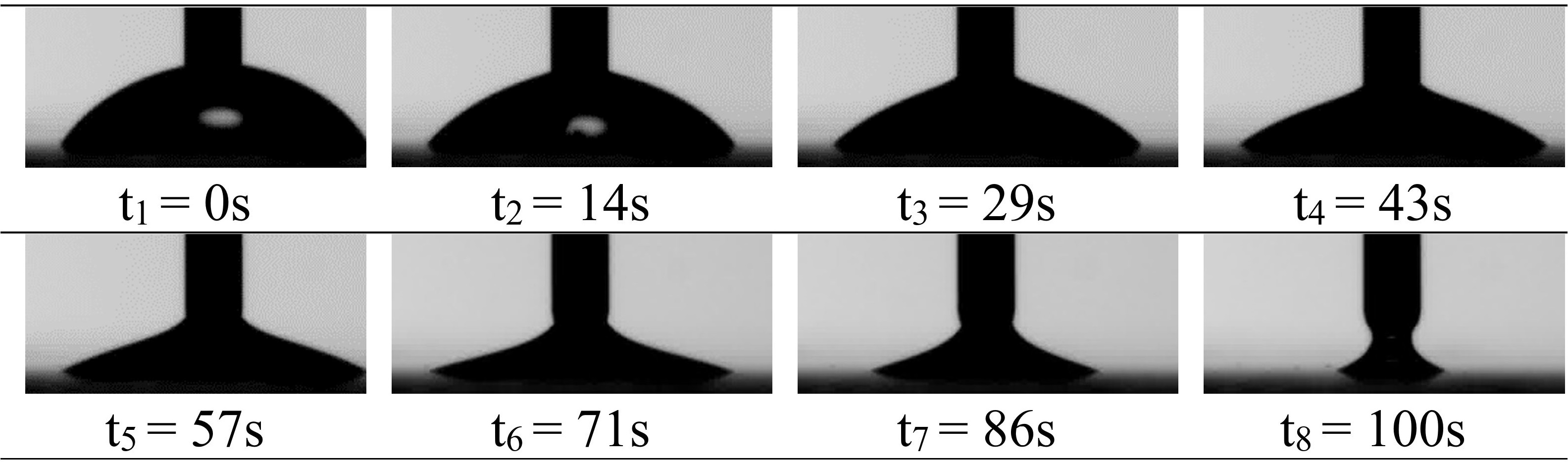}}
    }
    \;
    \subfloat[]{
        \raisebox{-4.5ex}{\includegraphics[scale=0.24]{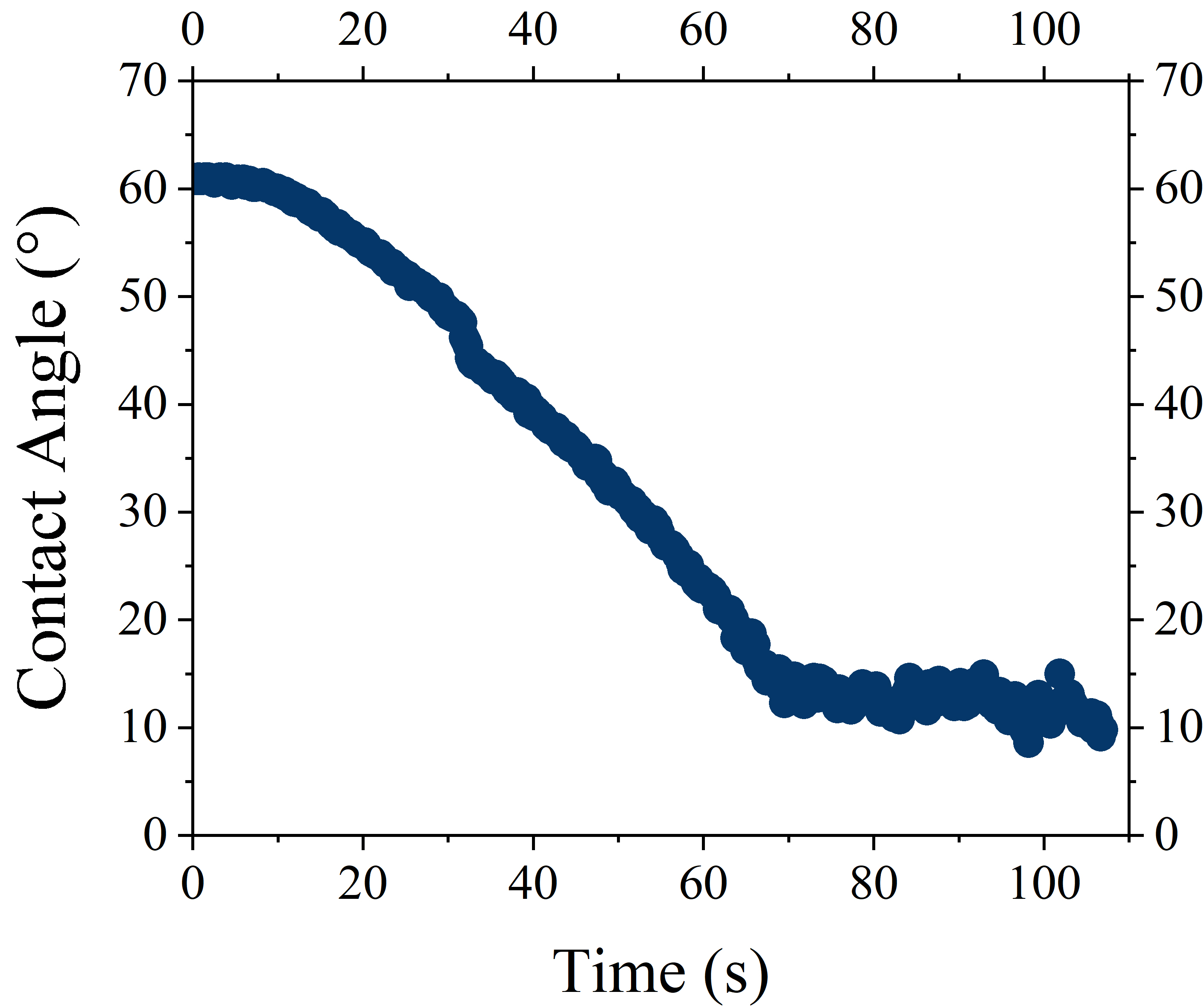}}
    }
    \caption{(a) Side-view images of a TDE solution droplet undergoing controlled deflation at \SI{0.1}{\micro\liter\per\second}. 
    Frames $t_1$ to $t_8$ denote the beginning and end of the experiment. 
    (b) Corresponding receding contact angle as a function of time.}
    \label{fig:Receding_Contact_Angle_TDE_Solution}
\end{figure*}


\begin{figure*}[t]
  \centerline{\includegraphics[scale=0.18]{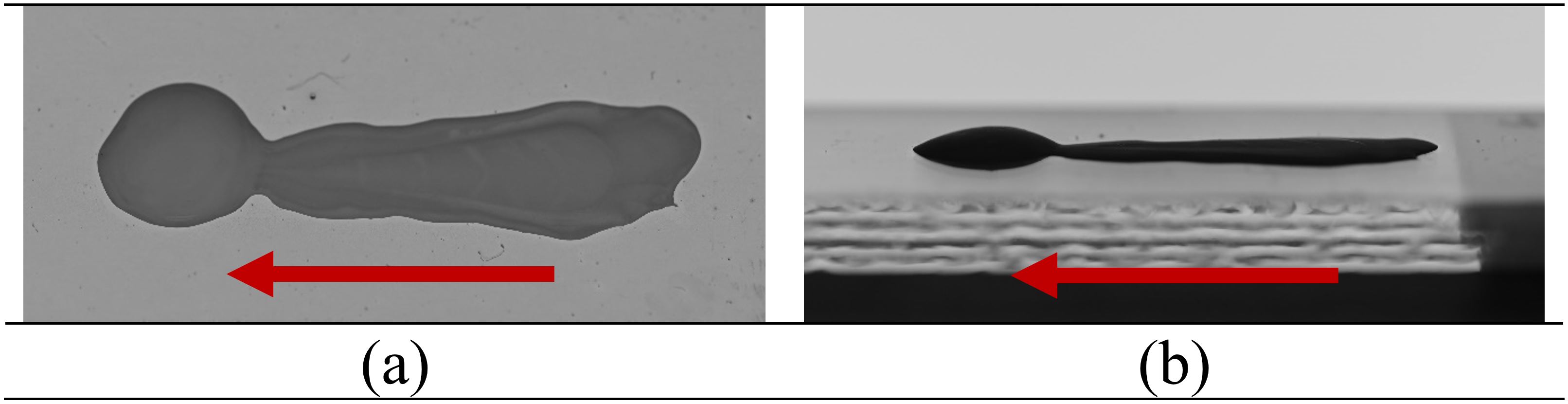}}
  \caption{(a) Top-view and (b) side-view images of the suspension layer deposited on the substrate following droplet motion. 
  The droplet consisted of a 33~wt.\% suspension of \SI{5}{\micro\metre} silica particles in TDE. 
  The substrate was translated at a velocity of \SI{200}{\micro\metre\per\second}. 
  Red arrows indicate the direction of motion from start to stop.}
\label{fig:Suspension_Layer_TDE_33}
\end{figure*}

%
Particle-particle interactions in normal direction were measured with Colloidal Probe-AFM (CP-AFM) using an MFP-3D Bio (Asylum Research Inc., Santa Barbara, USA) mounted on an inverted optical microscope (Axio Observer Z1, Zeiss, Oberkochen, Germany). 
Cantilevers (CSC37, MikroMasch, Sofia, Bulgaria) were calibrated ($k=\SI{0.25}{\newton\per\metre}$ and \SI{0.28}{\newton\per\metre}) via thermal noise method~\citep{hutter1993calibration}. 
A \SI{5}{\micro\metre} silica particle was glued with a two-component epoxy glue to the cantilever. 
The silica colloidal probe is from the same batch of silica particles used in the other experiments.  
To prevent sample particle movement during measurements glass substrates were modified with a PDMS pseudo brush to increase adhesiveness. 
The procedure followed a modified version of~\citep{Krumpfer:2011ab} and~\citep{eifert2014simple}. 
Ethanol cleaned substrates were heated to \SI{200}{\celsius}. 
\SI{100}{\micro\litre} of \SI{5}{\centi St} PDMS oil was dispersed on a substrate to react. 
The reaction is considered complete when no further visible emission of volatile byproducts from the silicone oil is observed, typically within \SIrange{2}{3}{\minute} of annealing. 
The substrates were subsequently rinsed with ethanol and then sonicated in acetone for \SI{10}{\minute} to remove residuals of unreacted PDMS. 
Afterwards, they were washed, sonicated, and kept in water. 
Before the measurement the substrates were dried with nitrogen and mounted into a fluid cell. 
Small droplets of the \SI{5}{\micro\metre} silica particle suspension were deposited on the now more adhesive substrate. 
The suspension was allowed to dry. 
The two indexed match liquids i.e. TDE and NaSCN are distinctive aqueous solutions. 
TDE being an organic solvent and NaSCN as highly concentrated solution have no ordinary water structure. 
The fluid cell was now flushed 3 times with the measurement solution, i.e. an index matched liquid (NaSCN, TDE), \SI{150}{\milli M} KCl or \SI{1}{\milli M} KCl, respectively. 
Measurements in \SI{1}{\milli M} and \SI{150}{\milli M} KCl provide a reference to typical particle–particle force measurements in aqueous media. 
In indexed matched liquids and \SI{150}{\milli M} we measured at least 16 different particles. 
Three force curves for each particle were taken at a measurement velocity of \SI{400}{\nano\metre\per\second}. 
Additional measurements were taken at \SI{20}{\nano\metre\per\second}, \SI{1}{\micro\metre\per\second}, \SI{2}{\micro\metre\per\second} and \SI{5}{\micro\metre\per\second} for indexed match liquids. 
Electrostatic forces were measured in \SI{1}{\milli M} KCl on 5 different particles with 100 force curves each. 
Measurement velocity was kept constant at \SI{400}{\nano\metre\per\second}. 
Force loads of up to \SI{58}{\nano\newton} were applied during approach. 
No significant time dependency is observed, as the force curves remain consistent across consecutive measurements taken over a few hours. 
All obtained force curves are baseline corrected and zero contact is defined as the constant compliance regime i.e. only cantilever deformation is detected. 
The force curves obtained from direct force measurements using CP-AFM are shown in Fig.~\ref{AFM-adhesion}.

\begin{figure*}[h]
  \centerline{\includegraphics[scale=0.3]{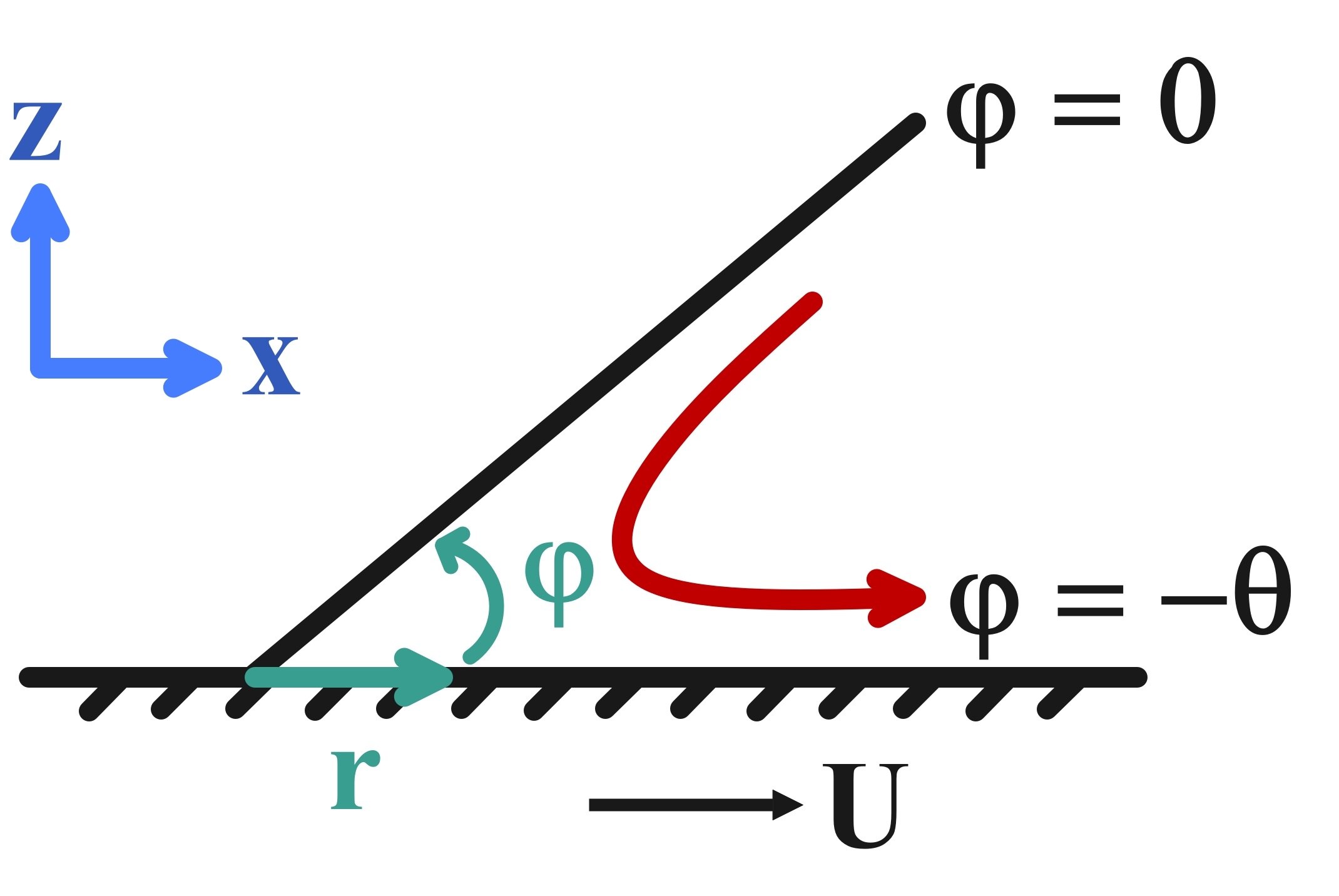}}
\caption{Schematic representation of the wedge-shaped geometry used in the Moffatt solution.}
\label{fig:Moffat_Scheme}
\end{figure*}

%
Interactions between silica particles exhibit the highest adhesion forces (minima in the force curves) in TDE. 
For measurements in different solutions, \numrange{10}{20} force curves are shown, except for KCl \SI{1}{\milli M}, where only five curves are displayed. 
Each force curve measured the interaction of the colloidal probe with a different particle fixed on the substrate. 
A comparison of averaged adhesion force values of individual experiments with standard deviation is given in Fig.~\ref{fig:AFM}. 
Particles in TDE have the highest adhesion force, \SI{-5.1 \pm 4.8}{\nano\newton}, compared to other solutions (Fig.~\ref{fig:AFM}). 
Adhesion force in \SI{11.65}{M} NaSCN, \SI{-1.5 \pm 1.3}{\nano\newton}, is similar to that in \SI{150}{\milli M} KCl, \SI{-1.8 \pm 1.3}{\nano\newton}. 
In \SI{1}{\milli M} KCl, \SI{-0.8 \pm 0.5}{\nano\newton}, particles exhibit the smallest adhesion force because electrostatic repulsive forces reduce attractive interaction. 
The large error bars indicate that particles have heterogeneous surface properties. 
This was verified by SEM images showing distinct surface topographies (Fig.~\ref{fig:Particles} of the main text). 
Adhesion forces scattering by almost the average value is common in particle-particle interaction measurements.

Despite the refractive index matching between particles and fluid, and the expectation that VdW forces are small in both cases, adhesion forces obtained by CP-AFM experiments in TDE solution are greater than those measured in \SI{11.65}{M} NaSCN solution. 
In general adhesion forces are in the range of previously reported values. 
Ranging from \SIrange{0.005}{1}{\milli\newton\per\meter} depending on electrolyte concentration and type of electrolyte~\citep{valmacco2016dispersion, valmacco2016forces, troncoso2014nanoscale, wang2013atomic}.  
However, comparisons on absolute values must be taken with care. Valmacco showed that reducing the RMS roughness of silica from roughly \qty{3}{\nano\metre} to \qty{1}{\nano\metre} changes Hamaker constant and as such adhesion more than \num{10}-fold~\citep{valmacco2016dispersion}.

The analysis of van der Waals forces highlights the crucial role of the dielectric function, which varies with frequency $\omega$~\citep{Cappella:1999aa}. 
Hamaker's approach simplifies this by assuming that the dielectric constant and other relevant properties remain constant across all frequencies~\citep{Hamaker:1937aa}, a simplification that does not hold true in reality. 
The dielectric function $\epsilon(\omega)$, which is intricately linked to the refractive index $n(\omega)$, varies across different frequencies. 
The dielectric constant has real and imaginary components related to the material's polarization response, and the Kramers-Kronig relations connect these real and imaginary parts~\citep{Dagastine:2002aa}. 
The frequency-dependent nature of the dielectric constant and refractive index influences the strength and nature of van der Waals forces, complicating attempts to nullify these forces through refractive index matching~\citep{Goodwin:2004ag}. 
Matching refractive indices at only one wavelength is insufficient to reduce the Hamaker constant to zero, necessitating consideration of the entire electromagnetic spectrum~\citep{hopkins2014dielectric}.

There is an electric double layer around charged particles, consisting of a layer of counter-ions closely bounded to the particle surface and a diffuse layer spreading into the medium. 
In a system of charged microparticles, the Debye length defines the characteristic distance over which electrostatic interactions or potentials decay in a medium~\citep{hunter2013zeta}. 
It reveals the screening effect of the medium on Coulomb forces, which shapes particle interactions and influences frictional behaviour during particle rearrangements~\citep{royer2016rheological, hribar2019electrostatic, yu2012boundary}. 
Screening length is directly related to the dielectric constant of the medium, which determines the attenuation of electrostatic potential, and inversely related to the square root of the ionic strength~\citep{hunter2013zeta}. 
With low ionic strength, the extended Debye lengths are expected to allow repulsive forces to act over long distances, thereby reducing contact and friction between particles. 
Conversely, with elevated ionic strength, the diminished Debye length is expected to reduce screening effects, facilitating closer proximity between particles. 
Reliable zeta potential values of the particles could not be measured in the highly concentrated NaSCN and TDE solutions, as they are far too conductive for standard measurement techniques. 
Under such conditions, the systems deviate from the basic assumptions underlying these methods.

\subsection{Transient dynamics of receding contact line} \label{Transient dynamics of receding contact line-Suppl}

To assess the wetting behaviour of the dispersion medium independently of the particulate phase, receding contact angle measurements were carried out using a DataPhysics OCA 35XL instrument. 
A sessile droplet of pure TDE solution (i.e., without particles) with a volume of \SIrange{10}{15}{\micro\liter\per\second} was deposited on a cleaned glass substrate, and subjected to controlled volume reduction at a low withdrawal rate of \SI{0.1}{\micro\liter\per\second}. 
The corresponding side-view image sequence and contact angle data are presented in Fig.~\ref{fig:Receding_Contact_Angle_TDE_Solution}a and b. 
The contact line steadily recedes as the droplet volume decreases, leaving no detectable liquid layer on the substrate. 
The receding contact angle gradually decreases from an initial value of approximately \SI{60}{\degree} to a plateau near \SI{15}{\degree} when the drop volume approaches zero. 
The absence of any deposited film behind the receding contact line in the pure fluid experiments demonstrates that the persistent particle-laden layer observed for concentrated suspensions originates from particle-induced dynamics, rather than from the wetting properties of the dispersion medium itself.

To visualize the suspension layer deposited behind the receding contact line, Fig.~\ref{fig:Suspension_Layer_TDE_33} presents both top view and side view images of the substrate following the cessation of droplet motion. 
The images were captured after a concentrated suspension droplet, composed of 33~wt.\% of \SI{5}{\micro\metre} silica particles dispersed in TDE, was translated at \SI{200}{\micro\metre\per\second} and came to rest.

\begin{figure*}[t]
    \centering
    \subfloat [\centering]
    {{\includegraphics[scale=0.16]{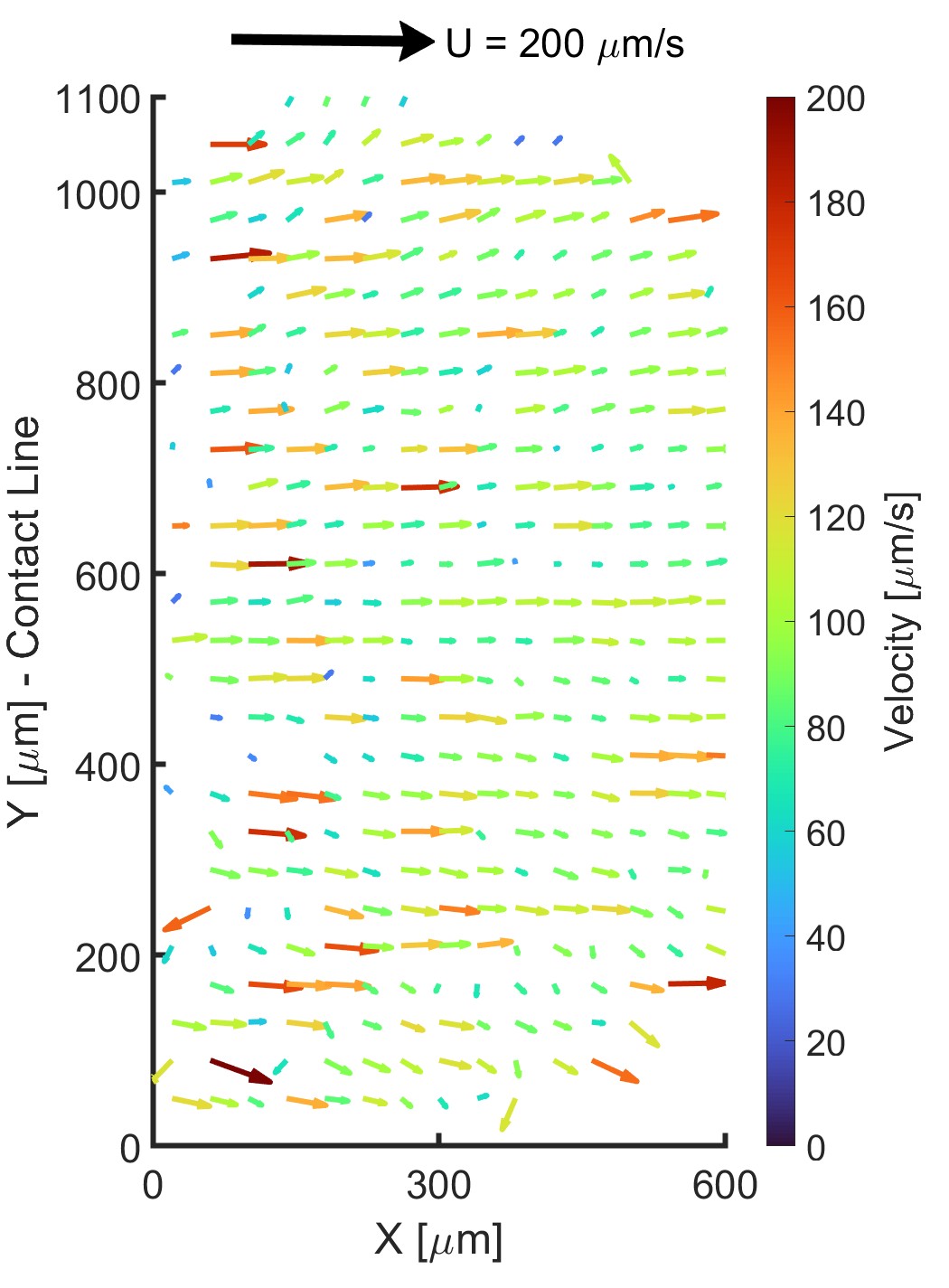}}}
    \subfloat [\centering]
    {{\includegraphics[scale=0.16]{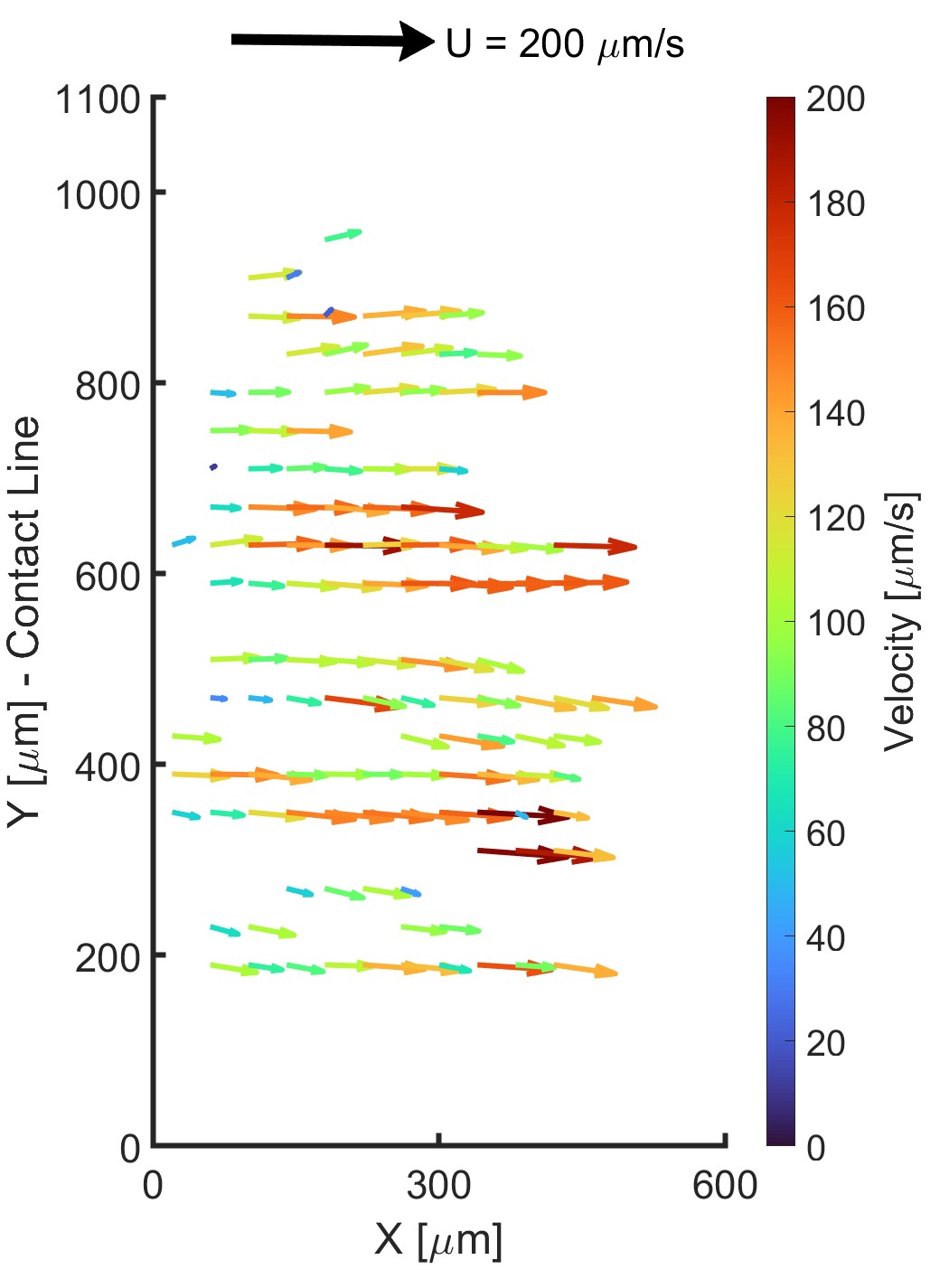}}}
    \subfloat [\centering]
    {{\includegraphics[scale=0.16]{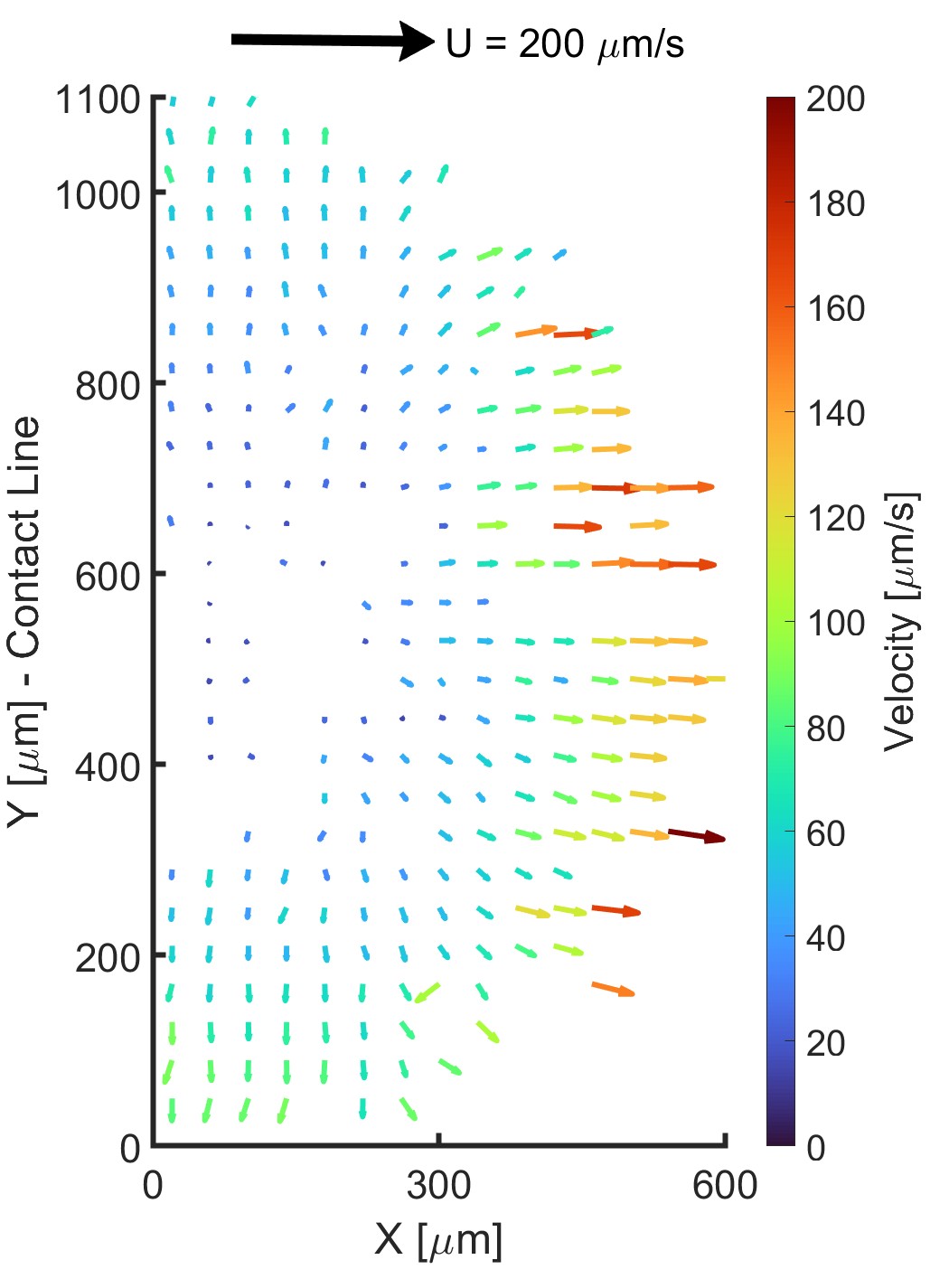}}}
\caption{Averaged flow fields in the substrate plane ($x$–$y$) for (a) TDE dispersion medium, (b) 30~wt.\% moderately concentrated suspension, and (c) 33~wt.\% dense suspension. For cases (a) and (b), velocity vectors are computed within a near-substrate region bounded by the geometric interface $z = x \cdot \tan(\theta/2)$ to exclude tracer motion near the free surface. 
For the dense suspension (c), all tracer data are included in the averaging to preserve the anisotropic flow structure. 
The $y$-axis corresponds to the contact line. 
The black arrows indicate the direction of substrate motion at $U = \SI{200}{\micro\metre\per\second}$.}
    \label{fig:Top_View_FlowField}
\end{figure*}

\subsection{Flow field measurements}
\subsubsection{Moffatt theory} \label{Moffatt theory-Suppl}

To compare our experimentally measured flow fields in the vicinity of the advancing contact line of droplets, we reference the classical similarity solution derived by Moffatt (1964)~\citep{moffatt1964viscous} for Stokes flow near a sharp corner. 
The local velocity field is governed by the biharmonic form of the Stokes equations, $\nabla^4 \psi = 0$, where $\psi(r,\varphi)$ is the stream function expressed in polar coordinates. 
Moffatt showed that self-similar separable solutions of the form $\psi(r, \varphi) = r^\lambda f(\varphi)$ exist, where $\lambda$ is an eigenvalue determined by the wedge angle and the imposed boundary conditions.

In our particular configuration relevant to the advancing contact line dynamics, the lower boundary ($\varphi = -\theta$) is modeled as a solid substrate moving tangentially at a constant velocity of $U = \SI{200}{\micro\metre\per\second}$, while the upper boundary ($\varphi = 0$) is considered to be an immobile interface and treated as a no-slip wall, illustrated in Fig.~\ref{fig:Moffat_Scheme}. 
The flow is then described by a similarity solution with eigenvalue $\lambda = 1$. 
For this case, the azimuthal similarity function $f_1(\varphi)$ is given by:

\begin{equation}
f_1(\varphi) = \frac{\varphi \cos\varphi \sin\theta - \theta \cos\theta \sin\varphi}{\sin\theta \cos\theta - \theta}
\end{equation}

By differentiating $f_1(\varphi)$ analytically with respect to $\varphi$, we obtain the expression required to evaluate the radial velocity component:

\begin{equation}
\frac{df_1}{d\varphi} = \frac{\theta \cos\theta \cos\varphi + \varphi \sin\theta \sin\varphi - \sin\theta \cos\varphi}{\theta - \sin\theta \cos\theta}
\end{equation}

Using these expressions, the radial ($v_r$) and azimuthal ($v_\varphi$) velocity components in polar coordinates are obtained as:

\begin{equation}
v_r = \frac{1}{r} \frac{\partial \psi}{\partial \varphi} = U \cdot \frac{df_1}{d\varphi}
\end{equation}

\begin{equation}
v_\varphi = -\frac{\partial \psi}{\partial r} = -U \cdot f_1(\varphi)
\end{equation}

In our experimental configuration, $v_x$ is parallel to the substrate (direction of relative motion), $v_y$ is parallel to the contact line, and $v_z$ denotes the out of plane component. 
The Moffatt solution is strictly two-dimensional, confined to the $v_x$–$v_z$ plane. To compare with experimental velocity fields captured in Cartesian coordinates, these components are transformed as:

\begin{equation}
v_{x,\;\mathrm{theo}} = - v_\varphi \sin\varphi + v_r \cos\varphi 
\end{equation}

\begin{equation}
v_{z,\;\mathrm{theo}} = v_r \sin\varphi + v_\varphi \cos\varphi
\end{equation}

The second column of Fig.~\ref{fig:Average_Flowfield_Deviation} in the main text presents the Moffatt-theory velocity fields for three representative cases: dispersion medium (b), moderately concentrated suspension (e), and dense suspension (h), corresponding to advancing contact angles $\theta$ of \SI{38}{\degree}, \SI{53}{\degree}, and \SI{80}{\degree}, respectively.

\subsubsection{Flow field reconstruction from APTV data} \label{Flow field reconstruction from APTV data-Suppl}
To construct spatially resolved average flow fields from APTV data, we segment the entire observation domain into uniform square grids of \SI{20}{\micro\metre} $\times$ \SI{20}{\micro\metre}, with the coordinate origin coinciding with the contact line. 
All velocity vectors whose positions fall within each grid cell over the entire duration of the experiment are identified. 
These local ensembles of velocity measurements are then spatially averaged to construct a mean flow field defined over fixed spatial coordinates, taken as the geometric centers of the corresponding grid cells. 
To remove spurious measurements, we adapt the outlier detection approach of Westerweel et al.~\citep{westerweelUniversalOutlierDetection2005} to operate on velocity vectors within each grid cell. 
For each velocity vector, its Euclidean deviation from the cell’s median vector is computed and normalized by the median of all such deviations in the same cell. 
Vectors with normalized residuals exceeding a threshold value of $r^* > 4$ are classified as outliers and excluded from the averaging process. 
The remaining velocity vectors within each cell are averaged to obtain the local mean. 
To minimize curvature effects near the droplet edges and ensure reliable statistics for comparison with theory, we restrict our analysis to a $\pm\SI{250}{\micro\metre}$ window along the $y$-axis (parallel to the contact line), around the droplet’s centerline in the direction of motion ($x$-axis). 
Grid cells containing fewer than three vectors are deemed unreliable and excluded from the plots. 
The resulting two-dimensional velocity fields are shown in Fig.~\ref{fig:Average_Flowfield_Deviation} of main text. 

\vspace{\baselineskip}
\begin{figure*}[t]
\centering
\includegraphics[scale=0.65]{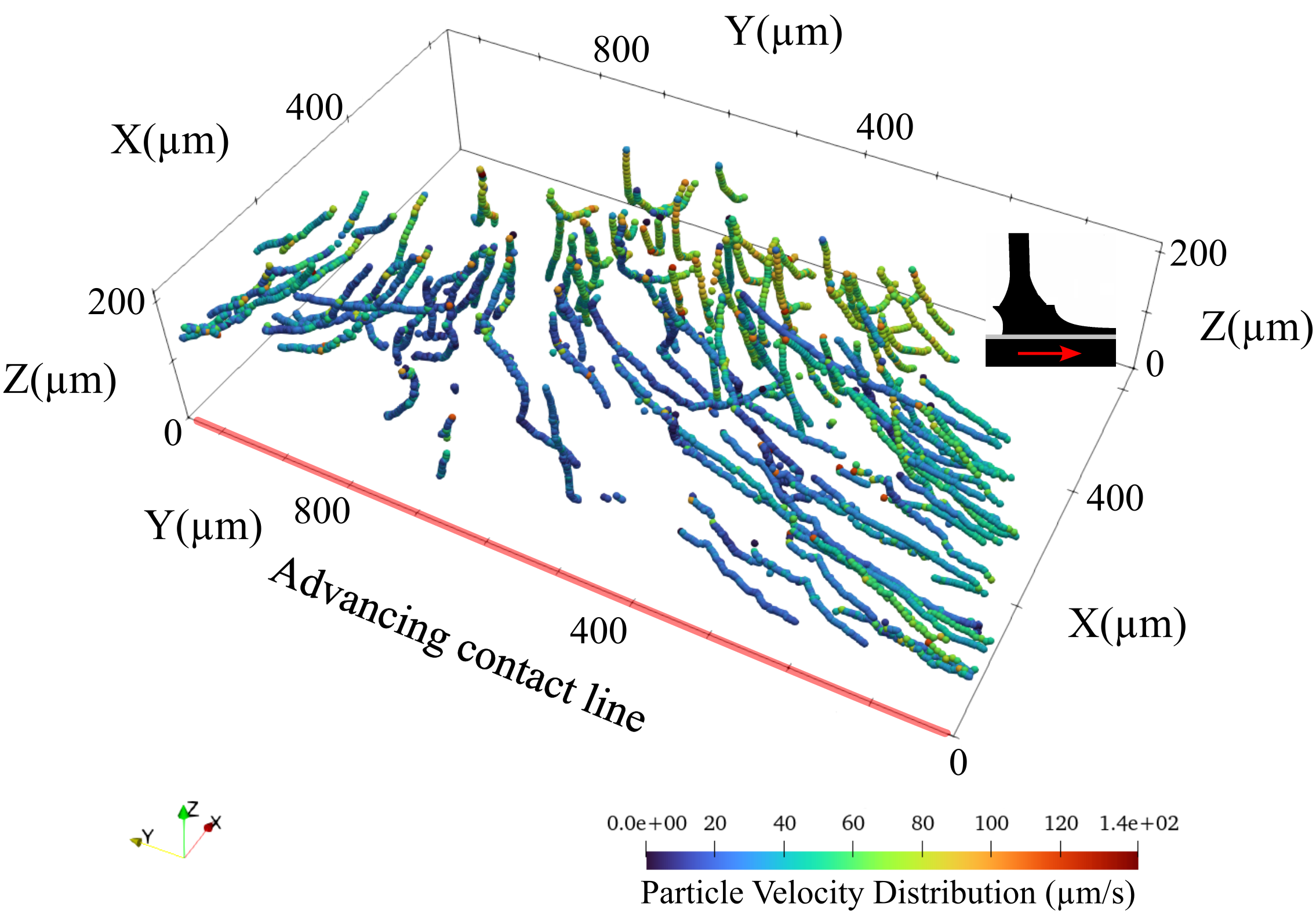}
\caption{Second run of flow field measurements near the advancing contact line of a suspension containing 33~wt.\% silica particles in TDE solution, confirming reproducibility.}
\label{Deviation-Hydrodynamics}
\end{figure*}

\subsubsection{Top view of the measurements} \label{Top view of the measurements-Suppl}

Fig.~\ref{fig:Top_View_FlowField} presents the spatially averaged flow fields projected onto the substrate plane ($x$–$y$) for three representative cases: (a) dispersion medium (TDE solution), (b) 30~wt.\% moderately concentrated suspension, and (c) 33~wt.\% dense suspension. 
For the cases that conform to hydrodynamic predictions (panels (a) and (b)), top-view averaging was restricted to the region near the substrate. 
Only tracer velocities below the geometric boundary defined by $z = x \cdot \tan(\theta/2)$, with $\theta/2$ half the advancing contact angle, were considered. 
This ensures that the averaged flow field reflects near-substrate motion and excludes tracer particles near the free surface, where counter flows dominate. 
Incorporating opposing velocity vectors leads to their cancellation, diminishing the meaningful features of the top-view flow field. 
For the dense suspension (panel (c)), all tracer data were included in the averaging. 
Since the flow strongly deviates from hydrodynamic predictions, restricting the analysis to half the advancing contact angle would artificially filter out relevant structural features. 
The spatial averaging was performed over square grid cells of \SI{20}{\micro\metre} $\times$ \SI{20}{\micro\metre}, ensuring local resolution while reducing noise. 
The magnitude of the velocity vectors is calculated as the Euclidean norm of the velocity components, given by $|\vec{v}_{xy,\;\mathrm{exp}}| = \sqrt{v_{x,\;\mathrm{exp}}^2 + v_{y,\;\mathrm{exp}}^2}$, thereby excluding any contributions from the out of plane ($z$) direction.

In the dispersion medium and in the moderately concentrated suspension (panels (a) and (b)), the flow is directed predominantly along the $x$-axis, consistent with unidirectional droplet motion. 
By contrast, the dense suspension (panel (c)) exhibits a markedly different topology: streamlines near the advancing contact line diverge laterally toward the droplet sides along the $y$-direction. 
Close to the contact line, these deviating flows show significantly reduced velocities, supporting the existence of mechanically arrested regions and highlighting the emergence of anisotropic flow structures within the droplet that are not captured by classical hydrodynamic models.

\subsubsection{Deviation from hydrodynamic solution} \label{Deviation from hydrodynamic solution-Suppl}
Flow field measurements were repeated at least three times for each experiment to guarantee reproducibility. 
The data showcased in Fig.~\ref{Deviation-Hydrodynamics} correspond to the second run of flow field measurements conducted near the advancing contact line, employing a dispersion of 33~wt.\% silica particles dispersed in a TDE solution. 
This confirms the deviation of the flow field from hydrodynamic predictions.

\begin{figure*}[t]
    \centering
    \includegraphics[width=0.5\linewidth]{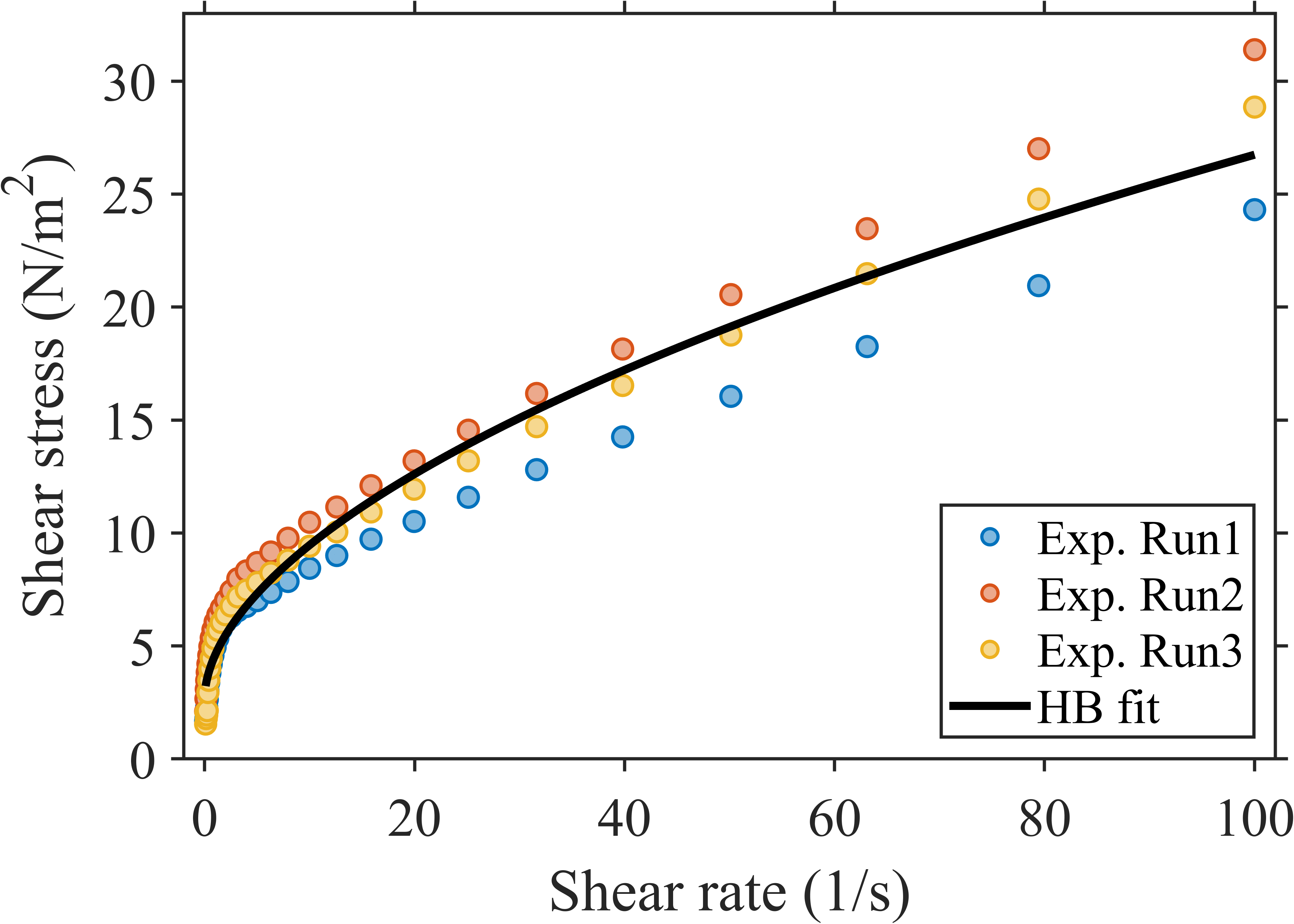}
\caption{Measured shear stress as a function of shear rate for the NaSCN-based suspension (28~wt.\% silica), together with a Herschel-Bulkley fit. 
Individual markers correspond to three independent experimental runs, and the solid line shows the fitted curve, $\tau = \tau_y + K\,\dot{\gamma}^{\,n}$.}
    \label{fig:HB-fit}
\end{figure*}

\subsection{Yielded-layer thickness in NaSCN-based wetting experiments} \label{Section: Yielded-layer thickness in NaSCN-based wetting experiments}

To quantify the thickness of the yielded-layer adjacent to the substrate, we fitted the NaSCN-based suspension rheology (Fig.~\ref{SRS-Rheology}) to a Herschel–Bulkley constitutive law \citep{herschel1926konsistenzmessungen, chambonExperimentalInvestigationViscoplastic2014}, equation~\ref{equation: Herschel–Bulkley}.
\begin{equation} \label{equation: Herschel–Bulkley}
\tau(\dot{\gamma})=\tau_y + K\,\dot{\gamma}^{\,n}
\end{equation}
In this equation, $\tau$ denotes the shear stress, $\tau_y$ the yield stress, $K$ the consistency index, $\dot{\gamma}$ the shear rate, and $n$ the flow index.

We use nonlinear least-squares optimization (trust-region reflective algorithm) implemented in MATLAB's \texttt{lsqcurvefit}.
The fit minimizes the squared deviation between measured and model stresses, and parameter uncertainties are obtained from the Jacobian-based covariance at the optimum. 
The fitted parameters are:
\begin{align*}
\tau_y &= \SI{2.69}{\pascal}
\quad (95\%~\mathrm{CI}:~\SIrange{1.95}{3.42}{\pascal}),\\
K &= \SI{1.91}{\pascal\second^{n}}
\quad (95\%~\mathrm{CI}:~\SIrange{1.35}{2.47}{\pascal\second^{n}}),\\
n &= \num{0.550}
\quad (95\%~\mathrm{CI}:~\numrange{0.486}{0.614}),
\end{align*}
with $R^{2}=0.958$, confirming a finite yield stress and shear-thinning response.

To relate this stress-based yielding criterion to a characteristic shear-rate scale, we rewrite the Herschel–Bulkley law as an excess stress above yield, $\tau-\tau_y = K\dot{\gamma}^{n}$.
In an ideal yield-stress material, flow requires $\tau>\tau_y$.
However, in the droplet geometry, we do not directly access the local stress field.
We therefore introduce the natural Herschel–Bulkley shear-rate scale at which the viscous stress becomes comparable to the yield stress.
Specifically, we define a yield-onset shear rate $\dot{\gamma}_0$ by the crossover condition $K{\dot{\gamma}_0}^{n}=\tau_y$, equation~\ref{equation: crossover condition}.
\begin{equation} \label{equation: crossover condition}
\dot{\gamma}_0 = \left(\frac{\tau_y}{K}\right)^{1/n}
\end{equation}
This $\dot{\gamma}_0$ is not an independent material constant but the shear-rate scale implied by the fitted parameters. 
It represents the minimal shear rate of the yielded region, such that any region which is genuinely yielded must satisfy $\dot{\gamma}\gtrsim\dot{\gamma}_0$. 
Using the best-fit parameters yields $\dot{\gamma}_0=\SI{1.86}{\per\second}$.

To estimate the thickness of the yielded region, we exploit the kinematic fact that the substrate-induced velocity drop $U$ across a thin shear band of thickness $\delta$ implies a characteristic shear rate $\dot{\gamma}\sim U/\delta$.
Although this does not describe the detailed vertical shear profile of a yield-stress fluid, combining it with the requirement $\dot{\gamma}\gtrsim\dot{\gamma}_0$ provides a conservative (maximal) estimate for the thickness of the sheared layer, equation~\ref{equation: thickness of the sheared layer}.
\begin{equation} \label{equation: thickness of the sheared layer}
\delta \approx \frac{U}{\dot{\gamma}_0}
\end{equation}
Using the experimental substrate velocity $U=\SI{0.2}{\milli\meter\per\second}$, the best-fit parameters give $\delta \approx \SI{1.076e-4}{\meter} \approx \SI{108}{\micro\meter}$.

Propagating the fitted parameter uncertainties through the same scaling, we performed Monte-Carlo error propagation by drawing $2\times10^{4}$ parameter triplets $(\tau_y, K, n)$ from the Jacobian-based covariance matrix, recomputing $\dot{\gamma}_0$ and hence $\delta$ for each draw.
This yields a median thickness of \SI{107.4}{\micro\meter} and a 95\% confidence interval of \SIrange{44.9}{306.9}{\micro\meter}.
Thus, while the precise numerical value is uncertain, its order of magnitude is robust, and only a thin near-substrate band on the order of tens of particle diameters is expected to be fully yielded.

\begin{figure*}[t]
\centering
\includegraphics[width=0.6\linewidth]{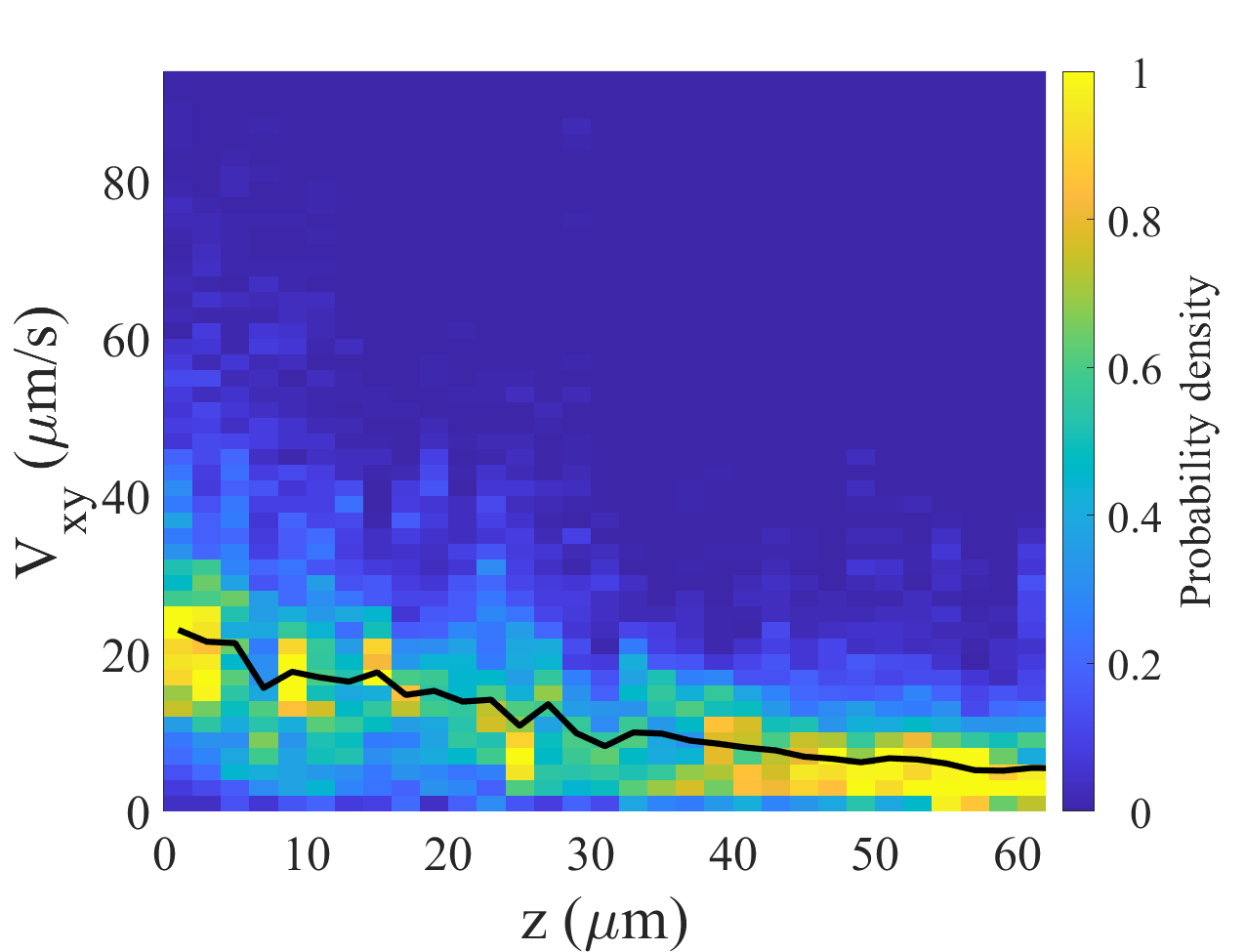}
\caption{Probability density of the instantaneous velocity magnitude $V_{xy}$ as a function of height $z$ above the substrate for the NaSCN-based suspension, obtained from APTV measurements. 
Colors indicate the normalized bin probability, and the black line shows the median profile $V_{xy}^{\mathrm{med}}(z)$. 
A narrow band of elevated velocities is confined to the first $\approx \SI{50}{\micro\meter}$ above the substrate, beyond which the flow decays to a low-velocity creep-like motion.}
\label{fig:Vxy-density}
\end{figure*}

To test this rheology-based estimate against the wetting experiments, we computed the velocity magnitude ($V_{xy}$) of each tracked tracer particle from APTV data and constructed two-dimensional histograms in the $(z, V_{xy})$ plane using uniform bins for height and velocity. 
The counts are normalized to obtain a probability density, and we overlay the median velocity profile $V_{xy}^{\mathrm{med}}(z)$ obtained from the same bins (Fig.~\ref{fig:Vxy-density}). 
This representation suppresses the point to point scatter in individual trajectories and reveals the statistically dominant flow pattern. 
A bright band of elevated velocities localized close to the substrate, above which both the probability density and the median velocity rapidly decay to a low, nearly height-independent plateau. 

As an operational definition of the yielded-layer in the experiment, we identify its outer edge at the height $z=\delta_{\mathrm{exp}}$ where the median velocity has dropped to roughly half of its near-wall value and the probability of observing velocities above this level becomes negligible. 
From the profile in Fig.~\ref{fig:Vxy-density}, this criterion places the extent of the strongly sheared region at $\delta_{\mathrm{exp}}\approx \SI{50}{\micro\meter}$, beyond which the material exhibits only weak creep. 

Comparing the two measures, the experimentally inferred thickness $\delta_{\mathrm{exp}}$ is smaller than the rheological upper bound $\delta \approx \SI{108}{\micro\meter}$ and lies close to the lower limit of the confidence interval obtained from the Herschel–Bulkley fit.  
As discussed above, equation~\ref{equation: thickness of the sheared layer} provides an upper bound on the thickness of the yielded zone. 
In addition, the rheological estimate was obtained for a \SI{28}{wt.\%} NaSCN-based suspension, whereas the wetting experiments use \SI{30}{wt.\%}. 
The higher solid fraction increases the yield stress and thus reduces the actual $\delta$ relative to the value inferred from rheology. 
Taken together, these considerations show that the two measures are physically consistent and support the picture that only a thin near-substrate layer is fully fluidized, while the bulk of the drop remains essentially unyielded.

\subsection{Silica density and sedimentation time scales} \label{Subsection Silica density and sedimentation-Suppl}
\subsubsection{Density of silica particles} \label{Subsubsection silica density-Suppl}
To determine the density of the silica particles used in our suspensions, we analyzed the fluid displacement after adding the particles into a \SI{5}{\milli\liter} measuring cylinder and allowing \SI[unit-mode=text]{72}{hours} for the fluid to permeate into the porous particles. 
We estimate the particle density to be around $\sim$\SI{1.9}{\gram\per\cubic\centi\meter}. 
Owing to intrinsic porosity, particles exhibit a lower effective density than bulk fused silica, which typically has a density of approximately \SI{2.2}{\gram\per\cubic\centi\meter}. 
These silica microspheres are specifically manufactured for use as porous stationary phases in high performance liquid chromatography (HPLC) columns. 
Due to the uncertainty inherent in such density measurements, particularly arising from the small sample volumes and the presence of particle porosity, we refrained from converting weight fractions to volume fractions. 
Nevertheless, using the estimated particle density of \SI{1.9}{\gram\per\cubic\centi\meter} as a reference, our suspensions reach their respective critical concentrations at comparatively low weight and volume fractions relative to values reported in the literature~\citep{guazzelli2018rheology}, as confirmed by our protorheology measurements.

\subsubsection{Sedimentation timescales in rheological and wetting experiments} \label{Subsubsection silica sedimentation-Suppl}
Although the silica particles are not density matched with the suspending media, sedimentation is not expected to significantly affect our rheological measurements. 
The characteristic time for a single spherical particle to sediment over a distance equal to its radius can be estimated by balancing gravitational and viscous drag forces in the Stokes regime:

\begin{equation}
    t_{sedmintation} = \frac{9\eta}{2 \Delta \rho g a}
\end{equation}

where $\eta$ is the fluid viscosity, $\Delta \rho$ is the density difference between the particle and the fluid, $g$ is gravitational acceleration, and $a$ is the particle radius.

For \SI{5}{\micro\metre} spherical silica particles dispersed in TDE and NaSCN solutions, the density differences are $\Delta \rho =$ \SI{0.76}{\gram\per\cubic\centi\meter} and \SI{0.63}{\gram\per\cubic\centi\meter}, respectively.  
Using low shear viscosities of approximately \SI{0.5}{\pascal\second} for 31~wt.\% silica in TDE and \SI{10}{\pascal\second} for 28~wt.\% silica in NaSCN (see Fig.~\ref{SRS-Rheology}), we estimate the sedimentation times. 
The resulting sedimentation times are approximately 1 minute per particle radius in TDE and 24 minutes per particle radius in NaSCN. 
Given the drop size of approximately \SIrange{2}{3}{\milli\meter} for protorheology experiments and the Couette cell height of \SI{27.93}{\milli\meter} for rheology experiments, these values imply sedimentation timescales far exceeding the duration of our rheological characterization experiments.

For the wetting experiments on the DropSlider setup, we allow the suspension droplets to rest for five minutes after deposition and before initiating the measurements to ensure that the collective motion of the particles dies out~\citep{voltzRayleighTaylorInstabilitySedimenting2001}.
The gap between the upper disk and the lower substrate is \SI{2}{\milli\meter} in the disk geometry, and the droplet height is $\approx$ \SI{3}{\milli\meter} in prism geometry. 
Considering these vertical length scales along with the calculated sedimentation timescales, we do not expect significant sedimentation induced flow to develop during the wetting dynamics experiments.

In addition to the sedimentation timescales discussed in the absence of any flow, it is crucial to consider that the suspension flow is primarily driven by the substrate motion at \SI{200}{\micro\metre\per\second}. 
This driven flow not only dominates over sedimentation timescales, which are on the order of minutes per particle diameter, but also promotes particle interactions and contact forces, such as friction, which further diminish the influence of sedimentation during the wetting experiments. 
This underscores that sedimentation does not influence the interpretation of our data, particularly for the dense systems studied here.

\section*{Author contributions}
R. Azizmalayeri (lead), P. Rostami (equal), T. Witzmann (equal), C. O. Klein (equal), and G. K. Auernhammer (equal) all contributed to: Data curation, Formal analysis, Investigation, Methodology, Software, Validation, Visualization, Writing – original draft, and Writing – review \& editing. 
G. K. Auernhammer additionally contributed to: Conceptualization, Funding acquisition, Project administration, Supervision, and Resources.

\section*{Conflicts of interest}
There are no conflicts of interest to declare.

\section*{Data availability}
The data supporting the findings of this study are provided in the Supplementary Information PDF accompanying the manuscript. 
Further data is available on Zenodo research repository at \href{https://doi.org/10.5281/zenodo.14587081}{https://doi.org/10.5281/zenodo.14587081}.

\section*{Acknowledgements}
This research was funded by the Deutsche Forschungsgemeinschaft (DFG) under Project No. 265191195–SFB 1194, “Interaction between Transport and Wetting Processes,” particularly through projects A06 and A02.


\balance


\bibliography{rsc} 

@article{Luding:2008ab,
	address = {{$[$}Luding, Stefan{$]$} UTwente, CTW, TS, NL-7500 AE Enschede, Netherlands. {$[$}Luding, Stefan{$]$} Delft Univ Technol, TNW, DelftChem Tech, NL-2628 BL Delft, Netherlands. Luding, S (reprint author), UTwente, CTW, TS, POB 217, NL-7500 AE Enschede, Netherlands s.luding@utwente.nl},
	an = {WOS:000256257700002},
	author = {Luding, S.},
	da = {Jun},
	doi = {10.1007/s10035-008-0099-x},
	id = {8976},
	isbn = {1434-5021},
	j2 = {Granul. Matter},
	journal = {Granular Matter},
	la = {English},
	m3 = {Article; Proceedings Paper},
	number = {4},
	pages = {235-246},
	st = {Cohesive, frictional powders: contact models for tension},
	title = {Cohesive, frictional powders: contact models for tension},
	ty = {JOUR},
	url = {<Go to ISI>://WOS:000256257700002},
	volume = {10},
	year = {2008}}

@article{Seto:2012aa,
	author = {Seto, Ryohei and Botet, Robert and Auernhammer, G{\"u}nter K. and Briesen, Heiko},
	c7 = {128},
	da = {2012/12/12},
	doi = {10.1140/epje/i2012-12128-4},
	id = {3474},
	isbn = {1292-8941},
	j2 = {Eur. Phys. J. E},
	journal = {Eur. Phys. J. E},
	la = {English},
	lb = {seto2012},
	number = {12},
	pages = {1-12},
	st = {Restructuring of colloidal aggregates in shear flow},
	title = {Restructuring of colloidal aggregates in shear flow},
	ty = {JOUR},
	url = {http://dx.doi.org/10.1140/epje/i2012-12128-4},
	volume = {35},
	year = {2012}}

@article{Cappella:1999aa,
	author = {Cappella, B. and Dietler, G.},
	booktitle = {Surf. Sci. Rep.},
	doi = {10.1016/S0167-5729(99)00003-5},
	id = {1309},
	isbn = {0167-5729},
	lb = {cappella1999},
	number = {1-3},
	pages = {1-104},
	st = {Force-distance curves by atomic force microscopy},
	title = {Force-distance curves by atomic force microscopy},
	ty = {JOUR},
	url = {http://www.sciencedirect.com/science/article/B6TVY-3XSJYCF-1/2/36c9a40233a5703475e1917f68af4281},
	volume = {34},
	year = {1999}}

@inbook{Goodwin:2004ag,
	author = {Goodwin, Jim W.},
	doi = {10.1002/0470093919.ch3},
	id = {4400},
	isbn = {9780470093917},
	journal = {Colloids and Interfaces with Surfactants and Polymers --An Introduction},
	pages = {61-93},
	publisher = {John Wiley \& Sons, Ltd},
	st = {Interactions between Colloidal Particles},
	title = {Interactions between Colloidal Particles},
	ty = {CHAP},
	url = {http://dx.doi.org/10.1002/0470093919.ch3},
	year = {2004}}

@article{Dagastine:2002aa,
	author = {Dagastine, Raymond R. and Prieve, Dennis C. and White, Lee R.},
	da = {5/1/},
	doi = {http://dx.doi.org/10.1006/jcis.2002.8239},
	id = {4133},
	isbn = {0021-9797},
	journal = {Journal of Colloid and Interface Science},
	number = {1},
	pages = {78-83},
	st = {Calculations of van der Waals Forces in 2-Dimensionally Anisotropic Materials and Its Application to Carbon Black},
	title = {Calculations of van der Waals Forces in 2-Dimensionally Anisotropic Materials and Its Application to Carbon Black},
	ty = {JOUR},
	url = {http://www.sciencedirect.com/science/article/pii/S0021979702982391},
	volume = {249},
	year = {2002}}

@article{Hamaker:1937aa,
	author = {Hamaker, H. C.},
	id = {2073},
	isbn = {0031-8914},
	journal = {Physica},
	m3 = {doi: DOI: 10.1016/S0031-8914(37)80203-7},
	number = {10},
	pages = {1058-1072},
	st = {The London--van der Waals attraction between spherical particles},
	title = {The London--van der Waals attraction between spherical particles},
	ty = {JOUR},
	url = {http://www.sciencedirect.com/science/article/B6X42-4KYW41H-P/2/35bd0b8b7cf3091b9bebc6cf31ef9974},
	volume = {4},
	year = {1937}}

@article{Krumpfer:2011ab,
	address = {{$[$}Krumpfer, JW; McCarthy, TJ{$]$} Univ Massachusetts, Polymer Sci \& Engn Dept, Amherst, MA 01003 USA. McCarthy, TJ (reprint author), Univ Massachusetts, Polymer Sci & Engn Dept, Amherst, MA 01003 USA tmccarthy@polysci.umass.edu},
	an = {WOS:000294790500031},
	author = {Krumpfer, J. W. and McCarthy, T. J.},
	da = {Sep},
	doi = {10.1021/la202583w},
	id = {8953},
	isbn = {0743-7463},
	j2 = {Langmuir},
	journal = {Langmuir},
	la = {English},
	m3 = {Article},
	number = {18},
	pages = {11514-11519},
	st = {Rediscovering Silicones: "Unreactive" Silicones React with Inorganic Surfaces},
	title = {Rediscovering Silicones: "Unreactive" Silicones React with Inorganic Surfaces},
	ty = {JOUR},
	url = {<Go to ISI>://WOS:000294790500031},
	volume = {27},
	year = {2011}}

@article{Butt:2005aa,
	author = {Butt, H. J. and Cappella, B. and Kappl, M.},
	booktitle = {Surf. Sci. Rep.},
	id = {1072},
	lb = {butt2005},
	pages = {1 -- 152},
	st = {Force measurements with the atomic force microscope: Technique, interpretation and applications},
	title = {Force measurements with the atomic force microscope: Technique, interpretation and applications},
	ty = {JOUR},
	volume = {59},
	year = {2005}}

@article{Tomas:2004aa,
	author = {Tomas, J{\"u}rgen},
	da = {2004/10/01},
	doi = {10.1007/s10035-004-0167-9},
	id = {4501},
	isbn = {1434-5021},
	j2 = {GM},
	journal = {Granular Matter},
	la = {English},
	lb = {tomas2004GranMatt},
	number = {2-3},
	pages = {75-86},
	st = {Fundamentals of cohesive powder consolidation and flow},
	title = {Fundamentals of cohesive powder consolidation and flow},
	ty = {JOUR},
	url = {http://dx.doi.org/10.1007/s10035-004-0167-9},
	volume = {6},
	year = {2004}}

@article{Wenzl:2013aa,
	author = {Wenzl, Jennifer and Seto, Ryohei and Roth, Marcel and Butt, Hans-J{\"u}rgen and Auernhammer, G{\"u}nterK},
	da = {2013/08/01},
	doi = {10.1007/s10035-012-0383-7},
	id = {4222},
	isbn = {1434-5021},
	j2 = {Granular Matter},
	journal = {Granular Matter},
	la = {English},
	lb = {wenzl2013a},
	number = {4},
	pages = {391-400},
	st = {Measurement of rotation of individual spherical particles in cohesive granulates},
	title = {Measurement of rotation of individual spherical particles in cohesive granulates},
	ty = {JOUR},
	url = {http://dx.doi.org/10.1007/s10035-012-0383-7},
	volume = {15},
	year = {2013}}

@article{Liu:1998aa,
	author = {Liu, A. L. and Nagel, S. R.},
	id = {1142},
	journal = {Nature},
	lb = {liu1998},
	pages = {21 -- 22},
	st = {Nonlinear dynamics - Jamming is not just cool any more},
	title = {Nonlinear dynamics - Jamming is not just cool any more},
	ty = {JOUR},
	volume = {396},
	year = {1998}}

@article{Huh_Mason_1977,
	author = {Huh, C. and Mason, S. G.},
	doi = {10.1017/S0022112077002134},
	journal = {Journal of Fluid Mechanics},
	number = {3},
	pages = {401--419},
	title = {The steady movement of a liquid meniscus in a capillary tube},
	volume = {81},
	year = {1977}}

@article{lu2016critical,
	author = {Lu, Gui and Wang, Xiao-Dong and Duan, Yuan-Yuan},
	journal = {Advances in colloid and interface science},
	pages = {43--62},
	publisher = {Elsevier},
	title = {A critical review of dynamic wetting by complex fluids: from Newtonian fluids to non-Newtonian fluids and nanofluids},
	volume = {236},
	year = {2016}}

@article{sudheer2023intertwined,
	author = {Sudheer, MVR and Yadav, Preeti and Thomas, Bincy and Ghosh, Udita U},
	journal = {The European Physical Journal Special Topics},
	number = {6},
	pages = {769--780},
	publisher = {Springer},
	title = {Intertwined roles of fluid--solid interactions and macroscopic flow geometry in dynamic wetting of complex fluids},
	volume = {232},
	year = {2023}}

@article{fuaad2021simulation,
	author = {Fuaad, PA and Swerin, Agne and Lundell, Fredrik and Toivakka, Martti},
	journal = {Journal of Coatings Technology and Research},
	pages = {1--10},
	publisher = {Springer},
	title = {Simulation of slot-coating of nanocellulosic material subject to a wall-stress dependent slip-velocity at die-walls},
	year = {2021}}

@article{m2017linking,
	author = {M'barki, Amin and Bocquet, Lyd{\'e}ric and Stevenson, Adam},
	journal = {Scientific reports},
	number = {1},
	pages = {6017},
	publisher = {Nature Publishing Group UK London},
	title = {Linking rheology and printability for dense and strong ceramics by direct ink writing},
	volume = {7},
	year = {2017}}

@article{timm2012investigation,
	author = {Timm, K and Myant, C and Nuguid, H and Spikes, HA and Grunze, Michael},
	journal = {International journal of cosmetic science},
	number = {5},
	pages = {458--465},
	publisher = {Wiley Online Library},
	title = {Investigation of friction and perceived skin feel after application of suspensions of various cosmetic powders},
	volume = {34},
	year = {2012}}

@article{karlsson2019characterization,
	author = {Karlsson, Mikael CF and {\'A}lvarez-Asencio, Rub{\'e}n and Bordes, Romain and Larsson, Anders and Taylor, Phil and Steenari, Britt-Marie},
	journal = {Journal of Coatings Technology and Research},
	pages = {607--614},
	publisher = {Springer},
	title = {Characterization of paint formulated using secondary TiO 2 pigments recovered from waste paint},
	volume = {16},
	year = {2019}}

@article{jeong2022dip,
	author = {Jeong, Deok-Hoon and Lee, Michael Ka Ho and Thi{\'e}venaz, Virgile and Bazant, Martin Z and Sauret, Alban},
	journal = {Journal of Fluid Mechanics},
	pages = {A36},
	publisher = {Cambridge University Press},
	title = {Dip coating of bidisperse particulate suspensions},
	volume = {936},
	year = {2022}}

@article{jorgensen2020deformation,
	author = {J{\o}rgensen, Loren and Forterre, Yo{\"e}l and Lhuissier, Henri},
	journal = {Journal of Fluid Mechanics},
	pages = {R2},
	publisher = {Cambridge University Press},
	title = {Deformation upon impact of a concentrated suspension drop},
	volume = {896},
	year = {2020}}

@article{huh1971hydrodynamic,
	author = {Huh, Chun and Scriven, Laurence E},
	journal = {Journal of colloid and interface science},
	number = {1},
	pages = {85--101},
	publisher = {Elsevier},
	title = {Hydrodynamic model of steady movement of a solid/liquid/fluid contact line},
	volume = {35},
	year = {1971}}

@article{moffatt1964viscous,
	author = {Moffatt, H Keith},
	journal = {Journal of Fluid Mechanics},
	number = {1},
	pages = {1--18},
	publisher = {Cambridge University Press},
	title = {Viscous and resistive eddies near a sharp corner},
	volume = {18},
	year = {1964}}

@article{dussan1979spreading,
	author = {Dussan, EB},
	journal = {Annual Review of Fluid Mechanics},
	number = {1},
	pages = {371--400},
	publisher = {Annual Reviews 4139 El Camino Way, PO Box 10139, Palo Alto, CA 94303-0139, USA},
	title = {On the spreading of liquids on solid surfaces: static and dynamic contact lines},
	volume = {11},
	year = {1979}}

@article{snoeijer2013moving,
	author = {Snoeijer, Jacco H and Andreotti, Bruno},
	journal = {Annual review of fluid mechanics},
	pages = {269--292},
	publisher = {Annual Reviews},
	title = {Moving contact lines: scales, regimes, and dynamical transitions},
	volume = {45},
	year = {2013}}

@article{li2023kinetic,
	author = {Li, Xiaomei and Bodziony, Francisco and Yin, Mariana and Marschall, Holger and Berger, R{\"u}diger and Butt, Hans-J{\"u}rgen},
	journal = {Nature Communications},
	number = {1},
	pages = {4571},
	publisher = {Nature Publishing Group UK London},
	title = {Kinetic drop friction},
	volume = {14},
	year = {2023}}

@article{grewal2015role,
	author = {Grewal, HS and Nam Kim, Hong and Cho, Il-Joo and Yoon, Eui-Sung},
	journal = {Scientific reports},
	number = {1},
	pages = {14159},
	publisher = {Nature Publishing Group UK London},
	title = {Role of viscous dissipative processes on the wetting of textured surfaces},
	volume = {5},
	year = {2015}}

@article{zhao2020spreading,
	author = {Zhao, Menghua and Ol{\'e}ron, Mathieu and Pelosse, Alice and Limat, Laurent and Guazzelli, Elisabeth and Roch{\'e}, Matthieu},
	journal = {Physical Review Research},
	number = {2},
	pages = {022031},
	publisher = {APS},
	title = {Spreading of granular suspensions on a solid surface},
	volume = {2},
	year = {2020}}

@article{pelosse2023probing,
	author = {Pelosse, Alice and Guazzelli, {\'E}lisabeth and Roch{\'e}, Matthieu},
	journal = {Journal of Fluid Mechanics},
	pages = {A7},
	publisher = {Cambridge University Press},
	title = {Probing dissipation in spreading drops with granular suspensions},
	volume = {955},
	year = {2023}}

@article{cox1986dynamics,
	author = {Cox, RG},
	journal = {Journal of fluid mechanics},
	pages = {169--194},
	publisher = {Cambridge University Press},
	title = {The dynamics of the spreading of liquids on a solid surface. Part 1. Viscous flow},
	volume = {168},
	year = {1986}}

@article{voinov1976hydrodynamics,
	author = {Voinov, OV},
	journal = {Fluid dynamics},
	number = {5},
	pages = {714--721},
	publisher = {Springer},
	title = {Hydrodynamics of wetting},
	volume = {11},
	year = {1976}}

@article{hocking1983spreading,
	author = {Hocking, LM},
	journal = {The Quarterly Journal of Mechanics and Applied Mathematics},
	number = {1},
	pages = {55--69},
	publisher = {Oxford University Press},
	title = {The spreading of a thin drop by gravity and capillarity},
	volume = {36},
	year = {1983}}

@article{butt2022contact,
	author = {Butt, Hans-J{\"u}rgen and Liu, Jie and Koynov, Kaloian and Straub, Benedikt and Hinduja, Chirag and Roismann, Ilia and Berger, R{\"u}diger and Li, Xiaomei and Vollmer, Doris and Steffen, Werner and others},
	journal = {Current Opinion in Colloid \& Interface Science},
	pages = {101574},
	publisher = {Elsevier},
	title = {Contact angle hysteresis},
	volume = {59},
	year = {2022}}

@article{henrich2016influence,
	author = {Henrich, Franziska and Fell, Daniela and Truszkowska, Dorota and Weirich, Marcel and Anyfantakis, Manos and Nguyen, Thi-Huong and Wagner, Manfred and Auernhammer, G{\"u}nter K and Butt, Hans-J{\"u}rgen},
	journal = {Soft Matter},
	number = {37},
	pages = {7782--7791},
	publisher = {Royal Society of Chemistry},
	title = {Influence of surfactants in forced dynamic dewetting},
	volume = {12},
	year = {2016}}

@article{Han2012Spreading,
	author = {Han, Jeongin and Kim, Chongyoup},
	doi = {10.1021/la204256z},
	eprint = {https://doi.org/10.1021/la204256z},
	journal = {Langmuir},
	note = {PMID: 22214216},
	number = {5},
	pages = {2680-2689},
	title = {Spreading of a Suspension Drop on a Horizontal Surface},
	url = {https://doi.org/10.1021/la204256z},
	volume = {28},
	year = {2012}}

@article{gans2019dip,
	author = {Gans, Adrien and Dressaire, Emilie and Colnet, B{\'e}n{\'e}dicte and Saingier, Guillaume and Bazant, Martin Z and Sauret, Alban},
	journal = {Soft matter},
	number = {2},
	pages = {252--261},
	publisher = {Royal Society of Chemistry},
	title = {Dip-coating of suspensions},
	volume = {15},
	year = {2019}}

@article{palma2019dip,
	author = {Palma, Sergio and Lhuissier, Henri},
	journal = {Journal of Fluid Mechanics},
	pages = {R3},
	publisher = {Cambridge University Press},
	title = {Dip-coating with a particulate suspension},
	volume = {869},
	year = {2019}}

@article{zhao2018rolling,
	author = {Zhao, Chuang and Li, Chengbo and Hu, Lin},
	journal = {Physica A: Statistical Mechanics and its Applications},
	pages = {181--191},
	publisher = {Elsevier},
	title = {Rolling and sliding between non-spherical particles},
	volume = {492},
	year = {2018}}

@article{santos2020granular,
	author = {Santos, Andrew Pablo and Bolintineanu, Dan S and Grest, Gary S and Lechman, Jeremy B and Plimpton, Steven J and Srivastava, Ishan and Silbert, Leonardo E},
	journal = {Physical Review E},
	number = {3},
	pages = {032903},
	publisher = {APS},
	title = {Granular packings with sliding, rolling, and twisting friction},
	volume = {102},
	year = {2020}}

@article{fuchs2014rolling,
	author = {Fuchs, Regina and Weinhart, Thomas and Meyer, Jan and Zhuang, Hao and Staedler, Thorsten and Jiang, Xin and Luding, Stefan},
	journal = {Granular matter},
	pages = {281--297},
	publisher = {Springer},
	title = {Rolling, sliding and torsion of micron-sized silica particles: experimental, numerical and theoretical analysis},
	volume = {16},
	year = {2014}}

@book{butt2018surface,
	author = {Butt, Hans-J{\"u}rgen and Kappl, Michael},
	publisher = {John Wiley \& Sons},
	title = {Surface and interfacial forces},
	year = {2018}}

@incollection{RANDALL20012733,
	address = {Oxford},
	author = {C. Randall and J. {Van Tassel}},
	booktitle = {Encyclopedia of Materials: Science and Technology},
	doi = {https://doi.org/10.1016/B0-08-043152-6/00487-3},
	editor = {K.H. J{\"u}rgen Buschow and Robert W. Cahn and Merton C. Flemings and Bernhard Ilschner and Edward J. Kramer and Subhash Mahajan and Patrick Veyssi{\`e}re},
	isbn = {978-0-08-043152-9},
	pages = {2733-2738},
	publisher = {Elsevier},
	title = {Electrophoretic Deposition},
	url = {https://www.sciencedirect.com/science/article/pii/B0080431526004873},
	year = {2001}}

@article{yadav2007effective,
	author = {Yadav, Preeti S and Dupre, Derek and Tadmor, Rafael and Park, Jennifer Shim and Katoshevski, David},
	journal = {Surface science},
	number = {19},
	pages = {4582--4585},
	publisher = {Elsevier},
	title = {Effective refractive index and intermolecular forces associated with a phase of functional groups},
	volume = {601},
	year = {2007}}

@article{wozniak2009highly,
	author = {Wozniak, Maciej and Graule, Thomas and de Hazan, Yoram and Kata, Dariusz and Lis, Jerzy},
	journal = {Journal of the European Ceramic Society},
	number = {11},
	pages = {2259--2265},
	publisher = {Elsevier},
	title = {Highly loaded UV curable nanosilica dispersions for rapid prototyping applications},
	volume = {29},
	year = {2009}}

@article{wozniak2011rheology,
	author = {Wozniak, Maciej and de Hazan, Yoram and Graule, Thomas and Kata, Dariusz},
	journal = {Journal of the European Ceramic Society},
	number = {13},
	pages = {2221--2229},
	publisher = {Elsevier},
	title = {Rheology of UV curable colloidal silica dispersions for rapid prototyping applications},
	volume = {31},
	year = {2011}}

@article{hopkins2014dielectric,
	author = {Hopkins, Jaime C and Dryden, Daniel M and Ching, Wai-Yim and French, Roger H and Parsegian, V Adrian and Podgornik, Rudolf},
	journal = {Journal of colloid and interface science},
	pages = {278--284},
	publisher = {Elsevier},
	title = {Dielectric response variation and the strength of van der Waals interactions},
	volume = {417},
	year = {2014}}

@article{guazzelli2018rheology,
	author = {Guazzelli, {\'E}lisabeth and Pouliquen, Olivier},
	journal = {Journal of Fluid Mechanics},
	pages = {P1},
	publisher = {Cambridge University Press},
	title = {Rheology of dense granular suspensions},
	volume = {852},
	year = {2018}}

@article{morris2020toward,
	author = {Morris, Jeffrey F},
	journal = {Physical Review Fluids},
	number = {11},
	pages = {110519},
	publisher = {APS},
	title = {Toward a fluid mechanics of suspensions},
	volume = {5},
	year = {2020}}

@article{trappe2001jamming,
	author = {Trappe, Veronique and Prasad, V and Cipelletti, Luca and Segre, PN and Weitz, David A},
	journal = {Nature},
	number = {6839},
	pages = {772--775},
	publisher = {Nature Publishing Group UK London},
	title = {Jamming phase diagram for attractive particles},
	volume = {411},
	year = {2001}}

@article{seto2013discontinuous,
	author = {Seto, Ryohei and Mari, Romain and Morris, Jeffrey F and Denn, Morton M},
	journal = {Physical review letters},
	number = {21},
	pages = {218301},
	publisher = {APS},
	title = {Discontinuous shear thickening of frictional hard-sphere suspensions},
	volume = {111},
	year = {2013}}

@book{mezger2020rheology,
	author = {Mezger, Thomas},
	publisher = {European Coatings},
	title = {The rheology handbook: for users of rotational and oscillatory rheometers},
	year = {2020}}

@article{otsuki2018coupling,
	author = {Otsuki, Akira},
	journal = {Electrophoresis},
	number = {5-6},
	pages = {690--701},
	publisher = {Wiley Online Library},
	title = {Coupling colloidal forces with yield stress of charged inorganic particle suspension: A review},
	volume = {39},
	year = {2018}}

@article{li2020numerical,
	author = {Li, Meng-Ge and Feng, Feng and Wu, Wei-Tao and Massoudi, Mehrdad},
	journal = {Energies},
	number = {24},
	pages = {6635},
	publisher = {MDPI},
	title = {Numerical simulations of the flow of a dense suspension exhibiting yield-stress and shear-thinning effects},
	volume = {13},
	year = {2020}}

@article{10.1063/5.0153614,
    author = {Lee, Yu-Fan and Whitcomb, Kevin and Wagner, Norman J.},
    title = "{Microstructure and rheology of shear thickening colloidal suspensions under transient flows}",
    journal = {Physics of Fluids},
    volume = {35},
    number = {7},
    pages = {073308},
    year = {2023},
    month = {07},
    issn = {1070-6631},
    doi = {10.1063/5.0153614},
    url = {https://doi.org/10.1063/5.0153614},
    eprint = {https://pubs.aip.org/aip/pof/article-pdf/doi/10.1063/5.0153614/18034188/073308\_1\_5.0153614.pdf},
}

@article{wiederseiner2011refractive,
	author = {Wiederseiner, S{\'e}bastien and Andreini, Nicolas and Epely-Chauvin, Ga{\"e}l and Ancey, Christophe},
	journal = {Experiments in fluids},
	pages = {1183--1206},
	publisher = {Springer},
	title = {Refractive-index and density matching in concentrated particle suspensions: a review},
	volume = {50},
	year = {2011}}

@article{wright2017review,
	author = {Wright, Stuart F and Zadrazil, Ivan and Markides, Christos N},
	journal = {Experiments in Fluids},
	pages = {1--39},
	publisher = {Springer},
	title = {A review of solid--fluid selection options for optical-based measurements in single-phase liquid, two-phase liquid--liquid and multiphase solid--liquid flows},
	volume = {58},
	year = {2017}}

@article{auernhammer2020transparent,
	author = {Auernhammer, G{\"u}nter K and Fataei, Shirin and Haustein, Martin A and Patel, Himanshu P and Schwarze, R{\"u}diger and Secrieru, Egor and Mechtcherine, Viktor},
	journal = {Materials \& Design},
	pages = {108673},
	publisher = {Elsevier},
	title = {Transparent model concrete with tunable rheology for investigating flow and particle-migration during transport in pipes},
	volume = {193},
	year = {2020}}

@article{straub2021flow,
	author = {Straub, Benedikt B and Schmidt, Henrik and Rostami, Peyman and Henrich, Franziska and Rossi, Massimiliano and K{\"a}hler, Christian J and Butt, Hans-J{\"u}rgen and Auernhammer, G{\"u}nter K},
	journal = {Soft Matter},
	number = {44},
	pages = {10090--10100},
	publisher = {Royal Society of Chemistry},
	title = {Flow profiles near receding three-phase contact lines: influence of surfactants},
	volume = {17},
	year = {2021}}

@article{rossi2014optimization,
	author = {Rossi, Massimiliano and K{\"a}hler, Christian J},
	journal = {Experiments in fluids},
	pages = {1--13},
	publisher = {Springer},
	title = {Optimization of astigmatic particle tracking velocimeters},
	volume = {55},
	year = {2014}}

@article{cierpka2010calibration,
	author = {Cierpka, Christian and Rossi, Massimiliano and Segura, Rodrigo and K{\"a}hler, Christian J},
	journal = {Measurement Science and Technology},
	number = {1},
	pages = {015401},
	publisher = {IOP Publishing},
	title = {On the calibration of astigmatism particle tracking velocimetry for microflows},
	volume = {22},
	year = {2010}}

@article{comtet2017pairwise,
	author = {Comtet, Jean and Chatt{\'e}, Guillaume and Nigues, Antoine and Bocquet, Lyd{\'e}ric and Siria, Alessandro and Colin, Annie},
	journal = {Nature communications},
	number = {1},
	pages = {15633},
	publisher = {Nature Publishing Group UK London},
	title = {Pairwise frictional profile between particles determines discontinuous shear thickening transition in non-colloidal suspensions},
	volume = {8},
	year = {2017}}

@article{Al-Hashemil2018review,
	author = {Al-Hashemi, Hamzah M Beakawi and Al-Amoudi, Omar S Baghabra},
	journal = {Powder technology},
	pages = {397--417},
	publisher = {Elsevier},
	title = {A review on the angle of repose of granular materials},
	volume = {330},
	year = {2018}}

@article{clavaud2017revealing,
	author = {Clavaud, C{\'e}cile and B{\'e}rut, Antoine and Metzger, Bloen and Forterre, Yo{\"e}l},
	journal = {Proceedings of the National Academy of Sciences},
	number = {20},
	pages = {5147--5152},
	publisher = {National Acad Sciences},
	title = {Revealing the frictional transition in shear-thickening suspensions},
	volume = {114},
	year = {2017}}

@article{hutter1993calibration,
	author = {Hutter, Jeffrey L and Bechhoefer, John},
	journal = {Review of scientific instruments},
	number = {7},
	pages = {1868--1873},
	publisher = {American Institute of Physics},
	title = {Calibration of atomic-force microscope tips},
	volume = {64},
	year = {1993}}

@article{eifert2014simple,
	author = {Eifert, Alexander and Paulssen, Dorothea and Varanakkottu, Subramanyan Namboodiri and Baier, Tobias and Hardt, Steffen},
	journal = {Advanced Materials Interfaces},
	number = {3},
	pages = {1300138},
	publisher = {Wiley Online Library},
	title = {Simple fabrication of robust water-repellent surfaces with low contact-angle hysteresis based on impregnation},
	volume = {1},
	year = {2014}}

@article{tomas2009energy,
	author = {Tomas, J},
	journal = {Particulate Science and Technology},
	number = {4},
	pages = {337--351},
	publisher = {Taylor \& Francis},
	title = {Energy absorption at particle contact, compression, and shear flow of dry ultrafine powder},
	volume = {27},
	year = {2009}}

@article{gallier2014rheology,
	author = {Gallier, Stany and Lemaire, Elisabeth and Peters, Fran{\c{c}}ois and Lobry, Laurent},
	journal = {Journal of Fluid Mechanics},
	pages = {514--549},
	publisher = {Cambridge University Press},
	title = {Rheology of sheared suspensions of rough frictional particles},
	volume = {757},
	year = {2014}}

@article{trulsson2017effect,
	author = {Trulsson, Martin and DeGiuli, Eric and Wyart, Matthieu},
	journal = {Physical Review E},
	number = {1},
	pages = {012605},
	publisher = {APS},
	title = {Effect of friction on dense suspension flows of hard particles},
	volume = {95},
	year = {2017}}

@article{melrose2004continuous,
	author = {Melrose, John R and Ball, Robin C},
	journal = {Journal of Rheology},
	number = {5},
	pages = {937--960},
	publisher = {The Society of Rheology},
	title = {Continuous shear thickening transitions in model concentrated colloids---The role of interparticle forces},
	volume = {48},
	year = {2004}}

@article{wagner2009shear,
	author = {Wagner, Norman J and Brady, John F},
	journal = {Physics Today},
	number = {10},
	pages = {27--32},
	publisher = {American Institute of Physics},
	title = {Shear thickening in colloidal dispersions},
	volume = {62},
	year = {2009}}

@article{fricke2020boundary,
  title={Boundary conditions for dynamic wetting-A mathematical analysis},
  author={Fricke, Mathis and Bothe, Dieter},
  journal={The European Physical Journal Special Topics},
  volume={229},
  number={10},
  pages={1849--1865},
  year={2020},
  publisher={Springer}
}

@article{ancey1999theoretical,
  title={A theoretical framework for granular suspensions in a steady simple shear flow},
  author={Ancey, Christophe and Coussot, Philippe and Evesque, Pierre},
  journal={Journal of Rheology},
  volume={43},
  number={6},
  pages={1673--1699},
  year={1999},
  publisher={The Society of Rheology}
}

@book{andreotti2013granular,
  title={Granular media: between fluid and solid},
  author={Andreotti, Bruno and Forterre, Yo{\"e}l and Pouliquen, Olivier},
  year={2013},
  publisher={Cambridge University Press}
}

@article{troncoso2014nanoscale,
  title={Nanoscale adhesive forces between silica surfaces in aqueous solutions},
  author={Troncoso, Paula and Saavedra, Jorge H and Acu{\~n}a, Sergio M and Jeldres, Ricardo and Concha, Fernando and Toledo, Pedro G},
  journal={Journal of colloid and interface science},
  volume={424},
  pages={56--61},
  year={2014},
  publisher={Elsevier}
}

@article{valmacco2016dispersion,
  title={Dispersion forces acting between silica particles across water: influence of nanoscale roughness},
  author={Valmacco, Valentina and Elzbieciak-Wodka, Magdalena and Besnard, C{\'e}line and Maroni, Plinio and Trefalt, Gregor and Borkovec, Michal},
  journal={Nanoscale Horizons},
  volume={1},
  number={4},
  pages={325--330},
  year={2016a},
  publisher={Royal Society of Chemistry}
}

@article{valmacco2016forces,
  title={Forces between silica particles in the presence of multivalent cations},
  author={Valmacco, Valentina and Elzbieciak-Wodka, Magdalena and Herman, David and Trefalt, Gregor and Maroni, Plinio and Borkovec, Michal},
  journal={Journal of colloid and interface science},
  volume={472},
  pages={108--115},
  year={2016b},
  publisher={Elsevier}
}

@article{wang2013atomic,
  title={Atomic force microscopy study of forces between a silica sphere and an oxidized silicon wafer in aqueous solutions of NaCl, KCl, and CsCl at concentrations up to saturation},
  author={Wang, Yuhua and Wang, Liguang and Hampton, Marc A and Nguyen, Anh V},
  journal={The Journal of Physical Chemistry C},
  volume={117},
  number={5},
  pages={2113--2120},
  year={2013},
  publisher={ACS Publications}
}

@article{chateau2018pinch,
  title={Pinch-off of a viscous suspension thread},
  author={Ch{\^a}teau, Joris and Guazzelli, {\'E}lisabeth and Lhuissier, Henri},
  journal={Journal of Fluid Mechanics},
  volume={852},
  pages={178--198},
  year={2018},
  publisher={Cambridge University Press}
}

@article{roche2024complexity,
  title={Complexity in wetting dynamics},
  author={Roch{\'e}, Matthieu and Talini, Laurence and Verneuil, Emilie},
  journal={Langmuir},
  volume={40},
  number={6},
  pages={2830--2848},
  year={2024},
  publisher={ACS Publications}
}

@article{bonn2009wetting,
  title={Wetting and spreading},
  author={Bonn, Daniel and Eggers, Jens and Indekeu, Joseph and Meunier, Jacques and Rolley, Etienne},
  journal={Reviews of modern physics},
  volume={81},
  number={2},
  pages={739--805},
  year={2009},
  publisher={APS}
}

@article{luu2009drop,
  title={Drop impact of yield-stress fluids},
  author={Luu, Li-Hua and Forterre, Yo{\"e}l},
  journal={Journal of Fluid Mechanics},
  volume={632},
  pages={301--327},
  year={2009},
  publisher={Cambridge University Press}
}

@article{rostami2024spreading,
  title={Spreading of a viscoelastic drop on a solid substrate},
  author={Rostami, Peyman and Fricke, Mathis and Schubotz, Simon and Patel, Himanshu and Azizmalayeri, Reza and Auernhammer, G{\"u}nter K},
  journal={Journal of Fluid Mechanics},
  volume={988},
  pages={A51},
  year={2024},
  publisher={Cambridge University Press}
}

@article{fall2009yield,
  title={Yield stress and shear banding in granular suspensions},
  author={Fall, Abdoulaye and Bertrand, Fran{\c{c}}ois and Ovarlez, Guillaume and Bonn, Daniel},
  journal={Physical review letters},
  volume={103},
  number={17},
  pages={178301},
  year={2009},
  publisher={APS}
}

@article{singh2023hidden,
  title={Hidden hierarchy in the rheology of dense suspensions},
  author={Singh, Abhinendra},
  journal={MRS Communications},
  volume={13},
  number={6},
  pages={971--979},
  year={2023},
  publisher={Springer}
}

@article{doan2023interactive,
  title={Interactive role of rolling friction and cohesion on the angle of repose through a microscale assessment},
  author={Doan, Thao and Indraratna, Buddhima and Nguyen, Thanh T and Rujikiatkamjorn, Cholachat},
  journal={International Journal of Geomechanics},
  volume={23},
  number={1},
  pages={04022250},
  year={2023},
  publisher={American Society of Civil Engineers}
}

@article{matsuo2014geometric,
  title={Geometric effect of angle of repose revisited},
  author={Matsuo, Miki Y and Nishiura, Daisuke and Sakaguchi, Hide},
  journal={Granular Matter},
  volume={16},
  pages={441--447},
  year={2014},
  publisher={Springer}
}

@article{hribar2019electrostatic,
  title={Electrostatic interactions between barium hexaferrite nanoplatelets in alcohol suspensions},
  author={Hribar Bo{\v{s}}tjan{\v{c}}i{\v{c}}, Patricija and Tom{\v{s}}i{\v{c}}, Matija and Jamnik, Andrej and Lisjak, Darja and Mertelj, Alenka},
  journal={The Journal of Physical Chemistry C},
  volume={123},
  number={37},
  pages={23272--23279},
  year={2019},
  publisher={ACS Publications}
}

@article{royer2016rheological,
  title={Rheological signature of frictional interactions in shear thickening suspensions},
  author={Royer, John R and Blair, Daniel L and Hudson, Steven D},
  journal={Physical review letters},
  volume={116},
  number={18},
  pages={188301},
  year={2016},
  publisher={APS}
}

@article{yu2012boundary,
  title={The boundary lubrication of chemically grafted and cross-linked hyaluronic acid in phosphate buffered saline and lipid solutions measured by the surface forces apparatus},
  author={Yu, Jing and Banquy, Xavier and Greene, George W and Lowrey, Daniel D and Israelachvili, Jacob N},
  journal={Langmuir},
  volume={28},
  number={4},
  pages={2244--2250},
  year={2012},
  publisher={ACS Publications}
}

@book{hunter2013zeta,
  title={Zeta potential in colloid science: principles and applications},
  author={Hunter, Robert J},
  volume={2},
  year={2013},
  publisher={Academic press}
}

@article{morrisProgressChallengesSuspension2023,
  title = {Progress and Challenges in Suspension Rheology},
  author = {Morris, Jeffrey F.},
  year = {2023},
  month = nov,
  journal = {Rheologica Acta},
  volume = {62},
  number = {11-12},
  pages = {617--629},
  issn = {0035-4511, 1435-1528},
  doi = {10.1007/s00397-023-01421-z},
  urldate = {2025-02-10},
  langid = {english}
}

@article{dinicPinchoffDynamicsDrippingontosubstrate2017,
  title = {Pinch-off Dynamics and Dripping-onto-Substrate ({{DoS}}) Rheometry of Complex Fluids},
  author = {Dinic, Jelena and Jimenez, Leidy Nallely and Sharma, Vivek},
  year = {2017},
  month = jan,
  journal = {Lab on a Chip},
  volume = {17},
  number = {3},
  pages = {460--473},
  publisher = {The Royal Society of Chemistry},
  issn = {1473-0189},
  doi = {10.1039/C6LC01155A},
  urldate = {2025-02-10},
  langid = {english}
}

@article{thievenazOnsetHeterogeneityPinchoff2022,
  title = {The Onset of Heterogeneity in the Pinch-off of Suspension Drops},
  author = {Thi{\'e}venaz, Virgile and Sauret, Alban},
  year = {2022},
  month = mar,
  journal = {Proceedings of the National Academy of Sciences},
  volume = {119},
  number = {13},
  pages = {e2120893119},
  issn = {0027-8424, 1091-6490},
  doi = {10.1073/pnas.2120893119},
  urldate = {2025-02-10},
  langid = {english}
}

@article{alexandrouBreakupCapillaryBridge2010,
  title = {Breakup of a Capillary Bridge of Suspensions},
  author = {Alexandrou, A. N. and Bazilevskii, A. V. and Entov, V. M. and Rozhkov, A. N. and Sharaf, A.},
  year = {2010},
  month = dec,
  journal = {Fluid Dynamics},
  volume = {45},
  number = {6},
  pages = {952--964},
  issn = {1573-8507},
  doi = {10.1134/S001546281006013X},
  urldate = {2025-02-10},
  langid = {english}
}

@article{bonnoitAcceleratedDropDetachment2012,
  title = {Accelerated Drop Detachment in Granular Suspensions},
  author = {Bonnoit, Claire and Bertrand, Thibault and Cl{\'e}ment, Eric and Lindner, Anke},
  year = {2012},
  month = apr,
  journal = {Physics of Fluids},
  volume = {24},
  number = {4},
  pages = {043304},
  issn = {1070-6631},
  doi = {10.1063/1.4704801},
  urldate = {2025-02-10}
}

@article{louvetNonuniversalityPinchOffYield2014,
  title = {Nonuniversality in the {{Pinch-Off}} of {{Yield Stress Fluids}}: {{Role}} of {{Nonlocal Rheology}}},
  shorttitle = {Nonuniversality in the {{Pinch-Off}} of {{Yield Stress Fluids}}},
  author = {Louvet, Nicolas and Bonn, Daniel and Kellay, Hamid},
  year = {2014},
  month = nov,
  journal = {Physical Review Letters},
  volume = {113},
  number = {21},
  pages = {218302},
  issn = {0031-9007, 1079-7114},
  doi = {10.1103/PhysRevLett.113.218302},
  urldate = {2025-06-25},
  copyright = {http://link.aps.org/licenses/aps-default-license},
  langid = {english}
}

@article{malbrancheScalingAnalysisShear2022,
  title = {Scaling {{Analysis}} of {{Shear Thickening Suspensions}}},
  author = {Malbranche, Nelya and Santra, Aritra and Chakraborty, Bulbul and Morris, Jeffrey F.},
  year = {2022},
  month = jul,
  journal = {Frontiers in Physics},
  volume = {10},
  pages = {946221},
  issn = {2296-424X},
  doi = {10.3389/fphy.2022.946221},
  urldate = {2025-06-26},
  langid = {english}
}

@article{morrisShearThickeningConcentrated2020a,
  title = {Shear {{Thickening}} of {{Concentrated Suspensions}}: {{Recent Developments}} and {{Relation}} to {{Other Phenomena}}},
  shorttitle = {Shear {{Thickening}} of {{Concentrated Suspensions}}},
  author = {Morris, Jeffrey F.},
  year = {2020},
  month = jan,
  journal = {Annual Review of Fluid Mechanics},
  volume = {52},
  number = {1},
  pages = {121--144},
  issn = {0066-4189, 1545-4479},
  doi = {10.1146/annurev-fluid-010816-060128},
  urldate = {2025-06-26},
  langid = {english}
}

@article{guyUnifiedDescriptionRheology2015,
  title = {Towards a {{Unified Description}} of the {{Rheology}} of {{Hard-Particle Suspensions}}},
  author = {Guy, B. M. and Hermes, M. and Poon, W. C. K.},
  year = {2015},
  month = aug,
  journal = {Physical Review Letters},
  volume = {115},
  number = {8},
  pages = {088304},
  issn = {0031-9007, 1079-7114},
  doi = {10.1103/PhysRevLett.115.088304},
  urldate = {2025-06-26},
  copyright = {http://link.aps.org/licenses/aps-default-license},
  langid = {english}
}

@article{atalikOccurrenceEvenHarmonics2004,
  title = {On the Occurrence of Even Harmonics in the Shear Stress Response of Viscoelastic Fluids in Large Amplitude Oscillatory Shear},
  author = {Atal{\i}k, Kunt and Keunings, Roland},
  year = {2004},
  month = sep,
  journal = {Journal of Non-Newtonian Fluid Mechanics},
  volume = {122},
  number = {1-3},
  pages = {107--116},
  issn = {03770257},
  doi = {10.1016/j.jnnfm.2003.11.012},
  urldate = {2025-06-30},
  copyright = {https://www.elsevier.com/tdm/userlicense/1.0/},
  langid = {english}
}

@article{saint-michelLocalOscillatoryRheology2016,
  title = {Local {{Oscillatory Rheology}} from {{Echography}}},
  author = {{Saint-Michel}, Brice and Gibaud, Thomas and Leocmach, Mathieu and Manneville, S{\'e}bastien},
  year = {2016},
  month = mar,
  journal = {Physical Review Applied},
  volume = {5},
  number = {3},
  pages = {034014},
  issn = {2331-7019},
  doi = {10.1103/PhysRevApplied.5.034014},
  urldate = {2025-06-30},
  copyright = {http://link.aps.org/licenses/aps-default-license},
  langid = {english}
}

@article{colosquiHydrodynamicallyDrivenColloidal2013,
  title = {Hydrodynamically {{Driven Colloidal Assembly}} in {{Dip Coating}}},
  author = {Colosqui, Carlos E. and Morris, Jeffrey F. and Stone, Howard A.},
  year = {2013},
  month = apr,
  journal = {Physical Review Letters},
  volume = {110},
  number = {18},
  pages = {188302},
  issn = {0031-9007, 1079-7114},
  doi = {10.1103/PhysRevLett.110.188302},
  urldate = {2025-07-01},
  copyright = {http://link.aps.org/licenses/aps-default-license},
  langid = {english}
}

@article{nessPhysicsDenseSuspensions2022,
  title = {The {{Physics}} of {{Dense Suspensions}}},
  author = {Ness, Christopher and Seto, Ryohei and Mari, Romain},
  year = {2022},
  month = mar,
  journal = {Annual Review of Condensed Matter Physics},
  volume = {13},
  number = {1},
  pages = {97--117},
  issn = {1947-5454, 1947-5462},
  doi = {10.1146/annurev-conmatphys-031620-105938},
  urldate = {2025-07-02},
  langid = {english}
}

@article{srivastavaViscometricFlowDense2021,
  title = {Viscometric Flow of Dense Granular Materials under Controlled Pressure and Shear Stress},
  author = {Srivastava, Ishan and Silbert, Leonardo E. and Grest, Gary S. and Lechman, Jeremy B.},
  year = {2021},
  month = jan,
  journal = {Journal of Fluid Mechanics},
  volume = {907},
  pages = {A18},
  issn = {0022-1120, 1469-7645},
  doi = {10.1017/jfm.2020.811},
  urldate = {2025-07-02},
  langid = {english}
}

@article{setoNormalStressDifferences2018,
  title = {Normal Stress Differences in Dense Suspensions},
  author = {Seto, Ryohei and Giusteri, Giulio G.},
  year = {2018},
  month = dec,
  journal = {Journal of Fluid Mechanics},
  volume = {857},
  pages = {200--215},
  issn = {0022-1120, 1469-7645},
  doi = {10.1017/jfm.2018.743},
  urldate = {2025-07-02},
  langid = {english}
}

@article{westerweelUniversalOutlierDetection2005,
  title = {Universal Outlier Detection for {{PIV}} Data},
  author = {Westerweel, Jerry and Scarano, Fulvio},
  year = {2005},
  month = dec,
  journal = {Experiments in Fluids},
  volume = {39},
  number = {6},
  pages = {1096--1100},
  issn = {0723-4864, 1432-1114},
  doi = {10.1007/s00348-005-0016-6},
  urldate = {2025-03-31},
  copyright = {http://www.springer.com/tdm},
  langid = {english}
}

@article{ahujaTwoStepYielding2020,
  title = {Two Step Yielding in Soft Materials},
  author = {Ahuja, Amit and Potanin, Andrei and Joshi, Yogesh M.},
  year = {2020},
  month = aug,
  journal = {Advances in Colloid and Interface Science},
  volume = {282},
  pages = {102179},
  publisher = {Elsevier BV},
  issn = {0001-8686},
  doi = {10.1016/j.cis.2020.102179},
  urldate = {2025-07-08},
  copyright = {https://www.elsevier.com/tdm/userlicense/1.0/},
  langid = {english}
}

@article{martouzetDynamicArrestSpreading2021,
  title = {Dynamic Arrest during the Spreading of a Yield Stress Fluid Drop},
  author = {Martouzet, Gr{\'e}goire and J{\o}rgensen, Loren and Pelet, Yoann and Biance, Anne-Laure and Barentin, Catherine},
  year = {2021},
  month = apr,
  journal = {Physical Review Fluids},
  volume = {6},
  number = {4},
  pages = {044006},
  issn = {2469-990X},
  doi = {10.1103/PhysRevFluids.6.044006},
  urldate = {2025-02-10},
  langid = {english}
}

@article{maronApplicationReeeyringGeneralized1956,
  title = {Application of Ree-Eyring Generalized Flow Theory to Suspensions of Spherical Particles},
  author = {Maron, Samuel H. and Pierce, Percy E.},
  year = {1956},
  month = feb,
  journal = {Journal of Colloid Science},
  volume = {11},
  number = {1},
  pages = {80--95},
  publisher = {Elsevier BV},
  issn = {0095-8522},
  doi = {10.1016/0095-8522(56)90023-x},
  urldate = {2025-07-12},
  copyright = {https://www.elsevier.com/tdm/userlicense/1.0/},
  langid = {english}
}

@article{setoShearJammingFragility2019,
  title = {Shear Jamming and Fragility in Dense Suspensions},
  author = {Seto, Ryohei and Singh, Abhinendra and Chakraborty, Bulbul and Denn, Morton M. and Morris, Jeffrey F.},
  year = {2019},
  month = aug,
  journal = {Granular Matter},
  volume = {21},
  number = {3},
  pages = {82},
  issn = {1434-5021, 1434-7636},
  doi = {10.1007/s10035-019-0931-5},
  urldate = {2025-02-10},
  langid = {english}
}

@article{houssaisAthermalSedimentCreep2021,
  title = {Athermal Sediment Creep Triggered by Porous Flow},
  author = {Houssais, M. and Maldarelli, Charles and Morris, Jeffrey F.},
  year = {2021},
  month = jan,
  journal = {Physical Review Fluids},
  volume = {6},
  number = {1},
  pages = {},
  publisher = {American Physical Society (APS)},
  issn = {2469-990X},
  doi = {10.1103/physrevfluids.6.l012301},
  urldate = {2025-07-12},
  copyright = {https://link.aps.org/licenses/aps-default-license},
  langid = {english}
}

@article{deboeufParticlePressureSheared2009,
  title = {Particle {{Pressure}} in a {{Sheared Suspension}}: {{A Bridge}} from {{Osmosis}} to {{Granular Dilatancy}}},
  shorttitle = {Particle {{Pressure}} in a {{Sheared Suspension}}},
  author = {Deboeuf, Ang{\'e}lique and Gauthier, Georges and Martin, J{\'e}r{\^o}me and Yurkovetsky, Yevgeny and Morris, Jeffrey F.},
  year = {2009},
  month = mar,
  journal = {Physical Review Letters},
  volume = {102},
  number = {10},
  pages = {},
  publisher = {American Physical Society (APS)},
  issn = {0031-9007, 1079-7114},
  doi = {10.1103/physrevlett.102.108301},
  urldate = {2025-07-12},
  copyright = {http://link.aps.org/licenses/aps-default-license},
  langid = {english}
}

@article{voltzRayleighTaylorInstabilitySedimenting2001,
  title = {Rayleigh-{{Taylor}} Instability in a Sedimenting Suspension},
  author = {V{\"o}ltz, C. and Pesch, W. and Rehberg, I.},
  year = {2001},
  month = dec,
  journal = {Physical Review E},
  volume = {65},
  number = {1},
  pages = {011404},
  issn = {1063-651X, 1095-3787},
  doi = {10.1103/PhysRevE.65.011404},
  urldate = {2025-03-27},
  copyright = {http://link.aps.org/licenses/aps-default-license},
  langid = {english}
}

@article{einsteinBerichtigungMeinerArbeit1911,
  title = {Berichtigung Zu Meiner {{Arbeit}}: ,,{{Eine}} Neue {{Bestimmung}} Der {{Molek{\"u}ldimensionen}}''︁},
  shorttitle = {Berichtigung Zu Meiner {{Arbeit}}},
  author = {Einstein, A.},
  year = {1911},
  month = jan,
  journal = {Annalen der Physik},
  volume = {339},
  number = {3},
  pages = {591--592},
  publisher = {Wiley},
  issn = {0003-3804, 1521-3889},
  doi = {10.1002/andp.19113390313},
  urldate = {2025-07-17},
  copyright = {http://onlinelibrary.wiley.com/termsAndConditions\#vor},
  langid = {english}
}

@article{einsteinNeueBestimmungMolekuldimensionen1906,
  title = {Eine Neue {{Bestimmung}} Der {{Molek{\"u}ldimensionen}}},
  author = {Einstein, A.},
  year = {1906},
  month = jan,
  journal = {Annalen der Physik},
  volume = {324},
  number = {2},
  pages = {289--306},
  publisher = {Wiley},
  issn = {0003-3804, 1521-3889},
  doi = {10.1002/andp.19063240204},
  urldate = {2025-07-17},
  copyright = {http://onlinelibrary.wiley.com/termsAndConditions\#vor},
  langid = {english}
}

@article{batchelorDeterminationBulkStress1972,
  title = {The Determination of the Bulk Stress in a Suspension of Spherical Particles to Order {\emph{C}}{\textsuperscript{2}}},
  author = {Batchelor, G. K. and Green, J. T.},
  year = {1972a},
  month = dec,
  journal = {Journal of Fluid Mechanics},
  volume = {56},
  number = {3},
  pages = {401--427},
  publisher = {Cambridge University Press (CUP)},
  issn = {0022-1120, 1469-7645},
  doi = {10.1017/s0022112072002435},
  urldate = {2025-07-17},
  copyright = {https://www.cambridge.org/core/terms},
  langid = {english}
}

@article{batchelorHydrodynamicInteractionTwo1972,
  title = {The Hydrodynamic Interaction of Two Small Freely-Moving Spheres in a Linear Flow Field},
  author = {Batchelor, G. K. and Green, J. T.},
  year = {1972b},
  month = nov,
  journal = {Journal of Fluid Mechanics},
  volume = {56},
  number = {2},
  pages = {375--400},
  publisher = {Cambridge University Press (CUP)},
  issn = {0022-1120, 1469-7645},
  doi = {10.1017/s0022112072002927},
  urldate = {2025-07-17},
  copyright = {https://www.cambridge.org/core/terms},
  langid = {english}
}

@article{aytounaDropFormationNonNewtonian2013,
  title = {Drop {{Formation}} in {{Non-Newtonian Fluids}}},
  author = {Aytouna, Mounir and Paredes, Jose and {Shahidzadeh-Bonn}, Noushine and Moulinet, S{\'e}bastien and Wagner, Christian and Amarouchene, Yacine and Eggers, Jens and Bonn, Daniel},
  year = {2013},
  month = jan,
  journal = {Physical Review Letters},
  volume = {110},
  number = {3},
  pages = {},
  publisher = {American Physical Society (APS)},
  issn = {0031-9007, 1079-7114},
  doi = {10.1103/physrevlett.110.034501},
  urldate = {2025-07-21},
  copyright = {http://link.aps.org/licenses/aps-default-license},
  langid = {english}
}

@article{morrisReviewMicrostructureConcentrated2009,
  title = {A Review of Microstructure in Concentrated Suspensions and Its Implications for Rheology and Bulk Flow},
  author = {Morris, Jeffrey F.},
  year = {2009},
  month = oct,
  journal = {Rheologica Acta},
  volume = {48},
  number = {8},
  pages = {909--923},
  issn = {0035-4511, 1435-1528},
  doi = {10.1007/s00397-009-0352-1},
  urldate = {2025-09-05},
  copyright = {http://www.springer.com/tdm},
  langid = {english}
}

@article{edensShearStressDependence2021,
  title = {Shear Stress Dependence of Force Networks in {{3D}} Dense Suspensions},
  author = {Edens, Lance E. and Alvarado, Enrique G. and Singh, Abhinendra and Morris, Jeffrey F. and Schenter, Gregory K. and Chun, Jaehun and Clark, Aurora E.},
  year = {2021},
  journal = {Soft Matter},
  volume = {17},
  number = {32},
  pages = {7476--7486},
  issn = {1744-683X, 1744-6848},
  doi = {10.1039/D1SM00184A},
  urldate = {2025-09-05},
  langid = {english}
}

@article{singhStressactivatedConstraintsDense2022,
  title = {Stress-Activated Constraints in Dense Suspension Rheology},
  author = {Singh, Abhinendra and Jackson, Grayson L. and Van Der Naald, Michael and De Pablo, Juan J. and Jaeger, Heinrich M.},
  year = {2022},
  month = may,
  journal = {Physical Review Fluids},
  volume = {7},
  number = {5},
  pages = {054302},
  issn = {2469-990X},
  doi = {10.1103/PhysRevFluids.7.054302},
  urldate = {2025-09-05},
  langid = {english}
}

@article{drelichContactAnglesHistory2020,
  title = {Contact Angles: History of over 200 Years of Open Questions},
  shorttitle = {Contact Angles},
  author = {Drelich, Jaroslaw W and Boinovich, Ludmila and Chibowski, Emil and Della Volpe, Claudio and Ho{\l}ysz, Lucyna and Marmur, Abraham and Siboni, Stefano},
  year = {2020},
  month = feb,
  journal = {Surface Innovations},
  volume = {8},
  number = {1-2},
  pages = {3--27},
  issn = {2050-6252, 2050-6260},
  doi = {10.1680/jsuin.19.00007},
  urldate = {2025-09-27},
  langid = {english}
}

@article{petersDirectObservationDynamic2016,
  title = {Direct Observation of Dynamic Shear Jamming in Dense Suspensions},
  author = {Peters, Ivo R. and Majumdar, Sayantan and Jaeger, Heinrich M.},
  year = 2016,
  month = apr,
  journal = {Nature},
  volume = {532},
  number = {7598},
  pages = {214--217},
  issn = {0028-0836, 1476-4687},
  doi = {10.1038/nature17167},
  urldate = {2025-11-20},
  langid = {english}
}

@article{herschel1926konsistenzmessungen,
  title={Konsistenzmessungen von gummi-benzoll{\"o}sungen},
  author={Herschel, Winslow H and Bulkley, Ronald},
  journal={Kolloid-Zeitschrift},
  volume={39},
  number={4},
  pages={291--300},
  year={1926},
  publisher={Springer}
}

@article{chambonExperimentalInvestigationViscoplastic2014,
  title = {Experimental Investigation of Viscoplastic Free-Surface Flows in a Steady Uniform Regime},
  author = {Chambon, Guillaume and Ghemmour, A. and Naaim, M.},
  date = {2014-09-10},
  journaltitle = {Journal of Fluid Mechanics},
  shortjournal = {J. Fluid Mech.},
  volume = {754},
  pages = {332--364},
  issn = {0022-1120, 1469-7645},
  doi = {10.1017/jfm.2014.378},
  url = {https://www.cambridge.org/core/product/identifier/S0022112014003784/type/journal_article},
  urldate = {2025-11-22},
  langid = {english}
}
\bibliographystyle{rsc} 

\end{document}